\documentclass[longauth]{aa}
\usepackage{euclid}
\usepackage{color}
\usepackage{graphicx}
\usepackage[fleqn]{amsmath}	
\usepackage{amssymb}	
\usepackage{mathtools}
\usepackage{txfonts}
\usepackage{booktabs}
\usepackage{natbib,twoopt}

\usepackage{float}
\usepackage[caption = false]{subfig}
\usepackage[breaklinks=true]{hyperref}
\hypersetup{
  colorlinks=true,   
  citecolor=blue,    
  urlcolor=blue,     
  linkcolor=blue,     
}
\bibpunct{(}{)}{;}{a}{}{,} 
\usepackage{lscape}

\usepackage[switch, modulo]{lineno}
\DeclareRobustCommand{\VAN}[3]{#2}
\let\VANthebibliography\thebibliography
\def\thebibliography{\DeclareRobustCommand{\VAN}[3]{##3}\VANthebibliography}

\usepackage{orcidlink}

\newcommand*{\spr}{SPRITZ}
\newcommand{\qsfit}{QSFIT}

\newcommand{\ebv}{{$E(B-V)~$}}
\newcommand{\fnz}{f(N_{\rm HI}, z)}
\newcommand{\Tl}{T_\lambda}
\newcommand{\mnhi}{N_{\rm HI}}

\newcommand{\mbh}{$M_{\rm BH}$}

\newcommand{\reve}[1]{{ #1}}
\newcommand{\rev}[1]{{ #1}}

\begin{document} 

\title{Euclid preparation.}
\subtitle{Spectroscopy of active galactic nuclei with NISP}
\newcommand{\orcid}[1]{\orcidlink{#1}} 
\author{Euclid Collaboration: E.~Lusso\orcid{0000-0003-0083-1157}$^{1,2}$\thanks{\email{elisabeta.lusso@unifi.it}}, S.~Fotopoulou\orcid{0000-0002-9686-254X}$^{3}$, M.~Selwood$^{3}$, V.~Allevato$^{4}$, G.~Calderone\orcid{0000-0002-7738-5389}$^{5}$, C.~Mancini\orcid{0000-0002-4297-0561}$^{6}$, M.~Mignoli\orcid{0000-0002-9087-2835}$^{7}$, M.~Scodeggio$^{6}$, L.~Bisigello\orcid{0000-0003-0492-4924}$^{8,9}$, A.~Feltre\orcid{0000-0003-0492-4924}$^{1}$
% ...
}

\institute{$^{1}$ INAF-Osservatorio Astrofisico di Arcetri, Largo E. Fermi 5, 50125, Firenze, Italy\\
$^{2}$ Dipartimento di Fisica e Astronomia, Universit\`{a} di Firenze, via G. Sansone 1, 50019 Sesto Fiorentino, Firenze, Italy\\
$^{3}$ School of Physics, HH Wills Physics Laboratory, University of Bristol, Tyndall Avenue, Bristol, BS8 1TL, UK\\
$^{4}$ INAF-Osservatorio Astronomico di Capodimonte, Via Moiariello 16, 80131 Napoli, Italy\\
$^{5}$ INAF-Osservatorio Astronomico di Trieste, Via G. B. Tiepolo 11, 34143 Trieste, Italy\\
$^{6}$ INAF-IASF Milano, Via Alfonso Corti 12, 20133 Milano, Italy\\
$^{7}$ INAF-Osservatorio di Astrofisica e Scienza dello Spazio di Bologna, Via Piero Gobetti 93/3, 40129 Bologna, Italy\\
$^{8}$ Dipartimento di Fisica e Astronomia "G. Galilei", Universit\`a di Padova, Via Marzolo 8, 35131 Padova, Italy\\
$^{9}$ INAF-Osservatorio Astronomico di Padova, Via dell'Osservatorio 5, 35122 Padova, Italy\\
$^{10}$ Department of Mathematics and Physics, Roma Tre University, Via della Vasca Navale 84, 00146 Rome, Italy\\
$^{11}$ INAF-Osservatorio Astronomico di Roma, Via Frascati 33, 00078 Monteporzio Catone, Italy\\
$^{12}$ INFN-Padova, Via Marzolo 8, 35131 Padova, Italy\\
$^{13}$ Aix-Marseille Universit\'e, CNRS, CNES, LAM, Marseille, France\\
$^{14}$ Dipartimento di Fisica e Astronomia "Augusto Righi" - Alma Mater Studiorum Universit\`a di Bologna, via Piero Gobetti 93/2, 40129 Bologna, Italy\\
$^{15}$ Max-Planck-Institut f\"ur Astronomie, K\"onigstuhl 17, 69117 Heidelberg, Germany\\
$^{16}$ Instituto de Astrof\'isica e Ci\^encias do Espa\c{c}o, Universidade do Porto, CAUP, Rua das Estrelas, PT4150-762 Porto, Portugal\\
$^{17}$ DTx -- Digital Transformation CoLAB, Building 1, Azur\'em Campus, University of Minho, 4800-058 Guimar\~aes, Portugal\\
$^{18}$ Department of Physics, Centre for Extragalactic Astronomy, Durham University, South Road, DH1 3LE, UK\\
$^{19}$ SISSA, International School for Advanced Studies, Via Bonomea 265, 34136 Trieste TS, Italy\\
$^{20}$ INFN, Sezione di Trieste, Via Valerio 2, 34127 Trieste TS, Italy\\
$^{21}$ Max Planck Institute for Extraterrestrial Physics, Giessenbachstr. 1, 85748 Garching, Germany\\
$^{22}$ Department of Physics and Astronomy, University of Southampton, Southampton, SO17 1BJ, UK\\
$^{23}$ INAF-Istituto di Astrofisica e Planetologia Spaziali, via del Fosso del Cavaliere, 100, 00100 Roma, Italy\\
$^{24}$ Jet Propulsion Laboratory, California Institute of Technology, 4800 Oak Grove Drive, Pasadena, CA, 91109, USA\\
$^{25}$ School of Physical Sciences, The Open University, Milton Keynes, MK7 6AA, UK\\
$^{26}$ Universit\'e Paris-Saclay, CNRS, Institut d'astrophysique spatiale, 91405, Orsay, France\\
$^{27}$ ESAC/ESA, Camino Bajo del Castillo, s/n., Urb. Villafranca del Castillo, 28692 Villanueva de la Ca\~nada, Madrid, Spain\\
$^{28}$ Institute of Cosmology and Gravitation, University of Portsmouth, Portsmouth PO1 3FX, UK\\
$^{29}$ INAF-Osservatorio Astronomico di Brera, Via Brera 28, 20122 Milano, Italy\\
$^{30}$ Aix-Marseille Universit\'e, CNRS/IN2P3, CPPM, Marseille, France\\
$^{31}$ Dipartimento di Fisica e Astronomia, Universit\`a di Bologna, Via Gobetti 93/2, 40129 Bologna, Italy\\
$^{32}$ INFN-Sezione di Bologna, Viale Berti Pichat 6/2, 40127 Bologna, Italy\\
$^{33}$ Universit\"ats-Sternwarte M\"unchen, Fakult\"at f\"ur Physik, Ludwig-Maximilians-Universit\"at M\"unchen, Scheinerstrasse 1, 81679 M\"unchen, Germany\\
$^{34}$ INAF-Osservatorio Astrofisico di Torino, Via Osservatorio 20, 10025 Pino Torinese (TO), Italy\\
$^{35}$ Dipartimento di Fisica, Universit\`a di Genova, Via Dodecaneso 33, 16146, Genova, Italy\\
$^{36}$ INFN-Sezione di Genova, Via Dodecaneso 33, 16146, Genova, Italy\\
$^{37}$ Department of Physics "E. Pancini", University Federico II, Via Cinthia 6, 80126, Napoli, Italy\\
$^{38}$ INFN section of Naples, Via Cinthia 6, 80126, Napoli, Italy\\
$^{39}$ Dipartimento di Fisica, Universit\`a degli Studi di Torino, Via P. Giuria 1, 10125 Torino, Italy\\
$^{40}$ INFN-Sezione di Torino, Via P. Giuria 1, 10125 Torino, Italy\\
$^{41}$ Institut de F\'{i}sica d'Altes Energies (IFAE), The Barcelona Institute of Science and Technology, Campus UAB, 08193 Bellaterra (Barcelona), Spain\\
$^{42}$ Port d'Informaci\'{o} Cient\'{i}fica, Campus UAB, C. Albareda s/n, 08193 Bellaterra (Barcelona), Spain\\
$^{43}$ Institute for Theoretical Particle Physics and Cosmology (TTK), RWTH Aachen University, 52056 Aachen, Germany\\
$^{44}$ Dipartimento di Fisica e Astronomia "Augusto Righi" - Alma Mater Studiorum Universit\`a di Bologna, Viale Berti Pichat 6/2, 40127 Bologna, Italy\\
$^{45}$ Institute for Astronomy, University of Edinburgh, Royal Observatory, Blackford Hill, Edinburgh EH9 3HJ, UK\\
$^{46}$ Jodrell Bank Centre for Astrophysics, Department of Physics and Astronomy, University of Manchester, Oxford Road, Manchester M13 9PL, UK\\
$^{47}$ European Space Agency/ESRIN, Largo Galileo Galilei 1, 00044 Frascati, Roma, Italy\\
$^{48}$ University of Lyon, Univ Claude Bernard Lyon 1, CNRS/IN2P3, IP2I Lyon, UMR 5822, 69622 Villeurbanne, France\\
$^{49}$ Institute of Physics, Laboratory of Astrophysics, Ecole Polytechnique F\'ed\'erale de Lausanne (EPFL), Observatoire de Sauverny, 1290 Versoix, Switzerland\\
$^{50}$ UCB Lyon 1, CNRS/IN2P3, IUF, IP2I Lyon, 4 rue Enrico Fermi, 69622 Villeurbanne, France\\
$^{51}$ Instituto de Astrof\'isica e Ci\^encias do Espa\c{c}o, Faculdade de Ci\^encias, Universidade de Lisboa, Campo Grande, 1749-016 Lisboa, Portugal\\
$^{52}$ Departamento de F\'isica, Faculdade de Ci\^encias, Universidade de Lisboa, Edif\'icio C8, Campo Grande, PT1749-016 Lisboa, Portugal\\
$^{53}$ Department of Astronomy, University of Geneva, ch. d'Ecogia 16, 1290 Versoix, Switzerland\\
$^{54}$ Department of Physics, Oxford University, Keble Road, Oxford OX1 3RH, UK\\
$^{55}$ Universit\'e Paris-Saclay, Universit\'e Paris Cit\'e, CEA, CNRS, AIM, 91191, Gif-sur-Yvette, France\\
$^{56}$ Istituto Nazionale di Fisica Nucleare, Sezione di Bologna, Via Irnerio 46, 40126 Bologna, Italy\\
$^{57}$ University Observatory, Faculty of Physics, Ludwig-Maximilians-Universit{\"a}t, Scheinerstr. 1, 81679 Munich, Germany\\
$^{58}$ Institute of Theoretical Astrophysics, University of Oslo, P.O. Box 1029 Blindern, 0315 Oslo, Norway\\
$^{59}$ Department of Physics, Lancaster University, Lancaster, LA1 4YB, UK\\
$^{60}$ von Hoerner \& Sulger GmbH, Schlo{\ss}Platz 8, 68723 Schwetzingen, Germany\\
$^{61}$ Technical University of Denmark, Elektrovej 327, 2800 Kgs. Lyngby, Denmark\\
$^{62}$ Cosmic Dawn Center (DAWN), Denmark\\
$^{63}$ Department of Physics and Helsinki Institute of Physics, Gustaf H\"allstr\"omin katu 2, 00014 University of Helsinki, Finland\\
$^{64}$ Universit\'e de Gen\`eve, D\'epartement de Physique Th\'eorique and Centre for Astroparticle Physics, 24 quai Ernest-Ansermet, CH-1211 Gen\`eve 4, Switzerland\\
$^{65}$ Department of Physics, P.O. Box 64, 00014 University of Helsinki, Finland\\
$^{66}$ Helsinki Institute of Physics, Gustaf H{\"a}llstr{\"o}min katu 2, University of Helsinki, Helsinki, Finland\\
$^{67}$ NOVA optical infrared instrumentation group at ASTRON, Oude Hoogeveensedijk 4, 7991PD, Dwingeloo, The Netherlands\\
$^{68}$ Universit\"at Bonn, Argelander-Institut f\"ur Astronomie, Auf dem H\"ugel 71, 53121 Bonn, Germany\\
$^{69}$ Department of Physics, Institute for Computational Cosmology, Durham University, South Road, DH1 3LE, UK\\
$^{70}$ Universit\'e Paris Cit\'e, CNRS, Astroparticule et Cosmologie, 75013 Paris, France\\
$^{71}$ Institut d'Astrophysique de Paris, 98bis Boulevard Arago, 75014, Paris, France\\
$^{72}$ Institut d'Astrophysique de Paris, UMR 7095, CNRS, and Sorbonne Universit\'e, 98 bis boulevard Arago, 75014 Paris, France\\
$^{73}$ European Space Agency/ESTEC, Keplerlaan 1, 2201 AZ Noordwijk, The Netherlands\\
$^{74}$ Department of Physics and Astronomy, University of Aarhus, Ny Munkegade 120, DK-8000 Aarhus C, Denmark\\
$^{75}$ Centre for Astrophysics, University of Waterloo, Waterloo, Ontario N2L 3G1, Canada\\
$^{76}$ Department of Physics and Astronomy, University of Waterloo, Waterloo, Ontario N2L 3G1, Canada\\
$^{77}$ Perimeter Institute for Theoretical Physics, Waterloo, Ontario N2L 2Y5, Canada\\
$^{78}$ Universit\'e Paris-Saclay, Universit\'e Paris Cit\'e, CEA, CNRS, Astrophysique, Instrumentation et Mod\'elisation Paris-Saclay, 91191 Gif-sur-Yvette, France\\
$^{79}$ Space Science Data Center, Italian Space Agency, via del Politecnico snc, 00133 Roma, Italy\\
$^{80}$ Centre National d'Etudes Spatiales -- Centre spatial de Toulouse, 18 avenue Edouard Belin, 31401 Toulouse Cedex 9, France\\
$^{81}$ Institute of Space Science, Str. Atomistilor, nr. 409 M\u{a}gurele, Ilfov, 077125, Romania\\
$^{82}$ Instituto de Astrof\'isica de Canarias, Calle V\'ia L\'actea s/n, 38204, San Crist\'obal de La Laguna, Tenerife, Spain\\
$^{83}$ Departamento de Astrof\'isica, Universidad de La Laguna, 38206, La Laguna, Tenerife, Spain\\
$^{84}$ Departamento de F\'isica, FCFM, Universidad de Chile, Blanco Encalada 2008, Santiago, Chile\\
$^{85}$ Institut d'Estudis Espacials de Catalunya (IEEC), Carrer Gran Capit\'a 2-4, 08034 Barcelona, Spain\\
$^{86}$ Institute of Space Sciences (ICE, CSIC), Campus UAB, Carrer de Can Magrans, s/n, 08193 Barcelona, Spain\\
$^{87}$ Satlantis, University Science Park, Sede Bld 48940, Leioa-Bilbao, Spain\\
$^{88}$ Centro de Investigaciones Energ\'eticas, Medioambientales y Tecnol\'ogicas (CIEMAT), Avenida Complutense 40, 28040 Madrid, Spain\\
$^{89}$ Infrared Processing and Analysis Center, California Institute of Technology, Pasadena, CA 91125, USA\\
$^{90}$ Instituto de Astrof\'isica e Ci\^encias do Espa\c{c}o, Faculdade de Ci\^encias, Universidade de Lisboa, Tapada da Ajuda, 1349-018 Lisboa, Portugal\\
$^{91}$ Universidad Polit\'ecnica de Cartagena, Departamento de Electr\'onica y Tecnolog\'ia de Computadoras,  Plaza del Hospital 1, 30202 Cartagena, Spain\\
$^{92}$ Institut de Recherche en Astrophysique et Plan\'etologie (IRAP), Universit\'e de Toulouse, CNRS, UPS, CNES, 14 Av. Edouard Belin, 31400 Toulouse, France\\
$^{93}$ Kapteyn Astronomical Institute, University of Groningen, PO Box 800, 9700 AV Groningen, The Netherlands\\
$^{94}$ INFN-Bologna, Via Irnerio 46, 40126 Bologna, Italy\\
$^{95}$ IFPU, Institute for Fundamental Physics of the Universe, via Beirut 2, 34151 Trieste, Italy\\
$^{96}$ INAF, Istituto di Radioastronomia, Via Piero Gobetti 101, 40129 Bologna, Italy\\
$^{97}$ Centre de Calcul de l'IN2P3/CNRS, 21 avenue Pierre de Coubertin 69627 Villeurbanne Cedex, France\\
$^{98}$ Institut f\"ur Theoretische Physik, University of Heidelberg, Philosophenweg 16, 69120 Heidelberg, Germany\\
$^{99}$ Universit\'e St Joseph; Faculty of Sciences, Beirut, Lebanon\\
$^{100}$ Junia, EPA department, 41 Bd Vauban, 59800 Lille, France\\
$^{101}$ ICSC - Centro Nazionale di Ricerca in High Performance Computing, Big Data e Quantum Computing, Via Magnanelli 2, Bologna, Italy\\
$^{102}$ Instituto de F\'isica Te\'orica UAM-CSIC, Campus de Cantoblanco, 28049 Madrid, Spain\\
$^{103}$ CERCA/ISO, Department of Physics, Case Western Reserve University, 10900 Euclid Avenue, Cleveland, OH 44106, USA\\
$^{104}$ Laboratoire Univers et Th\'eorie, Observatoire de Paris, Universit\'e PSL, Universit\'e Paris Cit\'e, CNRS, 92190 Meudon, France\\
$^{105}$ Dipartimento di Fisica e Scienze della Terra, Universit\`a degli Studi di Ferrara, Via Giuseppe Saragat 1, 44122 Ferrara, Italy\\
$^{106}$ Istituto Nazionale di Fisica Nucleare, Sezione di Ferrara, Via Giuseppe Saragat 1, 44122 Ferrara, Italy\\
$^{107}$ Universit\'e de Strasbourg, CNRS, Observatoire astronomique de Strasbourg, UMR 7550, 67000 Strasbourg, France\\
$^{108}$ Dipartimento di Fisica - Sezione di Astronomia, Universit\`a di Trieste, Via Tiepolo 11, 34131 Trieste, Italy\\
$^{109}$ NASA Ames Research Center, Moffett Field, CA 94035, USA\\
$^{110}$ Kavli Institute for Particle Astrophysics \& Cosmology (KIPAC), Stanford University, Stanford, CA 94305, USA\\
$^{111}$ Bay Area Environmental Research Institute, Moffett Field, California 94035, USA\\
$^{112}$ Minnesota Institute for Astrophysics, University of Minnesota, 116 Church St SE, Minneapolis, MN 55455, USA\\
$^{113}$ Universit\'e C\^{o}te d'Azur, Observatoire de la C\^{o}te d'Azur, CNRS, Laboratoire Lagrange, Bd de l'Observatoire, CS 34229, 06304 Nice cedex 4, France\\
$^{114}$ Institute Lorentz, Leiden University, PO Box 9506, Leiden 2300 RA, The Netherlands\\
$^{115}$ Institute for Astronomy, University of Hawaii, 2680 Woodlawn Drive, Honolulu, HI 96822, USA\\
$^{116}$ Department of Physics \& Astronomy, University of California Irvine, Irvine CA 92697, USA\\
$^{117}$ Department of Astronomy \& Physics and Institute for Computational Astrophysics, Saint Mary's University, 923 Robie Street, Halifax, Nova Scotia, B3H 3C3, Canada\\
$^{118}$ Departamento F\'isica Aplicada, Universidad Polit\'ecnica de Cartagena, Campus Muralla del Mar, 30202 Cartagena, Murcia, Spain\\
$^{119}$ Department of Computer Science, Aalto University, PO Box 15400, Espoo, FI-00 076, Finland\\
$^{120}$ Ruhr University Bochum, Faculty of Physics and Astronomy, Astronomical Institute (AIRUB), German Centre for Cosmological Lensing (GCCL), 44780 Bochum, Germany\\
$^{121}$ Univ. Grenoble Alpes, CNRS, Grenoble INP, LPSC-IN2P3, 53, Avenue des Martyrs, 38000, Grenoble, France\\
$^{122}$ Department of Physics and Astronomy, Vesilinnantie 5, 20014 University of Turku, Finland\\
$^{123}$ Serco for European Space Agency (ESA), Camino bajo del Castillo, s/n, Urbanizacion Villafranca del Castillo, Villanueva de la Ca\~nada, 28692 Madrid, Spain\\
$^{124}$ ARC Centre of Excellence for Dark Matter Particle Physics, Melbourne, Australia\\
$^{125}$ Centre for Astrophysics \& Supercomputing, Swinburne University of Technology, Victoria 3122, Australia\\
$^{126}$ W.M. Keck Observatory, 65-1120 Mamalahoa Hwy, Kamuela, HI, USA\\
$^{127}$ Oskar Klein Centre for Cosmoparticle Physics, Department of Physics, Stockholm University, Stockholm, SE-106 91, Sweden\\
$^{128}$ Astrophysics Group, Blackett Laboratory, Imperial College London, London SW7 2AZ, UK\\
$^{129}$ Dipartimento di Fisica, Sapienza Universit\`a di Roma, Piazzale Aldo Moro 2, 00185 Roma, Italy\\
$^{130}$ INFN-Sezione di Roma, Piazzale Aldo Moro, 2 - c/o Dipartimento di Fisica, Edificio G. Marconi, 00185 Roma, Italy\\
$^{131}$ Centro de Astrof\'{\i}sica da Universidade do Porto, Rua das Estrelas, 4150-762 Porto, Portugal\\
$^{132}$ Zentrum f\"ur Astronomie, Universit\"at Heidelberg, Philosophenweg 12, 69120 Heidelberg, Germany\\
$^{133}$ Dipartimento di Fisica, Universit\`a di Roma Tor Vergata, Via della Ricerca Scientifica 1, Roma, Italy\\
$^{134}$ INFN, Sezione di Roma 2, Via della Ricerca Scientifica 1, Roma, Italy\\
$^{135}$ Institute for Computational Science, University of Zurich, Winterthurerstrasse 190, 8057 Zurich, Switzerland\\
$^{136}$ ASTRON, the Netherlands Institute for Radio Astronomy, Postbus 2, 7990 AA, Dwingeloo, The Netherlands\\
$^{137}$ Anton Pannekoek Institute for Astronomy, University of Amsterdam, Postbus 94249, 1090 GE Amsterdam, The Netherlands\\
$^{138}$ Mullard Space Science Laboratory, University College London, Holmbury St Mary, Dorking, Surrey RH5 6NT, UK\\
$^{139}$ Department of Physics and Astronomy, University of California, Davis, CA 95616, USA\\
$^{140}$ Department of Astrophysical Sciences, Peyton Hall, Princeton University, Princeton, NJ 08544, USA\\
$^{141}$ Niels Bohr Institute, University of Copenhagen, Jagtvej 128, 2200 Copenhagen, Denmark\\
$^{142}$ Cosmic Dawn Center (DAWN)}
\date{\today}

\abstract{
The statistical distribution and evolution of key properties of active galactic nuclei (AGN), such as their accretion rate, mass, or spin, remain an open debate in astrophysics. The ESA \Euclid space mission, launched on July 1$^{\rm st}$ 2023, promises a breakthrough in this field. We create detailed mock catalogues of AGN spectra, from the rest-frame near-infrared down to the ultraviolet, including emission lines, to simulate what \Euclid will observe for both obscured (type 2) and unobscured (type 1) AGN. We concentrate on the red grisms of the NISP instrument, which will be used for the wide-field survey, opening a new window for spectroscopic AGN studies in the near-infrared. We quantify the efficiency in the redshift determination as well as in retrieving the emission line flux of the \ion{H}{$\alpha$}+[\ion{N}{ii}] complex as \Euclid is mainly focused on this emission line as it is expected to be the brightest one in the probed redshift range. 
Spectroscopic redshifts are measured for 83\% of the simulated AGN in the interval where the \ion{H}{$\alpha$} is visible (i.e., $0.89<z<1.83$ at a line flux $>2\times10^{-16}$\,erg s$^{-1}$ cm$^{-2}$, encompassing the peak of AGN activity at $z\simeq1-1.5$) within the spectral coverage of the red grism. Outside this redshift range, the measurement efficiency decreases significantly. 
Overall, a spectroscopic redshift is \reve{correctly} determined for about 90\% of type 2 AGN down to an emission line flux of roughly $3\times10^{-16}$\,erg s$^{-1}$ cm$^{-2}$, and for type 1 AGN down to $8.5\times10^{-16}$\,erg s$^{-1}$ cm$^{-2}$. Recovered black hole mass values show a small offset with respect to the input values by about 10\%, but the agreement is good overall.
With such a high spectroscopic coverage at $z<2$, we will be able to measure AGN demography, scaling relations, and clustering from the epoch of the peak of AGN activity down to the present-day Universe for hundreds of thousand AGN with homogeneous spectroscopic information.
}


\authorrunning{Euclid Collaboration}
\titlerunning{E. Lusso et al.}
\maketitle
\section{Introduction}
\label{Introduction}
Active galactic nuclei (AGN) are signposts of accretion of matter onto a supermassive black hole (SMBH) located at the centre of the majority of galaxies, and their activity appears to be one of the key mechanisms needed to quench star formation in massive galaxies \citep[e.g.,][]{gabor2010}, to reproduce observed galaxy properties (e.g., SMBH and stellar mass function, \citealt{granato2004,shankar2006,croton2006a,croton2006b}), and to circumvent overproduction of very massive galaxies in cosmological simulations \citep[e.g.,][]{ward2022}. Yet, the detailed conditions under which the feedback process initiates and is delivered to the host galaxy remain a topic of active research.
Moreover, the determination of the black hole mass for a very large sample of AGN can provide further hints on the idea that dark matter could also contribute to the SMBH growth \citep{delaurentis2022}. 
Among the many challenges associated with AGN studies is the fact that they are rare sources \citep[e.g.,][]{mullaney2012} and there is currently no universal selection criterion to identify them \citep[][]{lacy2004,stern2005,donley2012,stern2012,Kirkpatrick2012,berta2013,ji2022}, which often leads to controversial results.

\Euclid, the European Space Agency mission launched on July 1$^{\rm st}$ 2023 \citep{Laureijs11}, will observe $\sim$15\,000\,deg$^2$ of the extragalactic sky in the optical with VIS (the visible imager, \citealt{cropper2016}) and in the near-infrared (NIR) with NISP (the Near-Infrared Spectrometer and Photometer, \citealt{Maciaszek_2022}). 
About ten million AGN are expected to be observable with the \Euclid imager (Selwood et al., in preparation; Bisigello et al., in preparation), \reve{including over 100 high-redshift quasars at $7.0 < z < 7.5$, $\sim$25 quasars beyond $z \simeq 7$, and $\sim$8 
\citep{Barnett-EP5}. From this sample, several hundred thousand AGN will be observed spectroscopically across a wide redshift range, which will represent the largest AGN sample with NIR spectroscopy to-date.  }
Here we present detailed mocks of AGN spectra, from the rest-frame NIR up to the UV, inclusive of emission lines, to simulate what \Euclid will observe for both obscured (type 2) and unobscured (type 1) AGN. When properly analysed, they can reveal key properties of SMBHs, and our aim is to show that the foreseen quality of the \Euclid spectral data will be sufficient to extract such information.

In this paper, we forecast the performance of the NISP red grism regarding the observable parameters for the spectroscopically identified AGN with \Euclid.
The NISP's photometric channel  \citep{Maciaszek2016} will cover the 0.95--2.02\,$\mu$m range to a 5\,$\sigma$ point-source median depth of 24.4 AB mag \citep{Schirmer-EP18}. A spectrum will be extracted \reve{for} each 5\,$\sigma$ detection position with the NISP slitless spectrometer through two `red' grisms, i.e., RGS000 and RGS180, that can be tilted by $-4^{\circ}$ and $+4^{\circ}$ respectively to obtain four RGS orientations in total. The RGS cover the wavelength range 1.21--1.89 \,$\mu$m with a spectral resolution $R=450$ for a 0\farcs5 diameter source and an expected unresolved line flux at a 3.5\,$\sigma$ sensitivity of $2\times10^{-16}$\,erg s$^{-1}$ cm$^{-2}$ \citep{Scaramella-EP1,Paterson-EP32}. NISP is also equipped with an additional `blue' grism (0.93--1.37\,$\mu$m) at a similar spectral resolution \citep[see][for technical details on the NISP spectrograph]{Costille2018}, which will not be employed for the wide survey, \reve{but it will be utilised for calibration purposes in the deep one}.
We present the spectral properties (mainly focussing on the \ion{H}{$\alpha$} emission line) of the expected AGN spectra that will be observed by the NISP red grisms, which will open a new window for spectroscopic studies in the NIR. We created mock AGN spectra, from the rest-frame NIR to the UV, as expected to be observed by \Euclid for both type 2 and type 1 AGN. 

For type 1 AGN, we generated average (empirical) spectra in intervals of redshift, rest-frame equivalent width (EW), and full-width at half-maximum (FWHM), starting from broad-line (FWHM\,$>$\, 2000 km s$^{-1}$) blue quasars where \rev{we conservatively excluded} radio bright and broad absorption line (BAL) objects. As a result, the type 1 AGN sample is not \rev{fully} representative of the entire AGN population, but rather of its blue tail. \rev{Nonetheless, BALs would have to have very large equivalent widths \citep[e.g.,][]{hall2002} to impact on the final stack of type 1 AGN, thus most (classical) BALs should thus be accounted for in our templates. A very small population of broad iron absorption lines quasars (FeLoBAL) is missed, but this AGN class represents an already minor ($\sim$\,10\%) AGN sub-population \citep[e.g.,][]{McGraw2015,choi2022}, which tends to show strong reddening in the optical \citep[e.g.,][]{Villforth2019}. Likewise, radio-bright AGN spectral properties (e.g., continuum and emission lines) are similar to the radio-quiet ones in the bands we considered here, with the main differences arising at X-ray and, of course, radio energies, which are both not investigated in the present analysis.

Regarding the type 2 AGN, we build a set of (semi-empirical) templates following \citet{Bisigello2021}. We started from different host galaxies (e.g. elliptical, star forming, and star-burst) and emission line components where the number density of type 2 AGN class is derived from the observed galaxy luminosity function. Therefore, the final type 2 AGN sample that we considered to construct the stacks, has a realistic number of objects at different redshifts and luminosities.

Whilst we defer the reader to the relevant sections for details, we note that these two approaches are quite different, but motivated by the fact that, in this work, we are not aiming at recovering the AGN variety and classification starting from the NISP spectra, but rather to understand what the NISP performances are in terms of retrieving the emission line flux of a variety of AGN, and estimating the black hole mass for the type 1 AGN.}

The paper is structured as follows. We introduce the procedure utilised to build the empirical and semi-empirical AGN spectra for type 1 and type 2 AGN in Sect.~\ref{sec:spec_creation_1} and Sect.~\ref{sec:spec_creation_2}, respectively. Section~\ref{Spectra simulation} presents the simulation utilised to create the \Euclid NISP spectra from the incident AGN samples, whilst the efficiency and purity of the redshift measurements for the simulated AGN spectra is presented in Sect.~\ref{Redshift measurements}. Section~\ref{sec:analysisqsfit} details the spectral fitting procedure employed to compute the spectral properties.
Finally, we discuss the expected observable parameters and resolution limitations for the spectroscopically identified AGN with \Euclid in Sect.~\ref{sec:discussion}. \reve{All the type 1 and type 2 AGN templates built in this paper will be publicly available at the following link \url{https://drive.google.com/drive/folders/1tjryMUkHhD10NjH_lhfeEBnvYRY4uib7?usp=sharing}}.
Whenever luminosity values are reported, we have assumed a standard flat $\Lambda$CDM cosmology with $\Omega_{\rm m}=0.3$ and $H_0=70$\,km s$^{-1}$\,Mpc$^{-1}$.

\section{Empirical template spectra for type 1 AGN}\label{sec:spec_creation_1}
In this Section, we describe how the empirical spectra for type 1 AGN are built. We created a set of composite spectra over a wide range of redshifts ($0.3<z<6$) and emission line properties, which are then utilised to simulate \Euclid spectra.   

\subsection{Sample selection} \label{sample_selection_typeI}
The quasar sample we considered to construct the empirical incident AGN templates has been built following a similar approach as the one described in \citet[][]{LR16}. We started with the catalogue of quasar properties presented by \citet{shen2011}, which contains 105\,783 spectroscopically confirmed broad-line quasars, optically selected from the Sloan Digital Sky Survey Data Release 7 (SDSS DR7). 
We \rev{conservatively} removed from this catalogue all sources flagged as BAL quasars (i.e., sources with BAL\_FLAG=0 are non-BALs) and radio-loud (i.e., $F_{\nu}(6\,{\rm cm})/F_{\nu}(2500\,\AA)\geq10$, this removed 8257 quasars, or 8\% of the main SDSS sample). 
BAL quasars are often found in galaxies with red optical and UV colours, with broad absorption troughs (associated to the presence of winds or outflows) that could distort the continuum significantly. Radio bright objects seems to show steeper spectra than the ones for radio-quiet AGN \citep[e.g.,][]{zheng1997}, whilst other studies have found that the observed continuum slopes are flatter (redder) compared to the composite spectrum for typical (radio faint) SDSS quasars \citep[][]{kuzmicz2021}. 

Our selection resulted in 91\,732 SDSS quasars. We further excluded 136 quasars classified as BALs by \citet{gibson2009} and 17 quasars considered radio-loud in the catalogue published by \citet{mingo2016}, which is the largest available Mid-Infrared (WISE), X-ray (3XMM) and Radio (FIRST+NVSS) collection of AGN and star-forming galaxies.
This pre-cleaned SDSS quasar sample is thus composed of 91\,579 sources.

To create the first composite at low redshifts (low-$z$ stack hereafter), we selected all the AGN with $z<0.4$ FWHM of the \ion{H}{$\beta$} and \ion{H}{$\alpha$} emission lines larger than 2000 km s$^{-1}$, leading to 3505 sources. We then considered three different intervals for the FWHM (only the broad component of the modelled line, see \citealt{shen2011} for details) and for the rest-frame EW as given in Table \ref{tab:lowz}. 
Both EW and FWHM are 
measured using the \ion{H}{$\beta$} emission line. We then created 9 low-$z$ spectral stacks for each of the 9 different combinations of Table \ref{tab:lowz},
with an additional low-$z$ composite considering the whole sample. The number of available spectra within the 9 intervals as defined above are also summarised in Table~\ref{tab:lowz}. 
\begin{table}
\begin{center}
\caption{Number, $N$, of galaxies for the different AGN samples considered to create the spectral stacks at low redshifts ($z<0.4$). Rest-frame equivalent widths and full-width at half-maximum values \reve{for the \ion{H}{$\beta$} emission line} are in units of \AA\ and km s$^{-1}$, respectively.}
\begin{tabular}{ l l c  }
\hline
BinID & Intervals & $N$  \\ 
 \hline
1 & $15 \leq \text{EW} < 30$ and $2000 \leq \text{FWHM} < 3000$ & 47  \\  
2 & $15 \leq \text{EW} < 30$ and $3000 \leq \text{FWHM} < 5000$ & 94 \\ 
3 & $15 \leq \text{EW} < 30$ and \text{FWHM} $\geq$ 5000 & 176 \\ 
4 & $30 \leq \text{EW} < 60$ and $2000 \leq \text{FWHM} < 3000$ & 243  \\  
5 & $30 \leq \text{EW} < 60$ and $3000 \leq \text{FWHM} < 5000$ & 467 \\ 
6 & $30 \leq \text{EW} < 60$ and \text{FWHM} $\geq$ 5000 & 615 \\ 
7 & \text{EW} $\geq$ 60 and $2000 \leq \text{FWHM} < 3000$ & 323  \\  
8 & \text{EW} $\geq$ 60 and $3000 \leq \text{FWHM} < 5000$ & 766 \\ 
9 & \text{EW} $\geq$ 60 and \text{FWHM} $\geq$ 5000 & 716 \\ 
\hline
\label{tab:lowz}
\end{tabular}
\end{center}
\end{table}

To increase the coverage at bluer wavelengths and to keep the emission line properties homogeneous with respect to the low-$z$ sample, we considered AGN with the EW and the FWHM of the \ion{H}{$\beta$} emission line in the same intervals as listed in Table~\ref{tab:lowz}, with the additional requirement that the FWHM of the \ion{Mg}{ii} emission line be larger than 2000\,km s$^{-1}$ (mid-$z$ stacks hereafter). To speed up the calculation, all the following spectral composites are built by considering 100 random spectra in each single FWHM--EW bin and 1000 spectra for the total stack in the entire FWHM--EW selected region.

\begin{figure*}
 \includegraphics[width=\linewidth,clip]{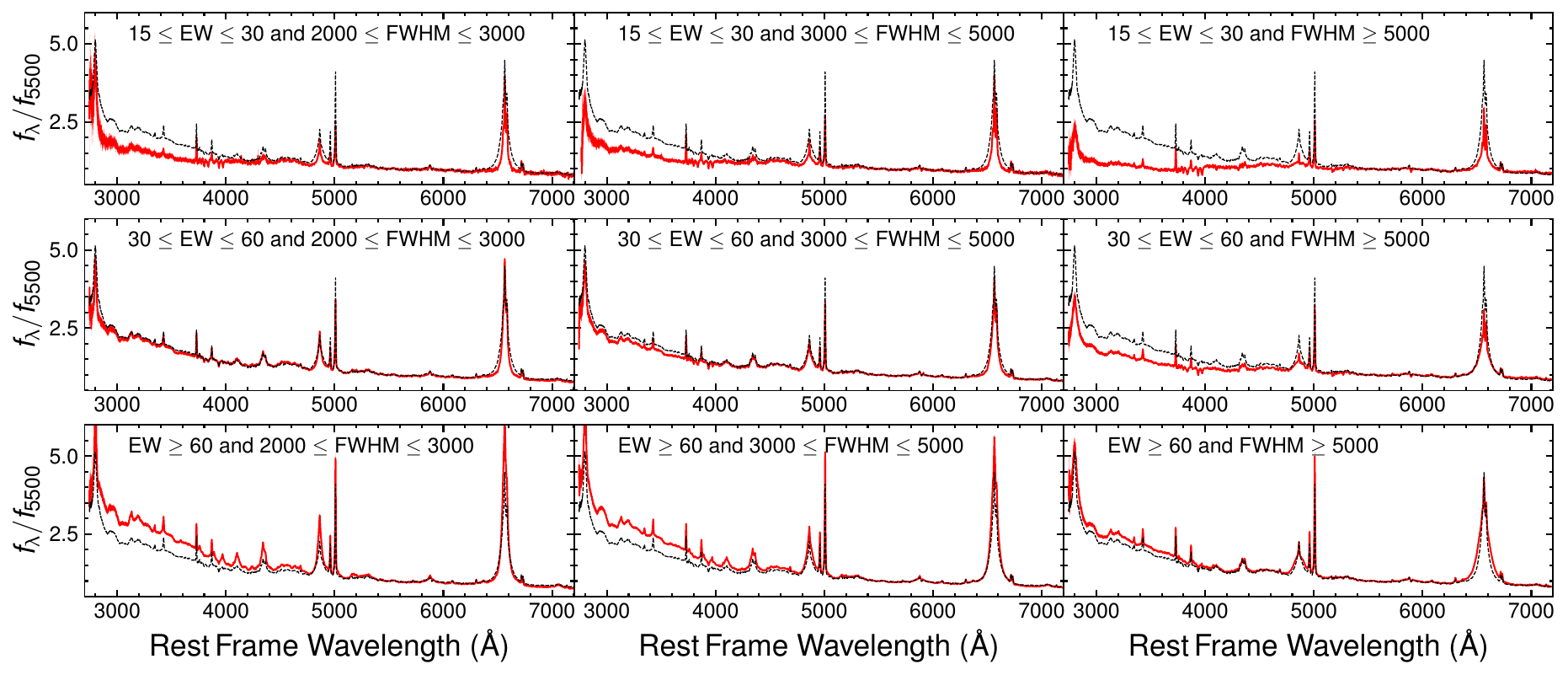}
 \caption{Composite spectra of the low-$z$ sample from ultraviolet to optical rest frame wavelengths. Black line: stack obtained from the entire sample, shown as a reference. \rev{The different composites (shown with the red lines) obtained in each bin are marked as in Table~\ref{tab:lowz} and are shown in separate panels. EW and FWHM are relative to the \ion{H}{$\beta$} emission line and are in units of \AA\ and km s$^{-1}$, respectively. } }
 \label{fig:composite-lowz}
\end{figure*}
To further extend the wavelength coverage (high-$z$ stack hereafter), thus including the \ion{C}{iv} emission line, we considered the \ion{Mg}{ii} line (only the broad component) as the reference since the \ion{H}{$\beta$} line is no longer available. We then considered three different intervals for the following combinations of FWHM (again, only the broad component of the modelled line) and EW, shown in Table \ref{highzbinsew}. 
\begin{table}
\begin{center}
\caption{AGN samples considered to create the spectral stacks at high redshifts. Rest-frame equivalent widths and full-width at half-maximum values \reve{for the broad component of the \ion{Mg}{ii} emission line} are in units of \AA\ and km s$^{-1}$, respectively.}
\begin{tabular}{ l l}
\hline
BinIDs & Intervals \\ 
 \hline
1/4/7 & $30\leq\text{EW}\leq40$, $3800\leq\text{FWHM}\leq4200$ \\
2/5/8 & $40\leq\text{EW}\leq45$, $4000\leq\text{FWHM}\leq4400$ \\
3/6/9 & $35\leq\text{EW}\leq45$, $5100\leq\text{FWHM}\leq6000$\\
\hline
\label{highzbinsew}
\end{tabular}
\end{center}
\end{table}
Again this has an additional requirement that the FWHM of the \ion{C}{iv} emission line be larger than 2000\,km s$^{-1}$.
The three FWHM--EW ranges listed in 
Table \ref{highzbinsew}
are chosen to closely match the values of the FWHM of the \ion{Mg}{ii} line in the mid-$z$ sample. Specifically, the first FWHM--EW interval in 
Table \ref{highzbinsew}
roughly matches the \ion{Mg}{ii} FWHM and EW values of the BinIDs 1, 4, and 7. 
The same considerations exist for the other two intervals.

The \ion{Ly}{$\alpha$} region and the \ion{Ly}{$\alpha$} forest are then included in the stack by selecting the quasars with 
    $40\leq\text{EW}(\text{\AA})\leq55$
    and 
    $5200\leq\text{FWHM}(\text{km s}^{-1})\leq6000$
for the \ion{C}{iv} emission line. The FWHM--EW interval above roughly matches the values of the FWHM of the \ion{C}{iv} line in the high-$z$ sample.
\rev{A summary of the redshift ranges spanned by the four different wavelength intervals considered to build the optical-UV stack is given in Table~\ref{tbl:zsum}.}
\begin{table}
\begin{center}
\caption{\rev{Summary of the redshift ranges spanned by the four different wavelength intervals (discussed in Sect.~\ref{sample_selection_typeI}) considered to build the optical-UV stack.}}
\begin{tabular}{lcc}
\hline
Sample & redshift range & $\langle z\rangle$\\ 
 \hline
low-$z$ & $0.08\leq z \leq 0.39$ & 0.30\\
mid-$z$ & $0.35\leq z \leq 0.90$ & 0.64\\
high-$z$ & $1.50\leq z \leq 2.25$ & 1.80 \\
far-UV (\ion{Ly}{$\alpha$}) & $1.50\leq z \leq 4.90$ & 2.20\\
\hline
\label{tbl:zsum}
\end{tabular}
\end{center}
\end{table}

\subsection{Average spectrum construction} 
\label{Average spectrum construction}
\begin{figure*}
 \includegraphics[width=\linewidth,clip]{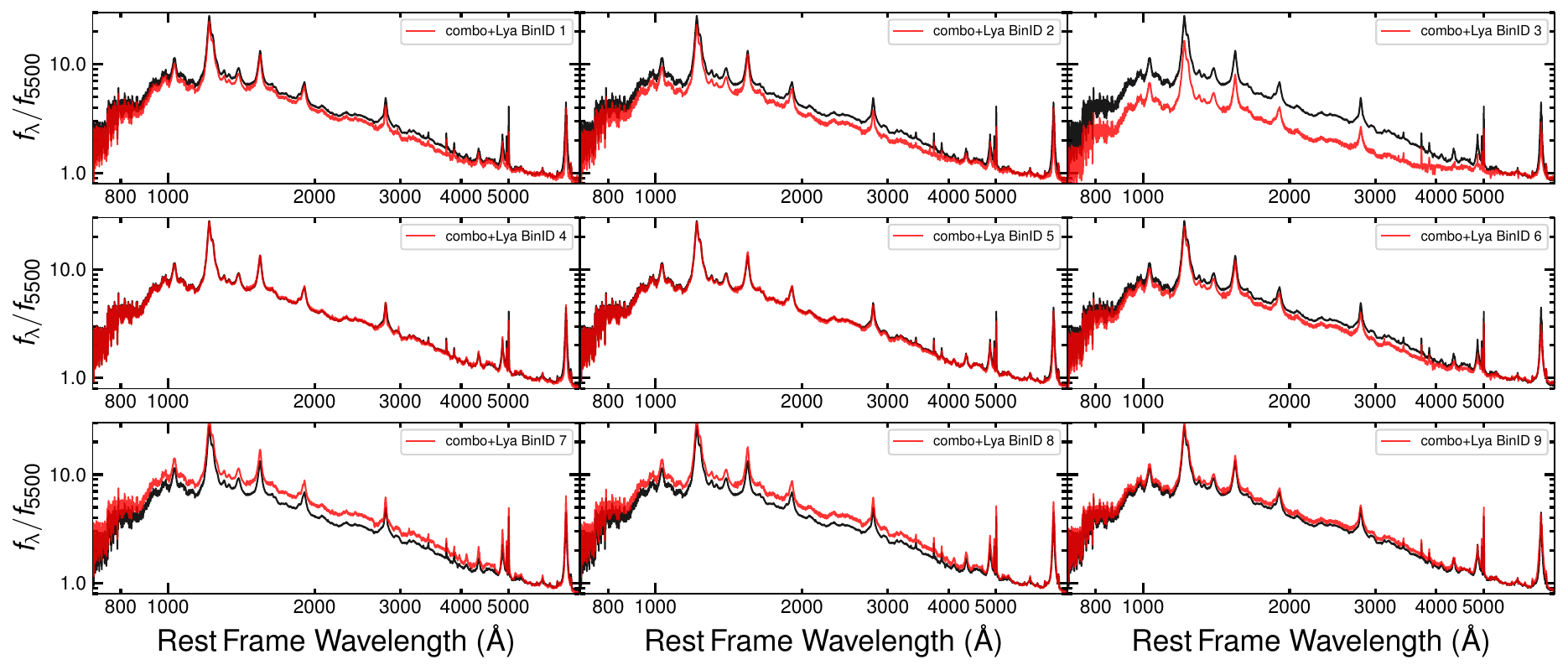}
 \caption{Composite spectra from ultraviolet to optical rest frame wavelengths. The black line represents the stack obtained from the entire sample, shown as a reference. The red lines in the different panels represent the final stack in each bin from the combined composites from blueward of \ion{Ly}{$\alpha$} to \ion{H}{$\alpha$}. The combined spectra covers the rest-frame wavelength interval 800--7000\,\AA.}
 \label{fig:composite}
\end{figure*}

We follow a procedure similar to that in \citet[][L15 hereafter]{lusso2015} to construct the average spectrum in each interval. Specifically, our steps are as follows. 
\begin{enumerate}
 \item We correct the quasar flux density\footnote{In the following we will use the word ``flux" to mean the flux density (i.e., flux per unit wavelength).} ($f_\lambda$) for Galactic reddening by adopting the \ebv estimates from \citet[SFD]{schlegel98} and the Galactic extinction curve from \citet{1999PASP..111...63F} with $R_V=3.1$. We do not correct the spectra for intrinsic dust absorption.

 \item We generate a rest-frame wavelength array with fixed dispersion $\Delta\lambda=0.3$\,\AA. The dispersion value was set to be large enough to include at least one entire pixel from the SDSS spectra around \ion{Ly}{$\alpha$}, where we assumed an average spectral resolution $R=2000$ across the entire wavelength range (i.e., $\Delta\lambda\simeq1250\,\text{\AA}/2R$). This assumption slightly oversamples the spectral stack at low-$z$ ($\Delta\lambda\simeq1.3$\,\AA\ at 5100\,\AA).

 \item Each quasar spectrum is shifted to the rest-frame 
 and linearly interpolated over the rest-frame wavelength array with the fixed dispersion $\Delta\lambda$ defined above.

 \item We normalize single spectra by their flux at rest $\lambda =5500$ \AA\ for the low-$z$, whilst the mid-$z$ and high-$z$ composites are both normalised at $2000$ \AA. The spectral stack at Ly$\alpha$ is normalized at $\lambda =1550$ \AA.

 \item All the flux values are then averaged (mean) to produce the stacked spectrum normalized to unity at the reference wavelengths above.  
\end{enumerate}
Uncertainties on each stack are computed as $\sigma/\sqrt{N-1}$ where $\sigma$ and $N$ are the standard deviation and the number of points in each spectral channel.
All the spectral stacks are finally combined together by taking the low-$z$ AGN composite (shown in Fig.~\ref{fig:composite-lowz}) as the reference and by matching the flux in the overlapping wavelength range between two nearby spectra.
The final optical/ultraviolet composite obtained from the entire sample, as well as in each bin, covers the rest-frame wavelength interval 800--7000\,\AA\ (see Fig.~\ref{fig:composite}).

\subsection{IGM transmission correction}
Blueward of Ly$\alpha$ emission in the quasar rest frame, absorption from
intergalactic \ion{H}{i} attenuates the quasar flux, both in the Lyman series
(creating the so-called \ion{H}{i} forest), and in the Lyman continuum at rest
$\lambda<912$\,\AA\ \citep[e.g.,][]{1990A&A...228..299M}.
Therefore, to correct the observed stacks (Fig.~\ref{fig:composite}) for the intergalactic medium (IGM) absorption by neutral hydrogen along the line-of-sight, we follow a similar (but very much simplified) approach as the one employed by L15 (see also Sect.~4 and 5 in \citealt{L2018}).

\reve{To recover the IGM-corrected quasar emission, we constructed the IGM transmission functions ($\Tl$) making use of the publicly available \texttt{pyigm} package for the analysis of the intergalactic medium\footnote{\url{https://github.com/pyigm/pyigm}} discussed in \citet[P14 and references therein]{Prochaska2014}. These functions have been computed through a cubic Hermite spline model which describes the \ion{H}{i} absorber distribution function which, in turn, depends on both redshift and column density, i.e.,  $\fnz=\partial^2n/\left(\partial \mnhi\partial z\right)$. For simplicity, we considered the default $\fnz$ which is currently the P14 formulation (see their section 2)\footnote{The interested reader should refer to section 1.7.3 at \url{https://pyigm.readthedocs.io/_/downloads/en/latest/pdf/}}. Briefly, this modelling assumes that the \ion{H}{i} forest is composed of discrete ``lines'' with Doppler parameter $b = 24$\,km\,s$^{-1}$ and that the normalization of $\fnz$ evolves as $\left(1+z\right)^{2}$ \citep{2009ApJ...705L.113P}.}
This redshift evolution is consistent, within the uncertainties, with the one obtained in high-redshift surveys 
\citep{Prochaska2014}. Opacity due to metal line transitions was ignored since they contribute negligibly to the total absorption in the Lyman continuum.

\begin{figure}
 \includegraphics[width=\linewidth,clip]{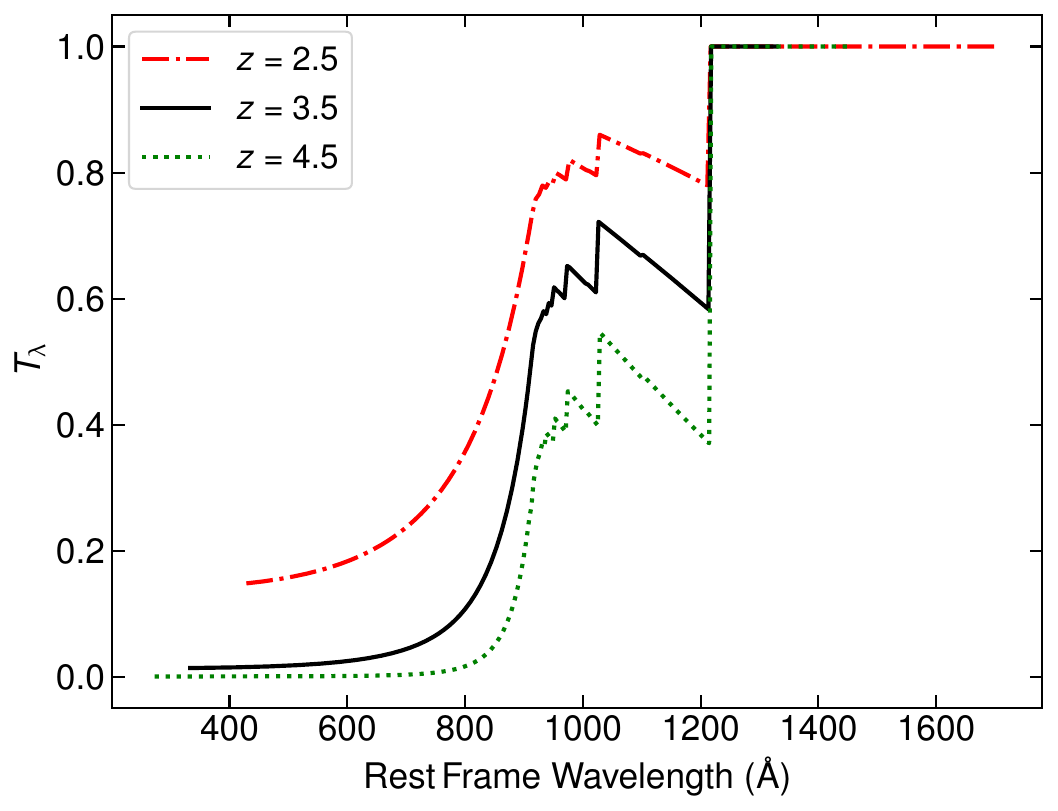}
 \caption{An example of our adopted Lyman series opacity ($T_{\lambda}$) at different redshifts. The lines represent the effective optical depth due to intervening absorbers and includes the contribution of absorption in the hydrogen Lyman series and the Lyman continuum at $\lambda<912$\,\AA.}
 \label{fig:igm}
\end{figure}

Here we consider 8 realisations of $\fnz$, and calculated $\Tl$ in the wavelength range 500--1250\,\AA\ for the redshifts 0.3, 0.7, 1.5, 2.5, 3.5, 4.5, 7.5. We also checked how many sources are contributing to the bluer part of the total stack and what is their average redshift. There are about 450 quasars contributing more than 25 spectral channels at $\lambda<1200$\,\AA. Their average redshift is 2.9, so an additional  IGM transmission function at $z=3$ was added to the final $\Tl$ library. 
Examples of the $\Tl$ function at $z=2.5$, $z=3.5$, and at $z=4.5$ are shown in Fig.~\ref{fig:igm}.
The observed spectral flux ($f_{\lambda,\rm obs}$) in the stack is finally divided by the IGM transmission curve to obtain the intrinsic corrected emission ($f_{\lambda,{\rm corr}}$), i.e.,  $f_{\lambda,{\rm corr}}=f_{\lambda,{\rm obs}}/T_{\lambda}$.

\begin{figure*}
\centering\includegraphics[width=0.8\linewidth,clip]{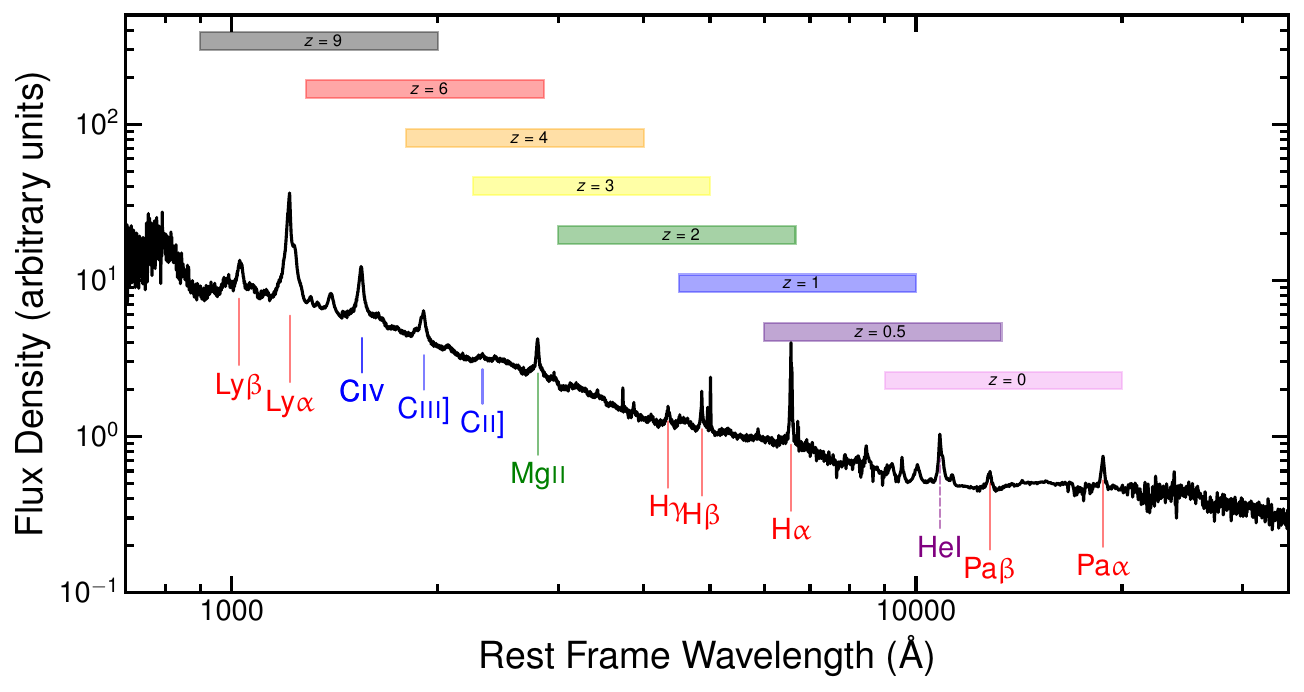} \includegraphics[width=0.8\linewidth,clip]{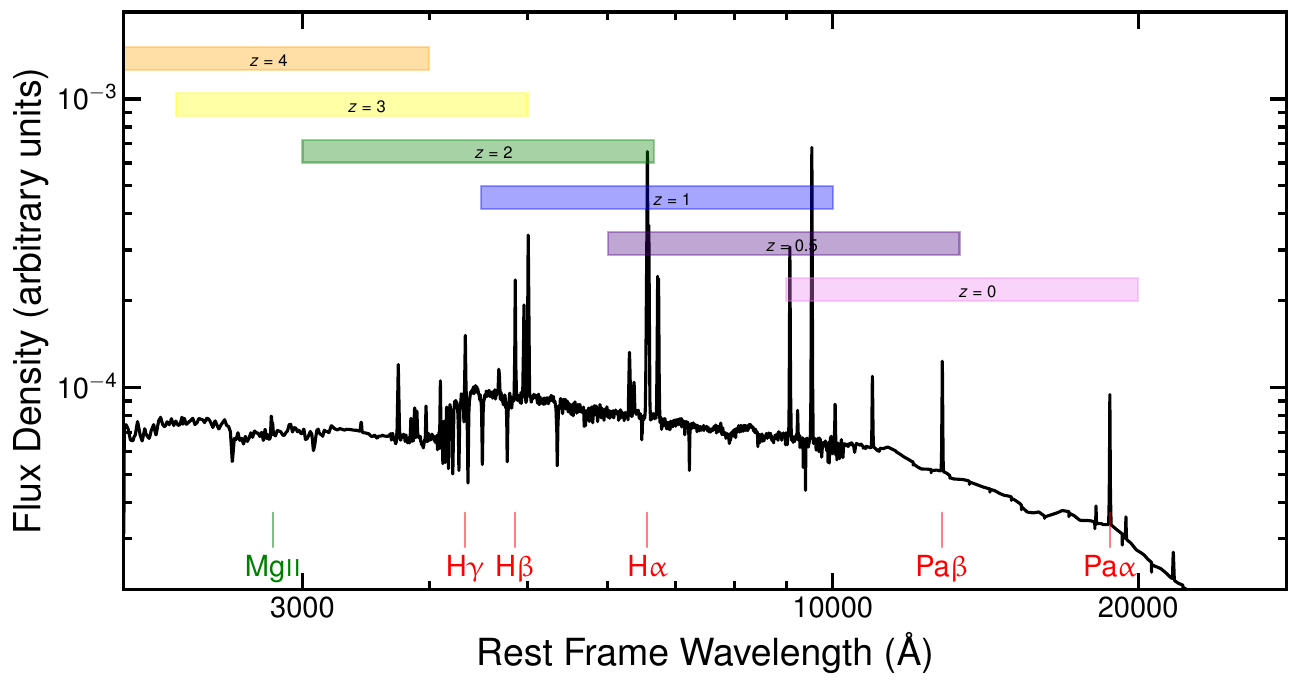}
 \caption{\emph{Top panel}: final empirical stack (flux density values are reported in units of erg s$^{-1}$ cm$^{-2}$ \AA$^{-1}$) for type 1 AGN covering the rest-frame wavelength range 700--36\,000\,\AA. The main emission lines are marked. Shaded coloured regions mark the portion of the spectrum that will be covered by the low-resolution \Euclid spectrograph up to redshift $z=9$. \reve{All the \rev{incident} spectra for the type 1 and type 2 AGN are available online at the link \url{https://drive.google.com/drive/folders/1tjryMUkHhD10NjH_lhfeEBnvYRY4uib7?usp=sharing}.} \emph{Bottom panel}: final empirical stack (flux density values are reported in arbitrary units) for type 2 AGN (SB-AGN, SF-AGN, and AGN2) covering the rest-frame wavelength range 2000--28\,000\,\AA. }
 \label{fig:stacktot}
\end{figure*}
\subsection{Extending the stack to the NIR}
\label{Extending the stack to the NIR}
As NISP will cover the rest-frame NIR for local AGN, we extended the composite to wavelengths longer than the SDSS spectral coverage by including 9 hard X-ray selected 
\rev{low-redshift type 1 AGN at $0.015<z<0.114$ ($K_{\rm s} < 14$)} from the Swift Burst Alert Telescope (BAT) \reve{published by \citet{Ricci22}. }
The AGN are observed with the folded-port infrared echellette \citep[FIRE,][]{fire08} instrument in the high-resolution echelle mode \citep[see, e.g.,][]{Ricci22}. FIRE covers the whole NIR bandpass in a single exposure, with nominal wavelength resolution of $R \approx 6000$ for a 0$\farcs$6-slit width, i.e., $\Delta v\simeq50$\,km s$^{-1}$.
In particular, \reve{from the starting sample of 13 broad-line AGN described in \citet{Ricci22}} we selected all sources classified as Seyfert~1 (Sy~1) to 1.5 \reve{(i.e. we excluded BAT ID 577)} 
and excluded, after visual inspection, those targets whose emission lines where particularly affected by low atmospheric transmission \reve{(BAT ID 1045)} 
and (or) showing a red spectral shape ($\alpha_\nu>0$, where $f_\nu \propto \nu^{\alpha_\nu}$, \reve{i.e. BAT IDs 411 and 657}) 
which is unusual for Sy1s in the NIR \citep{Glikman2006}. \reve{The list of sources is provided in Table~\ref{tab:sources-nir}}
These NIR spectra were then smoothed using a Savitzky–Golay filter, which preserves the average resolving power.

To \rev{increase the sample statistics and} extend both the redshift and the luminosity range, we complemented the data above with the sample of 23 infrared spectra of type 1 AGN published by \citet[][see their Table~2]{Glikman2006}. \reve{From the starting sample published by Glikman et al. (i.e. 27 quasars, we excluded SDSS J013418.1$+$001536.6 0 as the spectrum presents a strong absorption feature at 17 300 \AA\ close to the \ion{Pa}{$\alpha$}. We also removed SDSSJ130756.5$+$010709.6 and SDSSJ160507.9$+$483422.0 because both spectra do not have a common wavelength interval that can be used to normalise them to the FIRE data.} \rev{This data set covers the redshift range $0.118 < z < 0.418$ with $K_{\rm s} < 14.5$ and $M_i < -23$ ($\logten{(L_{\rm bol}/{\rm erg\,s}^{-1})}\simeq42.2-46.3$), and it is complementary both in terms of redshift range and luminosity with the low-$z$ SDSS sample ($\logten{(L_{\rm bol}/{\rm erg\,s}^{-1})}\simeq44.1-46.6$).} 

All the spectra are averaged following the procedure described in Sect.~\ref{Average spectrum construction} to produce the final stack down to \ion{Pa}{$\alpha$} (see the Appendix~\ref{ap:type1stackcomp} for further tests on additional NIR quasar datasets available in the literature).
The spectral stack is then combined by taking the low-$z$ AGN composite (shown in Fig.~\ref{fig:composite-lowz}) as the reference and by matching the flux in the overlapping wavelength range between two nearby spectra.
The final IGM corrected quasar composites, from the forest to the NIR obtained by stacking the entire quasar sample is shown in Fig.~\ref{fig:stacktot} (top panel). All the empirical spectra we constructed for the type 1 AGN (from UV to NIR) are available online as supporting material.

\subsection{Creation of the mock type 1 AGN sample}
Starting from the nine templates described in the previous sections, we constructed a grid of redshifts, $E(B-V)$, and bolometric luminosities to probe the expected observed parameter space of type 1 AGN: $0.1<z<7.0$ with a step width $\delta z=0.1$, $0<E(B-V)<1.25$ with $\delta E(B-V)=0.25$, and $42\le\logten{(L_{\rm bol}/{\rm erg\,s}^{-1})}\le47$, with $\delta\logten({L_{\rm bol}/{\rm erg\,s}^{-1})}=0.2$.

For each point in the grid, each of the nine templates is scaled to the bolometric luminosity using a bolometric correction value of 5.8 \reve{derived for the $B$-band} \citep{duras2020}. To account for intrinsically reddened type 1 quasars, we then apply an intrinsic extinction by assuming the SMC law (\citealt{prevot1984}, as appropriate for unobscured AGN; see also \citealt{2004AJ....128.1112H,salvato09}). The spectra are then redshifted and the appropriate IGM curve of \cite{Prochaska2014} is applied.
Finally, we measure the expected observed \Euclid magnitudes in \IE, \YE, \JE, and \HE. Each corresponding spectrum is saved and the resulting \Euclid magnitudes are stored in a catalog. 
We then keep a random subsample of 1248 spectra, across the grid, that satisfies the \Euclid depth of \IE $<24.5$. 

\begin{figure}
 \centering\includegraphics[width=\linewidth,clip]{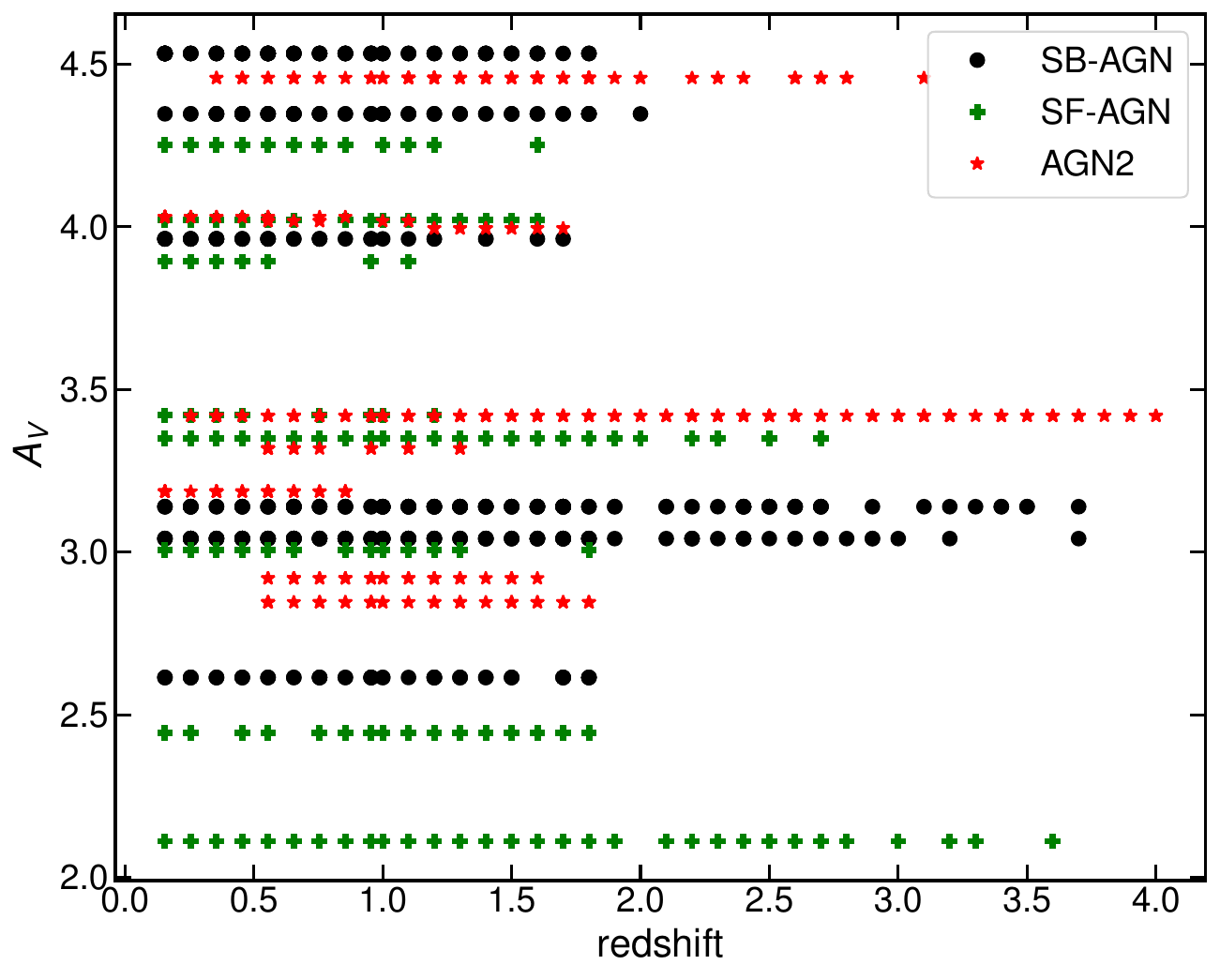}
 \caption{\rev{Total (ISM+birth clouds, see Sect.~\ref{sec:spec_creation_2}) dust attenuation in the $V$-band as a function of the (input) redshift considered to build the type 2 semi-empirical library of incident spectra. The three different AGN classes are marked with different symbols as in the legend.}}
 \label{fig:attenuation}
\end{figure}
\begin{figure*}
 \centering\includegraphics[width=0.8\linewidth,clip]{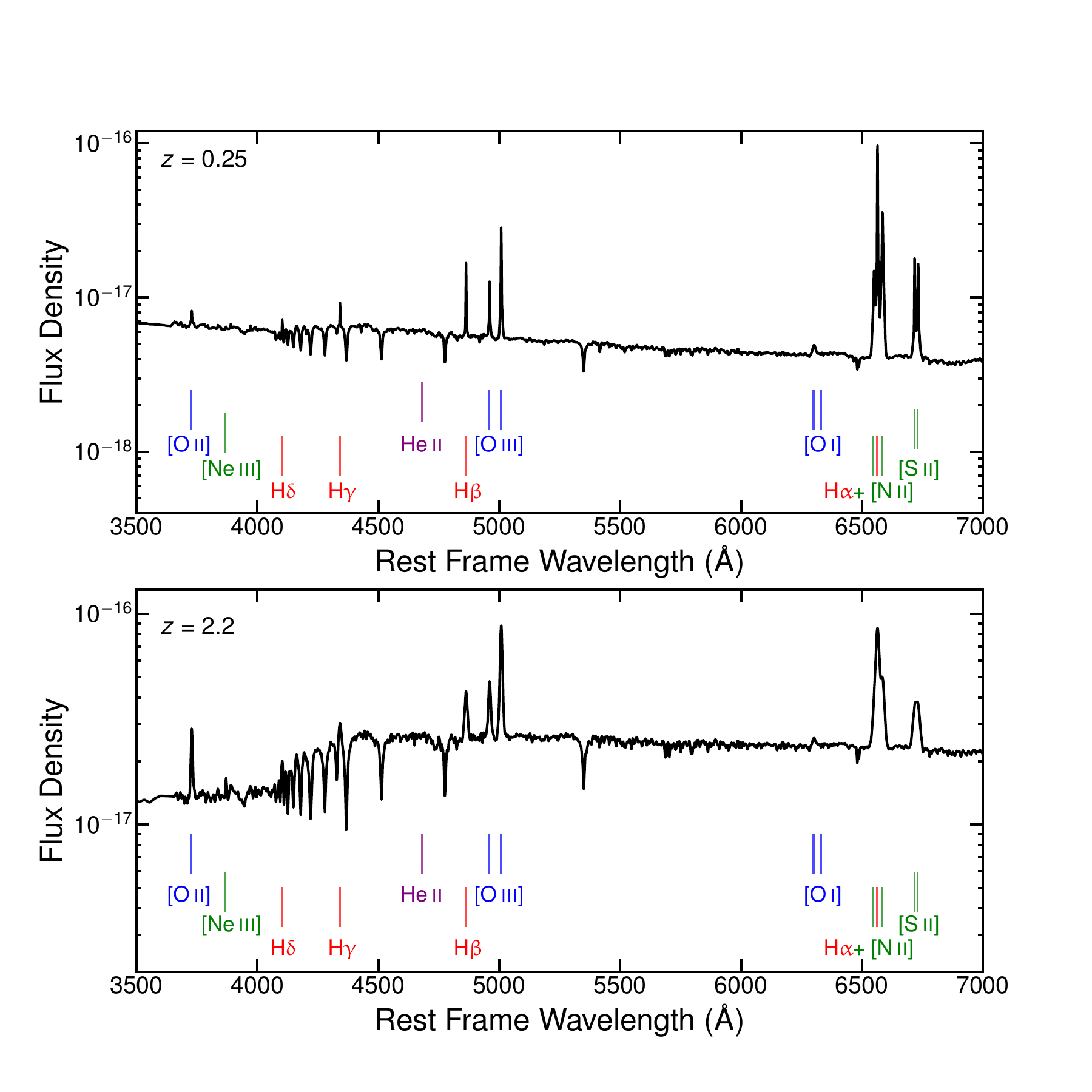}
 \caption{Examples of obscured type 2 AGN \rev{incident} spectra (flux density values are in units of erg s$^{-1}$ cm$^{-2}$ \AA$^{-1}$) \reve{at optical wavelengths, from 3500 \AA\ to 7000 \AA}. 
 \emph{Top panel}: 
 a $z=0.25$ type 2 AGN with a bolometric luminosity $\logten(L_{\rm bol} / {\rm erg\,s}^{-1})$ of $43.9$, hosted in a galaxy with a stellar mass $M_{\star}$ of $10^{9.2}$\,$M_{\odot}$ and SFR of $10^{0.9}$\,$M_{\odot}$ yr$^{-1}$. 
 \emph{Bottom panel}:     
 a massive ($M_{\star} = 10^{11.8}$\,$M_{\odot}$) galaxy at $z=2.2$ with intense star-formation (SFR\,$=$\,$10^{2.6}$\,$M_{\odot}$ yr$^{-1}$) hosting an AGN with $\logten(L_{\rm bol} / {\rm erg\,s}^{-1}) = 45.2$. The main emission lines are marked. }
 \label{fig:type2examples}
\end{figure*}

\section{Semi-empirical template spectra for type 2 AGN}\label{sec:spec_creation_2}
The number density, continuum and physical properties of type 2 AGN are derived starting from the 
spectro-photometric realisations of infrared-selected targets at all-$z$ 
\citep[\spr{},][]{Bisigello2021}. The simulation starts from the observed infrared luminosity function, complemented with the $K$-band luminosity function of elliptical galaxies and the galaxy stellar mass function of dwarf irregulars, 
and it is in agreement with a wide set of observables, including luminosity functions at several wavelengths, number counts (from radio to X-rays), the total galaxy stellar mass function, and the relation between star-formation rate (SFR) and stellar mass \citep[see][for further details]{Bisigello2021}. 

To build the templates for the type 2 AGN, we consider three AGN populations. Specifically, we consider a first population of star-forming galaxies hosting intrinsically faint AGN (SF-AGN), and two populations hosting a powerful AGN, namely a mildly obscured AGN (classical type 2 AGN, AGN2) and a heavily obscured AGN (hosted by a starburst galaxy, SB-AGN)\footnote{\rev{The interested reader on details regarding the SED classification in the three different classes should refer to Sect. 2 in \cite[][see their Fig. 1]{Gruppioni2013}.}}.
The number density of both SF-AGN and SB-AGN were derived from their observed IR luminosity functions \citep[\reve{8$-$1000 $\mu$m;}][]{Gruppioni2013}. The number density of AGN2 were similarly derived from the same IR luminosity function, but obtained by performing a simultaneous fit of the \textit{Herschel} \citep{Gruppioni2013} and UV observed luminosity functions \citep{Croom2009, McGreer2013, Ross2013, Akiyama2018, Schindler2019}, as derived by \cite{Bisigello2021}.

The spectral energy distributions (SEDs) of the three AGN populations were described with a set of 12 empirical templates \citep{polletta2007, Gruppioni2010}, chosen as they describe the multi-wavelength observations of the \textit{Herschel} AGN used to derived the starting luminosity functions. These SED templates, with the addition of emission lines (see Sect.~\ref{Inclusion of emission lines}), were used to \rev{classify the type 2 AGN population and to} derive the photometric fluxes in different \Euclid ancillary filters. \rev{The host galaxy continuum is derived from SED fitting using the  MAGPHYS code \citep{dacunha2008}, as discussed in \cite[see their Sec. 3.2]{Bisigello2021}.} The physical properties, such as AGN accretion luminosity and optical depth in the $V$-band \citep[and the relative contribution of the diffuse ISM and the birth clouds to dust attenuation, as described in][]{CharlotFall2000}, were derived by fitting the mentioned empirical templates with the code {\sc sed3fit} \citep{berta2013}, which assumes two sets of equally possible AGN libraries with smooth \citep{Fritz2006, Feltre2012} and clumpy tori \citep{Nenkova2008a, Nenkova2008b}. Finally, the total SFR was derived by summing the UV and IR component, derived from the 1600\,\AA\ and the IR stellar continuum, assuming the conversions by \cite{Kennicutt1998a, Kennicutt1998b}.  

\subsection{Inclusion of emission lines}
\label{Inclusion of emission lines}
We used the AGN accretion luminosity and the SFR as normalization factors to incorporate the emission line fluxes due to both AGN activity and star-formation.  
We briefly describe the main aspects related to the emission features of interest below \citep[see Sect. 2.3 of][for further details]{Bisigello2021}.
The nebular emission was incorporated into the \spr{} galaxy templates as follows.
\begin{itemize}
\item 
First, we considered the line emission due to star-formation.
The \ion{H}{${\alpha}$} flux was derived from the SFR \citep{Kennicutt1998a, Kennicutt1998b}. All the other hydrogen lines were, in turn, derived by assuming the case-B hydrogen recombination coefficient for an electron temperature of $T_{\rm e} $\,$=$\,$10^4$\,K and an electron density of $N_{\rm e}$\,$ =$\,$100$\,cm$^{-3}$. 
Other metal optical lines, namely [\ion{O}{ii}]$\lambda$3727, [\ion{Ne}{iiii}]$\lambda$3869, [\ion{N}{ii}]$\lambda\lambda$6548, 6584, [\ion{O}{iii}]$\lambda$5007, [\ion{S}{ii}]$\lambda\lambda$6717, 6731 and [\ion{S}{iii}]$\lambda$9069 were included using a set of empirical relations \citep[i.e.,][]{Kennicutt1998b, PettiniPagel2004, Jones2015, Kewley2013, Dopita2016, Kashino2019, Proxauf2014, Mingozzi2020}.
\item 
Next, we accounted for the line emission due to AGN activity. 
We considered theoretical predictions from photoionisation models developed by \cite{Feltre2016} using the code \texttt{CLOUDY} \citep[version c13.3;][]{Ferland2013}. These models reproduce the emission from the gas in the narrow-line emitting regions of AGN. Among the adjustable parameters of the entire model grid \citep[see Sect. 2.1 in][for details]{Feltre2016}, we considered an ionisation parameter at the Str\"{o}mgren radius between $\logten(U_{\rm S})=-1.5$ and $-3.5$, metallicity from $\approx 0.5$ to 2 times solar ($Z= 0.008$, 0.017, 0.03), a dust-to-metal ratio (which rules the dust depletion onto dust grains) of 0.3, a UV spectral index of the ionising radiation $\alpha =-1.4$, and an internal micro-turbulence velocity $v=100$\,km s$^{-1}$. 
The amplitude of the line fluxes were scaled to the AGN accretion luminosity derived as described above.
\item 
Next, we considered the scattering of the line fluxes. 
For each line, we either include the scatter associated to each considered empirical relation referenced above (when quoted in the corresponding papers) or a generic scatter of 0.1 dex. 
\item 
We considered dust attenuation next. 
The \reve{total} line fluxes \reve{(AGN$+$host)} were attenuated for the presence of dust by considering the two-components (one describing the diffuse ISM and another for the dust in the birth-clouds, see \citealt{Bisigello2021} for details) using the model of \cite{CharlotFall2000}. \reve{We thus assumed that the broad-line component is completely extincted, whilst the narrow AGN lines have the same dust attenuation as the stellar population. In this step, we also assumed that the AGN continuum has been totally extincted, therefore the only continuum visible is the one from the host galaxy, which has been determined from the SED3fit\footnote{\url{http://steatreb.altervista.org/alterpages/sed3fit.html}} code combining simple stellar populations from \citet{bruzalcharlot2003} with the approach presented in MAGPHYS code \citep{dacunha2008}. \rev{The dust attenuation is in the range $A_{V}\simeq2.1-4.5$, with a distribution as a function of redshift shown in Fig.~\ref{fig:attenuation} for the three different AGN classes.}}
\item 
Finally we considered the line widths. 
For the line emission due to star-formation, we derived the gas velocity dispersion from the stellar mass by taking into account the broadening of the nebular emission lines \citep{Bezanson2018}. For the AGN nebular emission, we assumed a FWHM in the typical range of type 2 AGN, between 500 and 800 km s$^{-1}$. \reve{The total line width thus consider both star-formation and AGN contributions, which have been added to the continuum at rest.} We finally assumed a Gaussian profile for the emission lines when creating the simulated spectra \citep{Bisigello2021}. 
\end{itemize}
A few examples of the resulting type 2 templates are shown in Fig.~\ref{fig:type2examples}, \reve{zooming-in at optical wavelengths}. The first is a $z=0.25$ type 2 AGN with a bolometric luminosity $\logten(L_{\rm bol} / {\rm erg\,s}^{-1})$ of $43.9$, hosted in a galaxy with a stellar mass $M_{\star}$ of $10^{9.2}$\,$M_{\odot}$ and SFR of $10^{0.9}$\,$M_{\odot}$\,yr$^{-1}$. The second is a massive ($M_{\star} = 10^{11.8}$\,$M_{\odot}$) galaxy at $z=2.2$ with intense star-formation (SFR\,$=$\,$10^{2.6}$\,$M_{\odot}$\,yr$^{-1}$) hosting an AGN with $\logten(L_{\rm bol} / {\rm erg\,s}^{-1}) = 45.2$.

\subsection{Sub-sample selection}\label{sample_selection_typeII}

\begin{figure*}
 \centering\includegraphics[width=0.8\linewidth,clip]{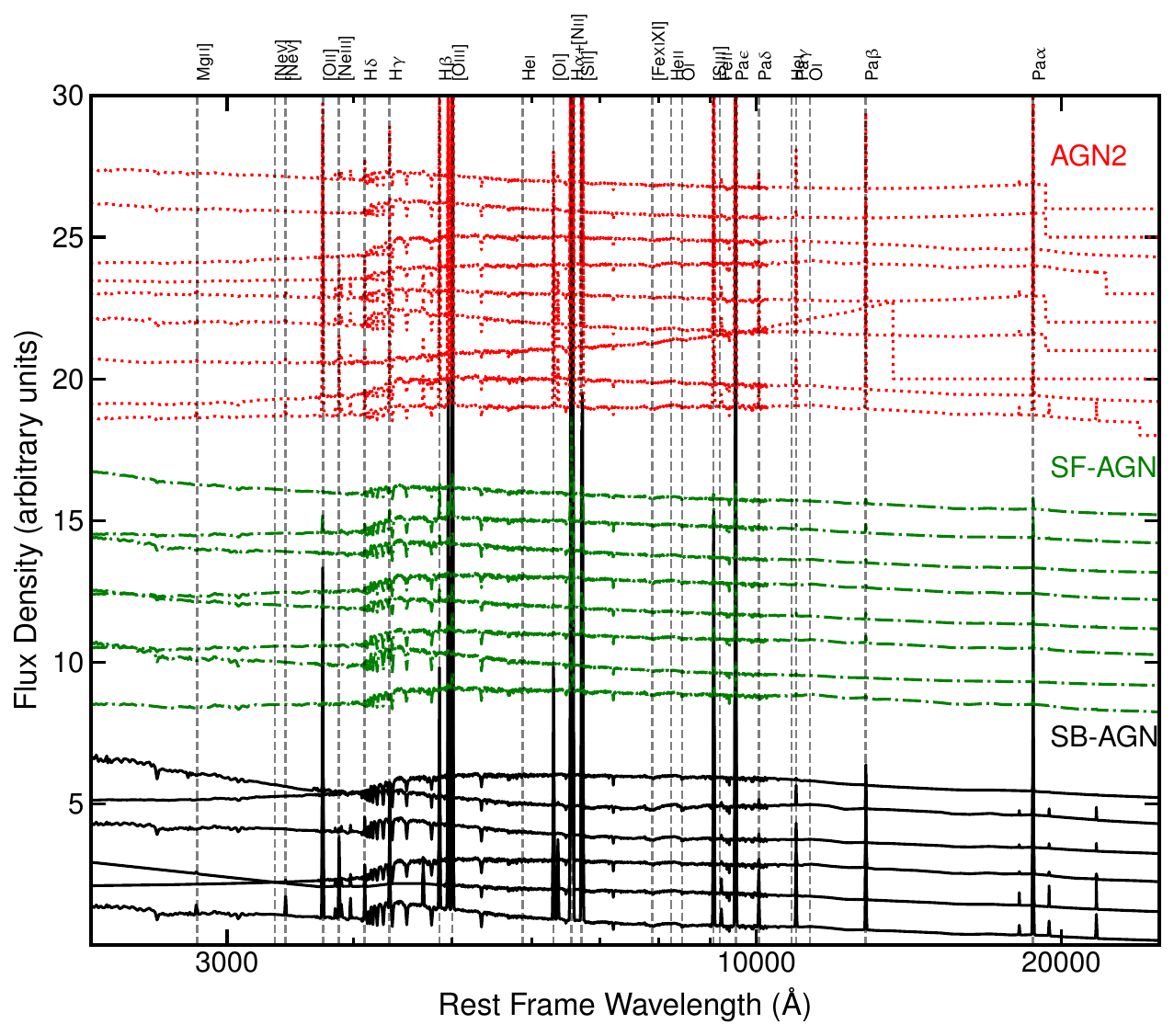}
 \caption{Type 2 \rev{incident} templates employed for the cross-correlation analysis in the redshift determination (i.e. {\tt xcoor}, see Sect. \ref{Redshift measurements}). Black solid line: 6 templates representing SB-AGN. Green dot-dashed line: 8 SF-AGN templates. Red dotted line: 9 AGN2 templates.}
 \label{fig:templates2}
\end{figure*}

For the selection of the type 2 AGN templates, we proceed as follows.
\begin{itemize}
\item We considered all the galaxies, among the AGN2, SF-AGN, and SB-AGN \spr{} populations, with AGN bolometric luminosity above a given threshold, $\logten(L_{\rm bol} / {\rm erg\,s}^{-1}) >$ 42 (see Sect.~\ref{sec:spec_creation_2}).

\item We then selected all the templates with a line flux above $\logten(F_{\rm line}/ {\rm erg\,s^{-1}cm^{-2}}) > -16.2$. For the purposes of this selection, we considered specific emission lines depending on the redshift (see Table \ref{tab:emissionlines}).

\item We applied a number density cut for a survey of 15\,000\,deg$^{2}$ resulting in 2561 templates from the \spr{} simulations. Finally, we randomly selected \reve{(i.e. random uniform)} 1248 spectra from the above sample. \reve{This random sample corresponds to the number of spectra fitted in the NISP simulation, which is sufficient for the paper's objectives. }
\end{itemize}
\reve{The 1248 spectra cover 23 classes representative of different AGN and host galaxies combinations as discussed above \citep[see Sect. 2.1.1 in][for further details]{Bisigello2021}. }
Figure~\ref{fig:stacktot} (bottom panel) shows the spectral composite, from the \ion{Mg}{ii} to the NIR obtained by stacking the entire sample of AGN2, SF-AGN, and SB-AGN, whilst the 23 templates separated by the three classifications are presented in Figure~\ref{fig:templates2}.
\begin{table}
\begin{center}
\caption{Emission line per redshift bin chosen for the selection of AGN2 templates}
\begin{tabular}{ c c  }
\hline
 Redshift interval & Emission line  \\ 
 \hline
$0.15 < z \leq 0.44$ & \ion{Pa}{$\beta$}  \\  
$0.44 < z \leq 0.90$ & [\ion{S}{iii}]$\lambda9069$ \\ 
$0.90 < z \leq 1.80$ & \ion{H}{$\alpha$} \\ 
$1.80 < z \leq 2.80$ & \ion{H}{$\beta$} \\ 
$2.80 < z \leq 4.00$ & [\ion{O}{ii}]$\lambda3727$ \\ 
\hline
\end{tabular}
\end{center}
\label{tab:emissionlines}
\end{table}
\section{Spectra simulation}
\label{Spectra simulation}
To generate a sample of simulated AGN spectra, we randomly picked 1248 sources from those selected with the previous steps described in Sect. \ref{sample_selection_typeI} for type 1 AGN and the 1248 type 2 AGN described in Sect. \ref{sample_selection_typeII}. We then simulated two \Euclid pointings with the NISP spectrometer, one pointing contains type 1 AGN while the other pointing contains type 2 AGN. We positioned the sources on a grid with sufficient spacing to ensure clear separation of individual spectra, while also avoiding the edges of the NISP detectors. This choice was made with the intention of maximising the number of AGN per pointing, in order to examine the NISP's ability to detect AGN based on their physical properties. It is important to note that the primary focus of this study does not involve the decontamination of spectra. However, the decontamination process is expected to be effectively addressed by the \Euclid reference observation sequence (ROS), which incorporates various grism orientations and high-resolution direct images \citep[further details on spectra decontamination in slitless spectroscopy are in][]{Ryan2018}.

\subsection{Configuration of the NISP-S simulator: TIPS}
The NISP spectroscopic channel (NISP-S) simulator (\texttt{TIPS}; \citealt{Zoubian2014}; see also \citealt{gabarra2023} for the pilot run of the simulation campaign) is part of OU-SIM (Organization Unit for the simulations of all instruments; Euclid Collaboration: Serrano et al., in prep.), which also includes the NISP photometric (NISP-P) simulator. The NISP-S channel simulator considers additional features with respect to the photometric channel. \reve{The detector noise, point spread function, and spectral dispersion were measured during ground tests \citep[][Kubik et al. in prep for the detector on-ground characterization]{Waczynski2016, Barbier2018, 2019costille, Maciaszek_2022} and were parameterised into \texttt{TIPS}. The detector noise, which is dominated by the dark current, is quite homogeneous for each of the 16 NISP detectors with a spatial mean level on the order of 0.011 ADU per frame per pixel. The maps of detector noise at the pixel-level were incorporated into \texttt{TIPS}. The spectral dispersion characterised at nine positions of the focal plane during the NISP ground test (Gillard et al. in prep for the spectrometer on-ground calibration) has been averaged at 13.4 $\AA$\,px$^{-1}$ in the simulations.}
The astrophysical background simulated by \texttt{TIPS} includes the zodiacal and out-of-field stray light \citep[see][for details on background estimation for the Euclid Wide Survey]{Scaramella-EP1} and are considered constant across the field of view. 

To perform the simulation, we chose the coordinates $({\rm RA},{\rm Dec})=(228\,394,6.590)$ degrees where the background is dominated by zodiacal light. At these coordinates, the zodiacal light is estimated to be 2.2$\,$photon$\,$s$^{-1}\,$px$^{-1}$. The zodiacal light in the Euclid Wide Survey is expected to vary between 1.1 and 3.0$\,$photon$\,$s$^{-1}\,$px$^{-1}$ with a median at 1.6$\,$photon$\,$s$^{-1}\,$px$^{-1}$. The out-of-field stray light is estimated at 0.3$\,$photon$\,$s$^{-1}\,$px$^{-1}$ in our coordinates and the median estimated value for the Euclid Wide Survey is 0.4$\,$photon$\,$s$^{-1}\,$px$^{-1}$. 

In terms of data acquisition, we replicated the ROS for the Euclid Wide spectroscopic survey \citep{Scaramella-EP1}. This strategy involved a total integration time of 2212 seconds divided into four exposures. Each exposure was dithered, with an approximate offset of three pixels between them. Additionally, we simulated the four RGS grism orientations specified in the \Euclid ROS, namely $+0^{\circ}$, $-4^{\circ}$, $+180^{\circ}$, and $+184^{\circ}$, in order to minimize the effects of cosmic ray impacts and faulty pixels. This approach will also serve the purpose of decontamination.

\begin{figure*}
\centering\includegraphics[width=0.8\linewidth,clip]{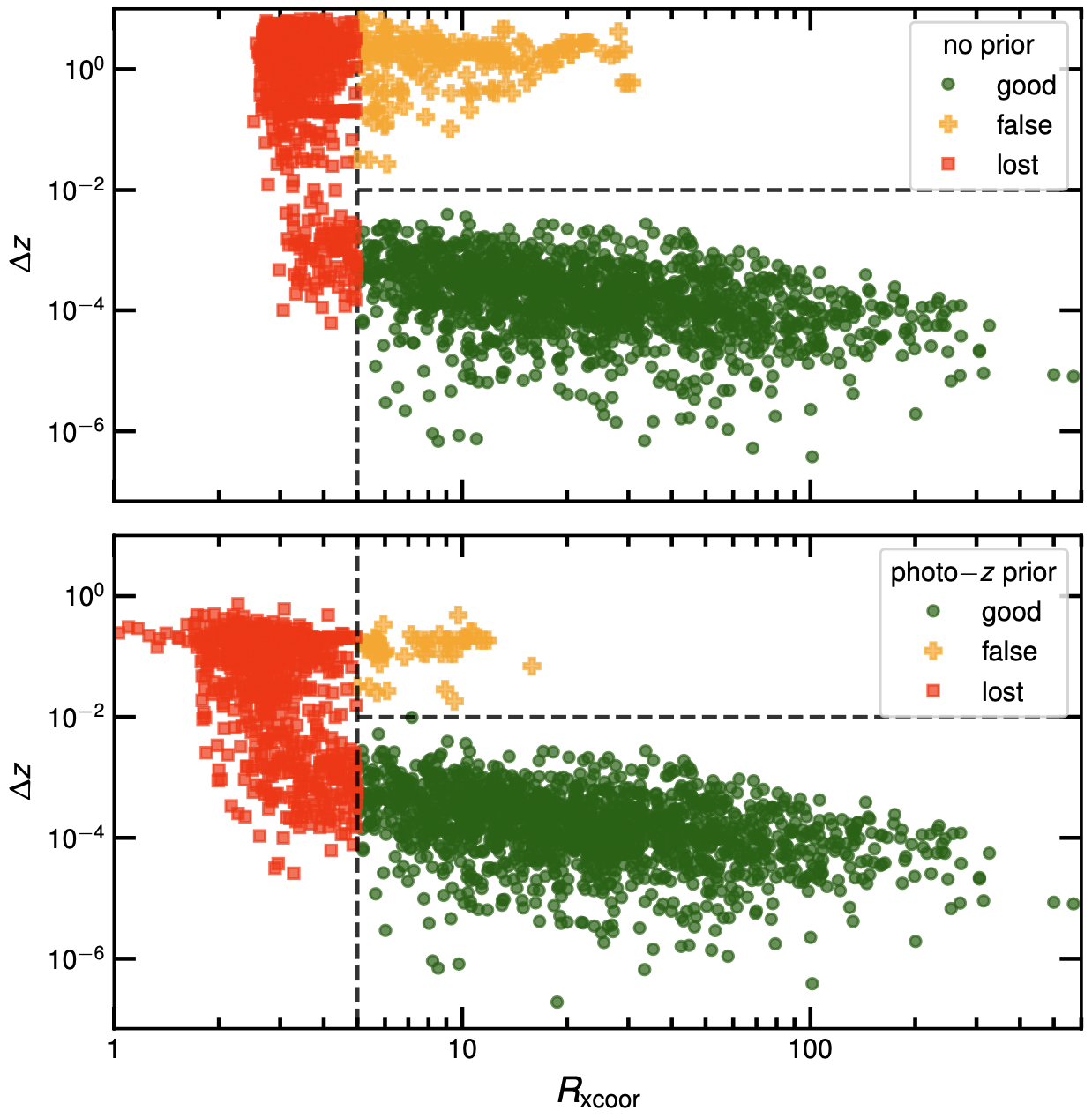}
 \caption{Redshift uncertainty $\Delta z$ (defined as $|{z_{\rm out} - z_{\rm in}}|$) \reve{for the total sample of 2496 AGN} as a function of the parameter $R_{\rm xcorr}$ (that is related to the shape of the correlation peak, and is inversely proportional to the redshift measurement uncertainty) when using no prior (\emph{upper panel}) and photometric redshift prior (\emph{lower panel}). Different colours refer to different classifications of the redshift estimates (see Sect. \ref{Redshift measurements} for more details).} 
 \label{fig:zestimates}
\end{figure*}

\subsection{Extraction of the 1D spectra}
The extraction of the 1D spectra from the simulated images was performed using the \Euclid Spectroscopic Image Reduction pipeline (SIR) which has been combined with the simulation pipeline by  \texttt{SIR\_SPECTROSIM\_RUNNER} \citep{Paganin2022mqg}. SIR is the official \Euclid pipeline for image reduction, wavelength and flux calibration, and extraction of the 1D spectra. The four dithered frames are processed separately by the SIR pipeline, extracting the first-order spectra using an aperture in pixels which is set at five pixels for point-like sources, as is the case for the AGN simulated in this work. The four resulting 1D spectra are then combined using inverse-variance weighting.
\reve{Examples of simulated \Euclid-like spectra for both type 1 and type 2 AGN are shown in Appendix~\ref{Sim atlas}.}

\section{Redshift measurements}
\label{Redshift measurements}
To estimate the efficiency and purity of the redshift measurements for the simulated AGN spectra described in Sect.~\ref{Spectra simulation}, we performed a cross-correlation between simulated and optimised templates created from the input AGN spectra. This choice allows us to better understand the effects on the redshift measurements due to the limited spectral coverage provided by the red grism, as well as by its spectral resolution, which introduces a severe blending between adjacent spectral lines.
The set of templates is composed by 9 type~1 composites (see Fig. \ref{fig:composite-lowz}) along with 23 type~2 semi-empirical spectra \reve{(see Fig. \ref{fig:templates2})} that cover all the different combinations of AGN and host galaxy presented in Sect. \ref{sec:spec_creation_2}. 

To automatically measure the redshift, we use the IRAF task {\tt xcorr}, which performs a cross-correlation between the Fourier-transformed simulated spectrum and the set of template spectra \citep{td1979}. The routine was adapted to uniformly cover the full redshift range $0\le z \le 6$ when searching for the correlation peak.
We considered the $R_{\rm xcorr}$ parameter, which is related to the shape of the correlation peak provided by {\tt xcorr}, to assess the goodness of the correlation between the input and template spectra. This quantity is expected to be inversely proportional to the redshift measurement uncertainties \citep[see Eq. 23 in][]{td1979}. \reve{We note that {\tt xcoor} strongly depends upon the choice of the templates utilised to fit the data but, in our experiment, we have considered the same AGN templates used to build the simulations. The template library we employed is thus representative of the data we fit. The redshift distribution determined from {\tt xcoor} is thus obtained in the most favourable condition. Moreover, since our main aim does not require a complete distribution of AGN, the redshift determination we present here only provides a test bed for the future OU-SPE pipeline, without having the ambitious goal of being complete.}
To quantify the quality of the redshift measurement, we
considered the following definitions.
\begin{itemize}
    \item Measurements are termed \textit{good} if $|{z_{\rm out} - z_{\rm in}}| < 0.01$, where $z_{\rm in}$ and $z_{\rm out}$ are the redshifts of the input and simulated spectrum and $R_{\rm xcorr}\ge5$, respectively. 
    \item Measurements are termed \textit{false} if $|{z_{\rm out} - z_{\rm in}}| > 0.01$ even if the parameter $R_{\rm xcorr}\ge5$. The fraction of redshift estimates flagged as \textit{false} gives us information about the purity of the AGN spectroscopic mock sample.
    \item \reve{Measurements are termed \textit{lost} if 
    $R_{\rm xcorr}<5$. This regime also includes AGN with a low signal-to-noise ratio (SNR) spectrum and (or) absence of strong emission lines. This is especially true at low redshift and high redshift where \ion{H}{$\alpha$} and \ion{H}{$\beta$}+[\ion{O}{iii}] are outside the wavelength coverage of the grism.}
\end{itemize}
As a first test, we adopted no prior assumption on either the redshift or the AGN classification of the simulated source. By running the 
cross-correlation completely blind, we measure the correct redshifts for almost 60\% of the AGN spectra (1487/2496), but with a 
relatively high contamination (7\%), i.e., 176 objects in which the 
redshift is measured with good confidence but with an incorrect 
redshift estimate (\reve{flagged as ``false''}, see Fig. \ref{fig:zestimates}, top panel). 
We do not find a significant improvement in the redshift measurements when adopting a type 1 or 2 template depending on the classification of the analysed AGN spectrum.

We then included a simulated set of photometric redshifts, with an uncertainty assumed to be $\delta z/(1+z)=0.2$, as a prior in the cross-correlation, by considering the photometric redshift value as the centre of a redshift window with a width of 0.5.  
The search for the correlation peak is carried out only within this redshift range. With this prior of the photometric redshift, we significantly improved the efficiency in measuring correct redshifts\footnote{We caution that the effect of catastrophic errors is not included in our analysis.}. The objects with a redshift measurement flagged \textit{good} ($|z_{\rm out} - z_{\rm in}|<0.01$ and $R_{\rm xcorr}\ge5$) constitute 65\% of the total \reve{(1622/2496)}, with an increase of the  efficiency that is due to the significant decrease of \textit{false} measurements (49/2496): only 2\% of objects have a confident but incorrect redshift estimate (i.e., $R_{\rm xcorr} \ge$ 5).

In terms of redshift intervals, we have identified three ranges. A low-redshift ($z<0.89$) interval where the \ion{H}{$\alpha$} emission line is not yet redshifted into the NISP spectral range. A second range where the \ion{H}{$\alpha$} emission line is within the NISP spectral coverage of the red grism ($0.89<z<1.83$) and a third range at high redshift ($z>1.83$) where the \ion{H}{$\alpha$} is, again, redshifted outside the observed NISP spectral range. 

Our results are very encouraging in the redshift interval where \ion{H}{$\alpha$} is visible. \reve{In this range, considering the results obtained including the photo-$z$ prior, our simulation comprises 844 AGN (307 type 1 and 537 type 2 AGN), and 689 of them (82\%) have \textit{good} and highly ranked redshifts, with further 144 spectra (17\%) with redshifts determined from {\tt xcorr} but they are not reliable based on our adopted criterion ($R_{\rm xcorr}<5$). The purity in this redshift range is very high (98\%), with only 11 objects classified as \textit{false}. 
We caution that the fraction of AGN with \textit{good} redshifts will be lower in the real data, since we are missing faint AGN in the simulated sample and the success rate is a function of flux (as it will be shown in Fig.~\ref{fig:ha_percent_12}).
}

Outside this redshift range, the measurement efficiency decreases significantly. 
This is expected at high redshift, where we have fainter objects and a wider redshift interval, and the emission lines are weak if not absent. At low redshift, the low efficiency is related to the lack of intense emission lines. 

We have also analysed the redshift measurements obtained from the latest version of the OU-SPE (organization Unit for the spectroscopic redshifts) pipeline. \reve{The OU-SPE pipeline estimates the correct redshift for 30\% considering the full sample. This fraction becomes 48\% when we limit the AGN sample to the redshift interval where \ion{H}{$\alpha$} is visible.} The principal cause for this inconsistency lies in the set of templates implemented in the pipeline, which are tailored for non-active galaxies (e.g., broad lines are not considered).
As the pipeline is currently under development (Euclid Collaboration: Le Brun et al., in preparation), we prefer to restrict the analysis to the redshift values as discussed above.

\section{Spectral analysis}\label{sec:analysisqsfit}
As the calibrated \Euclid spectra will be available to the scientific community online, we detail here a ``user-like'' spectral analysis of the simulated data mainly focussing on the emission line fluxes. 
Emission line fluxes for both incident and simulated \Euclid-like mock AGN spectra are measured using a modified version of the Quasar Spectral Fitting library (\qsfit, \citealt{calderone2017}). The updated \qsfit{} is a Julia language\footnote{\url{https://julialang.org/}} based automatic spectral fitting tool \citep{selwood2023} for both type 1 and type 2 AGN spectra from the optical to the ultraviolet, \rev{using the redshift values estimated in Sect. \ref{Redshift measurements}}. Properties of the AGN continuum, Balmer continuum and pseudo-continuum, iron complexes, host galaxy contributions, and emission lines are derived in the rest-frame wavelength range 1215--7300\,\AA\ \reve{\citep[see also e.g.,][]{shen2011,rakshit2020}}.
Scattered light and nebular continuum are not included in the spectral fit of type 2 AGN. These two contributions are nonetheless negligible with respect to the continuum and star formation \citep[e.g.,][]{vg2022}.
\begin{table}
  \caption{List of the optical and UV emission lines considered in \qsfit{} analysis of type 1 and type 2 incident spectra and their wavelengths (rest-frame, vacuum). B refers to broad lines, N to narrow lines, and VB to very broad lines. }
 \begin{center}
 \begin{tabular}{llcc}
  \hline
  Line & Wavelength (\AA) & type 1 & type 2 \\
  \hline
  \ion{C}{iii}] & 1908.53 & B+N & N \\
  \ion{C}{ii}] & 2327.64 & B & -- \\
  \ion{Mg}{ii} & 2796.352 & B+N & N \\
  $[$\ion{Ne}{v}$]$ & 3426.5 & N & N \\
  $[$\ion{O}{ii}$]$ & 3727 & N & N \\
  $[$\ion{Ne}{iii}$]$ & 3870.16 & N & N \\
  \ion{H}{$\delta$} & 4102.892 & B & - \\
  \ion{H}{$\gamma$} & 4341.684 & B & N \\
  $[$\ion{O}{iii}$]$ & 4364.436 & N & N \\
  \ion{He}{ii} & 4687.02 & B & - \\
  \ion{H}{$\beta$} & 4862.683 & B+N & N \\
  $[$\ion{O}{iii}$]$ & 4960.295 & N & N \\
  $[$\ion{O}{iii}$]$ & 5008.240 & N & N \\
  $[$\ion{O}{iii}$]$ (blue wing) & 5008.240 & N & N \\
  \ion{He}{i} & 5877.25 & B & - \\
  $[$\ion{O}{i}$]$ & 6302.046 & N & N \\
  $[$\ion{O}{i}$]$ & 6365.536 & N & N \\
  $[$\ion{N}{ii}$]$ & 6549.85 & N & N \\
  \ion{H}{$\alpha$} & 6564.61 & B+N+VB & N \\
  $[$\ion{N}{ii}$]$ & 6585.28 & N & N \\
  $[$\ion{S}{ii}$]$ & 6718.29 & N & N\\
  $[$\ion{S}{ii}$]$ & 6732.67 & N & N\\

  \hline
 \end{tabular}
 \end{center}
  \label{tab:qsfit_knownemlines}
\end{table}
\begin{table}
\caption{Default known emission line type properties available to be implemented with \qsfit{}. }
\begin{center}
\begin{tabular}{lc}
 \hline
 Line Type & FWHM (km s$^{-1}$)  \\ 
 \hline
Narrow & 100--2000  \\  
Broad & 900--15\,000 \\ 
VeryBroad & 10\,000--30\,000 \\ 
 \hline
\end{tabular}
\end{center}
\label{tab:qsfit_known_emission_lines_properties}
\end{table}
\qsfit{} utilizes a flexible fitting recipe which includes several physically motivated AGN spectral components to model the data. Whilst we refer the reader to the relative publication for details, we summarise below the main components considered to fit the data.

The first major component is the AGN continuum. 
In the general case for both type 1 and type 2 AGN, the AGN nuclear continuum is modelled as a single power-law model of the form 
\begin{equation}\label{eqn:simple_PL}
 L_{\lambda} = A \left( \frac{\lambda}{\lambda_{\rm 0}} \right) ^{\alpha_{\lambda}},
\end{equation}
across the available (rest-frame) wavelength range, where $\lambda_{0}$ is a reference wavelength, $A$ is the luminosity density at $\lambda = \lambda_{0}$ and $\alpha_{\lambda}$ is the spectral slope at $\lambda \ll \lambda_{0}$.
The parameter $\lambda_{0}$ is fixed at the median wavelength in the available spectral range, $A$ is constrained to be positive and $\alpha_{\lambda}$ is constrained to be in the range [$-$5, 5].

The next major component is the Balmer continuum and pseudo-continuum. 
The Balmer continuum for type 1 AGN is modelled in \qsfit{} according to \citet{grandi1982} and \citet{dietrich2002}:
\begin{equation}
 L_{\lambda} = A \, B_{\lambda}(T_{\rm e}) \left\{ 1 - \exp\left[ \tau_{\rm BE} \left( \frac{\lambda}{\lambda_{\rm BE}}\right)^{3}\right]\right\},
\end{equation}
where $A$ is the luminosity density at 3000\,\AA, $B_{\lambda}(T_{\rm e})$ is the blackbody curve at the electron temperature $T_{\rm e}$ (fixed at $T_{\rm e}$ = 15\,000\,K), $\tau_{\rm BE}$ is the optical depth at the Balmer edge (fixed at $\tau_{\rm BE}$ = 1), and $\lambda_{\rm BE}$ is the Balmer edge wavelength, fixed at 3645\,\AA. Higher order Balmer components (7 $\leq$ $n$ $\leq$ 50, i.e., the Balmer pseudo-continuum), are accounted for in the \qsfit{} model using the line ratios from \cite{sh1995} with a fixed electron density of 10$^{9}$\,cm$^{-3}$ and a temperature of 15\,000\,K. A free parameter in the fit determines the ratio of the higher-order blended line complex to the Balmer continuum at the Balmer edge. The whole component is finally broadened by a convolution with a Gaussian profile with a FWHM equal to 5000\,km s$^{-1}$.

Another significant component is the host galaxy template, especially for low-luminosity AGN. 
\qsfit{} models the contribution of the host galaxy emission to the AGN spectrum using a host galaxy template. A library of host galaxy SEDs is incorporated in \qsfit{}, consisting of the templates from SWIRE (Spitzer wide-area infrared extragalactic survey) collated by \citet{polletta2007} supplemented with twelve starburst galaxy templates generated to model cosmological evolution survey (COSMOS) AGN in \citet{ilbert2009}. This provides a selection of SEDs ranging from quiescent elliptical galaxies through to starbursts. 
By default, the template of an elliptical galaxy with an age of 5 Gyr drawn from the SWIRE library is used when the host galaxy is enabled in a fit.

We then considered the UV and optical iron templates. 
For type 1 AGN, broad and narrow (permitted and forbidden) iron lines are modelled using the UV and the optical iron complex templates of \citet{vestergaard&wilkes2001} and \citet{veroncetty&veron2004}, respectively. 

Known emission lines represent a subset of spectral transitions that are considered in a given \qsfit{} fitting recipe. Known emission lines can be defined as narrow, broad or very broad lines (referred to in Table \ref{tab:qsfit_knownemlines} as N, B and VB respectively), which carry different limits on their allowed parameter ranges as summarised in Table \ref{tab:qsfit_known_emission_lines_properties}. Known emission lines can be modelled with a Gaussian, a Lorentzian, or a Voigt profile. 

\reve{The integrated line luminosity ratios are fixed to a ratio 1:3 only for the oxygen doublet, [\ion{O}{iii}]$\lambda$4959 and [\ion{O}{iii}]$\lambda$5007, and (separately) for the nitrogen doublet, [\ion{N}{ii}]$\lambda$6549 and [\ion{N}{ii}]$\lambda$6583. Also, velocity offsets with respect to the redshift are fixed for the above mentioned lines. The only constraint for the narrow line widths is to be $< 10^3$ km s$^{-1}$.
No other constraint involving [\ion{O}{iii}]$\lambda\lambda$4959, 5007 and [\ion{N}{ii}]$\lambda\lambda$6549, 6583 is applied since they would be simultaneously available in the same spectrum only in a limited redshift range.}

Finally, Gaussian-profiled ``unknown'' lines are placed at areas of highest residuals in the latter stages of a spectral fit, with the aim of modelling additional emission lines features that are present in the spectrum but have not been listed in the current spectral lines library. Unknown lines are able to be placed at any wavelength contained in the spectrum, but are excluded from being placed within the \ion{H}{$\alpha$}+[\ion{N}{ii}] and \ion{H}{$\beta$}+[\ion{O}{iii}] complexes. The number of considered unknown lines is a parameter that should be adjusted based on the redshift and the spectral coverage of the spectra being analysed, to strike a balance between under and over utilization. We constrain unknown line FWHM to the range 500--10\,000\,km s$^{-1}$ for type 1 AGN and 10--2000\,km s$^{-1}$ for type 2 AGN. 
\begin{figure*}
 \includegraphics[width=0.49\textwidth,clip]{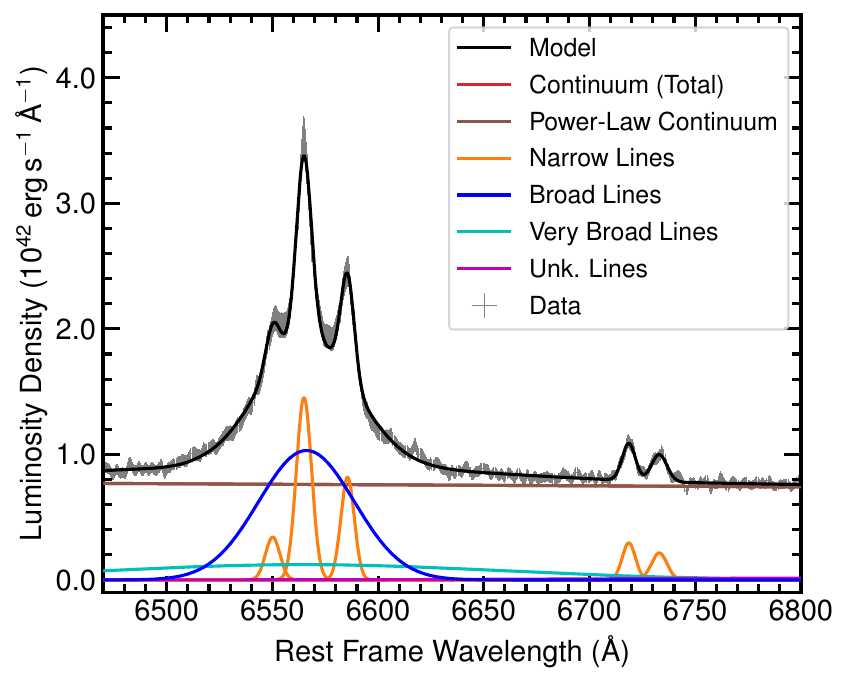}
 \hfill
 \includegraphics[width=0.49\textwidth,clip]{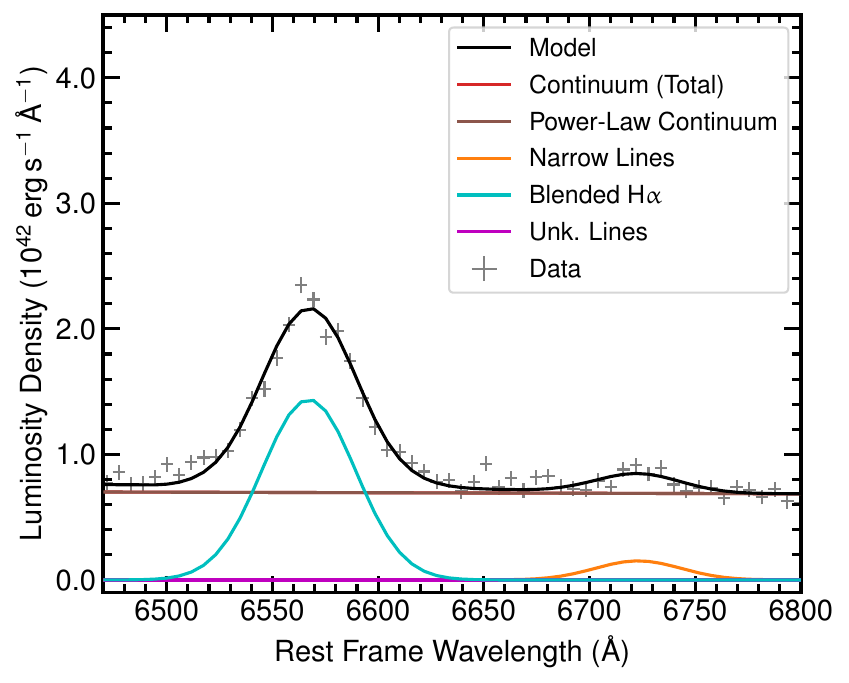}
 \vfill
 \includegraphics[width=0.49\textwidth,clip]{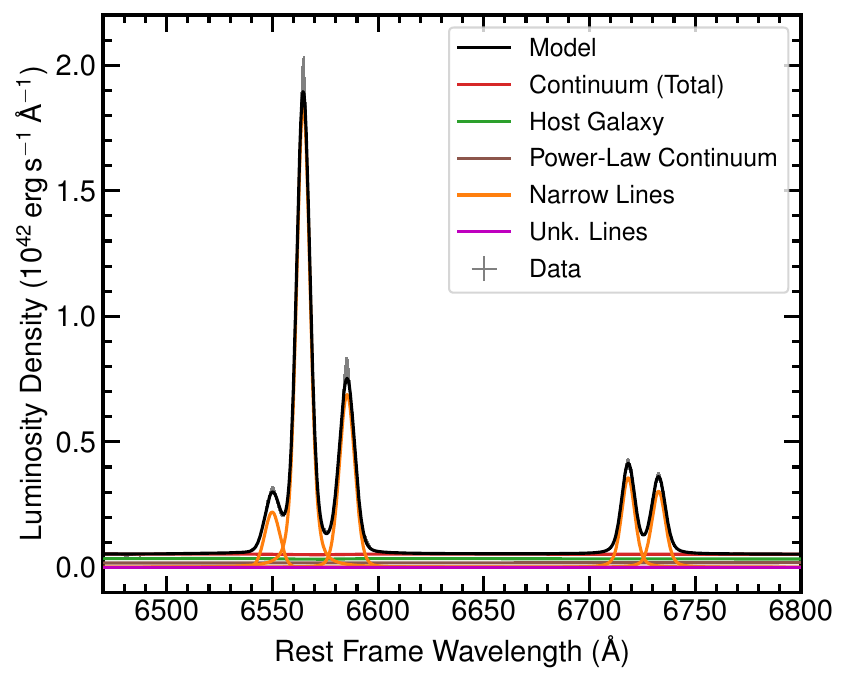}
 \hfill
 \includegraphics[width=0.49\textwidth,clip]{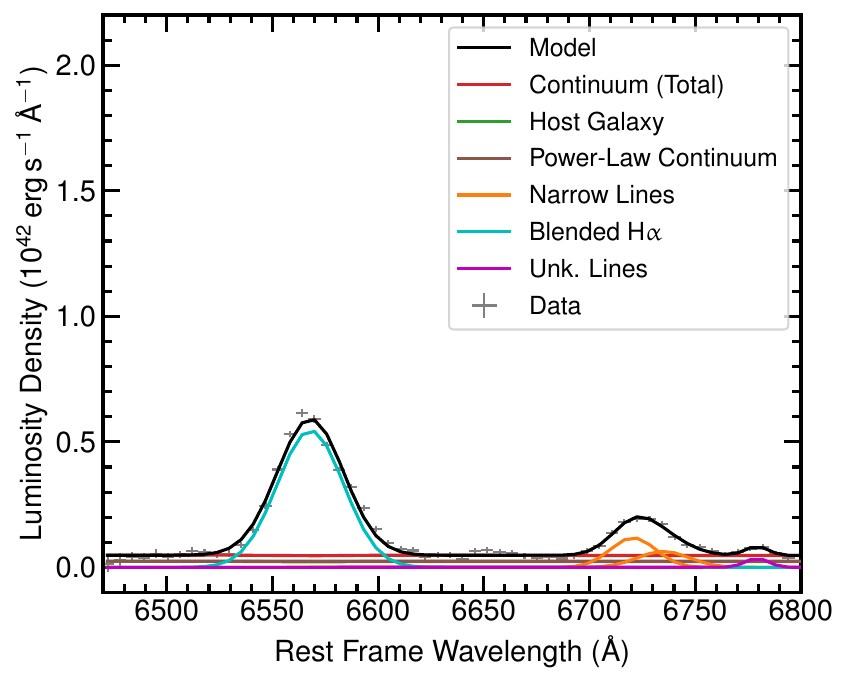}
 \caption{Example showing type 1 (\emph{top row}) and type 2 (\emph{bottom row}) AGN \ion{H}{$\alpha$} complex of incident (\emph{left}) and \Euclid-like simulated spectra (\emph{right}). The sources shown in this plot have redshifts $z = 1.7$ (type 1) and $z = 1.2$ (type 2).}
 \label{fig:ha_blending}
\end{figure*}
\begin{table}
  \caption{Summary of model components considered in \qsfit{} analysis of type 1 and type 2 incident spectra. PL means power law.}
 \begin{tabular*}{\columnwidth}{@{}l@{\hspace*{15pt}}l@{\hspace*{15pt}}l@{}}
  \hline
  Model Component & type 1 & type 2 \\
  \hline
  AGN continuum & Single PL & Single PL\\[2pt] 
  Balmer (pseudo-)continuum & Yes & -- \\[2pt] 
  Host galaxy template & -- & Yes \\[2pt] 
  UV and optical iron templates & Yes & -- \\[2pt] 
  Known emission lines & Yes & Yes \\[2pt] 
  Known emission line profiles & Gaussian & Voigt \\[2pt] 
  Max number of unknown lines & 5 & 4 \\[2pt] 
  \hline
 \end{tabular*}
\label{tab:qsfit_inc_model_summary}
\end{table}
\subsection{Spectral analysis of the incident spectra}
\label{sec:qsfit_inc_analysis}
We limit our analysis of incident spectra to the NISP-S red grism observed-frame wavelength range of 12\,500--18\,500\,\AA. Lines outside of this range will not be present in the simulated spectra which are generated for the red grism observed-frame wavelength range only. Due to this observed-frame wavelength range cut, any samples that subsequently fall outside \qsfit's rest-frame wavelength analysis range (1215--7300\,\AA) are not considered in our analysis.\footnote{We do not include emission lines shorter than \ion{C}{iii}] in the spectral fit, as we are only analysing the spectra in the red grism wavelength range (for which the simulated spectra are created). Since the highest redshift AGN in the catalog are at $z\simeq7$ and the red grism coverage starts at 12\,500\,\AA\ observed-frame, the shortest rest-frame wavelength we can possibly observe is at 1563\,\AA.} 
Table \ref{tab:qsfit_knownemlines} provides an overview of the known optical and UV emission lines considered in our analysis of type 1 and type 2 AGN incident spectra. 

Prior to the \qsfit{} analysis, we re-sampled the spectra to have evenly spaced wavelength samples in logarithmic space, thus ensuring a constant resolution of 100\,km~s$^{-1}$.
We found five unknown lines to be an appropriate maximum amount for modelling unconsidered elements in type 1 incident spectra in the NISP-S wavelength range. Since type 2 spectra exhibit less complex features with respect type 1 AGN, we find that a maximum number of four unknown lines are appropriate to model type 2 incident spectra in the NISP-S wavelength range. The maximum number of unknown lines is utilised in 54\% of type 1 incident spectra fits and 96\% of type 2 incident spectra fits with \qsfit{}. \reve{To model known emission lines in the type 2 incident spectra, we considered a Gaussian spectral profile (a test considering different spectral profiles is discussed in Appendix~\ref{Line profile modelling}), to be consistent with the simulated data (see the next Section). }
Table \ref{tab:qsfit_inc_model_summary} provides a summary of the individual recipes used to fit type 1 and type 2 incident mock spectra. 
\begin{table*}
 \caption{Summary of model components considered in \qsfit{} analysis of type 1 and type 2 simulated spectra. PL = power law.}
 \centering\begin{tabular}{lll}
  \hline
  Model Component & type 1 & type 2 \\
  \hline
  AGN continuum & Single PL & Single PL\\ 
  Balmer (pseudo-)continuum & Yes & -- \\ 
  Host galaxy template & -- & 5\,Gyr elliptical \\ 
  UV and optical iron templates & Yes & -- \\ 
  Known emission lines & Yes (modified \ion{H}{$\alpha$}) & Yes (modified \ion{H}{$\alpha$}) \\ 
  Known emission line profiles & Gaussian & Gaussian \\ 
  Max number of unknown lines & 2 & 2 \\ 
  \hline
 \end{tabular}
 \label{tab:qsfit_sim_model_summary}
\end{table*}

\subsection{Spectral analysis of the simulated spectra}\label{sec:qsfit_sim_analysis}
\begin{figure}
 \includegraphics[width=\linewidth,clip]{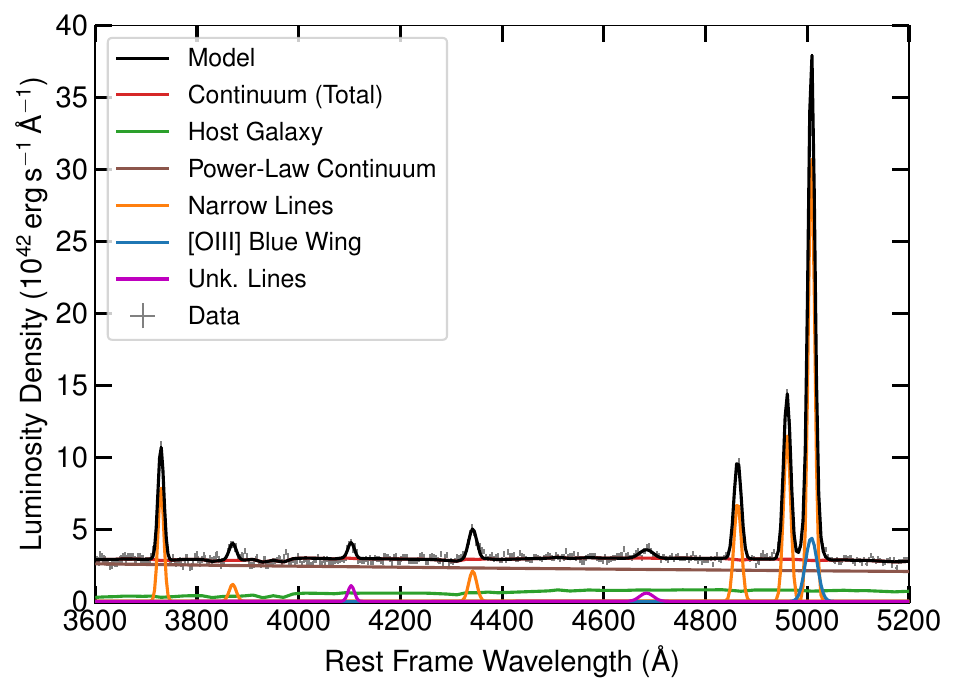}
 \caption{Example showing a \Euclid-like simulated type 2 AGN spectrum where the maximum number of unknown lines are used.}
 \label{fig:simt2_unknownex}
\end{figure}

Identical to the treatment of the incident spectra, we resample the linearly-spaced simulated spectra to be logarithmically spaced in wavelength. We deconvolve the simulated spectra using a Gaussian NISP-S instrumental resolution of 667\,km s$^{-1}$ (corresponding to a spectral resolution of $R = 450$). 
The maximum number of unknown lines allowed to perform the spectral fitting for type 1 and type 2 simulated spectra is reduced with respect to the incident spectra models due to the decreased wavelength range covered by the simulated spectra, which are generated for the NISP-S red grism only. A total of two unknown lines is found to be sufficient for the modelling of unconsidered elements in type 1 and type 2 simulated spectra. \reve{We recall here that the unknown lines are ``nuisance lines" useful to deal with minor spectral features in the data which, if neglected by our model, would produce an overall bad fit. An example is shown in Figure \ref{fig:simt2_unknownex} for a simulated type 2 spectrum. We remind that these lines are never used to model spectral features such as the \ion{H}{$\alpha$}+[\ion{N}{ii}] complex, for instance. The maximum number of unknown lines is utilised in 77\% of type 1 simulated spectra fits and 95\% of type 2 simulated spectra fits.
The fraction in type 2 AGN is the same as for the incident data, whilst it is higher than in type 1 AGN, but we must consider that there are only two unknown lines included in the spectral fit recipe and also that many of the spectra have a low SNR which sometimes results in an unknown line (or two) being used for unphysical noise features. }

Due to the low spectral resolution of NISP-S, we find that the \ion{H}{$\alpha$} complex (\ion{H}{$\alpha$}+[\ion{N}{ii}]) cannot be reasonably deblended in simulated spectra. The blending of the lines results in our standard three-component fitting of the \ion{H}{$\alpha$}+[\ion{N}{ii}] complex becoming highly degenerate. The individual components included for the complex therefore no longer follow their intended use and will blend together to fill the residual of the feature. Figure \ref{fig:ha_blending} presents an example showing the difference between the \ion{H}{$\alpha$} complex in a type 1 incident spectrum and its \Euclid simulation. We compared two alternative methods of modelling the blended \ion{H}{$\alpha$} complex in type 1 simulated spectra: (1) a single broad Gaussian component and (2) both a narrow and broad Gaussian component. We drew a comparison between the total \ion{H}{$\alpha$}+[\ion{N}{ii}] complex flux measured for the incident spectra against the flux measured using our two models for simulated spectra. A single Gaussian profile provided a median percentage difference compared to incident measurements of 26\%, with a median flux measurement difference (incident flux minus simulated flux) of $-6.5\times10^{-18}$\,erg~s$^{-1}$~cm$^{-2}$. By employing a narrow and broad component, we found a median percentage difference compared to incident measurements of 24\%, with a median flux measurement difference of $-3.8\times10^{-16}$\,erg~s$^{-1}$~cm$^{-2}$. Hence, we find a roughly even percentage difference compared to our incident spectra flux measurements using both models, but a lower median flux measurement difference with the single Gaussian model. We therefore opt to use the more simplistic single Gaussian model for type 1 simulated spectra \ion{H}{$\alpha$} complexes as there are no significant gains in measurement performance using a more complex model.

The same blending effect of the \ion{H}{$\alpha$} complex is observed in type 2 AGN simulated spectra (see Fig. \ref{fig:ha_blending}). We find the blended \ion{H}{$\alpha$} complex for type 2 AGN to be well modelled with a broad Gaussian component for which the FWHM is constrained in the range 900--3500\,km s$^{-1}$. We found that Gaussian emission line profiles most effectively capture the flux that is present in type 2 simulated spectra (90\% of that measured in the incident spectra, see Sect. \ref{sec:qsfit_comparisons}). We therefore used Gaussian emission line profiles to model known emission lines in these spectra.

Other than the changes discussed above, all the other components considered in the spectral fit of the type 1 and type 2 AGN simulated spectra models are identical to those used to analyse the type 1 and type 2 incident spectra.\footnote{\url{https://github.com/MattSelwood/QSFit-Euclid-NISP-AGN}} Table \ref{tab:qsfit_sim_model_summary} provides a summary of the individual recipes used to fit type 1 and type 2 simulated \Euclid-like spectra. 
\begin{figure*}
 \centering\includegraphics[width=0.49\textwidth,clip]{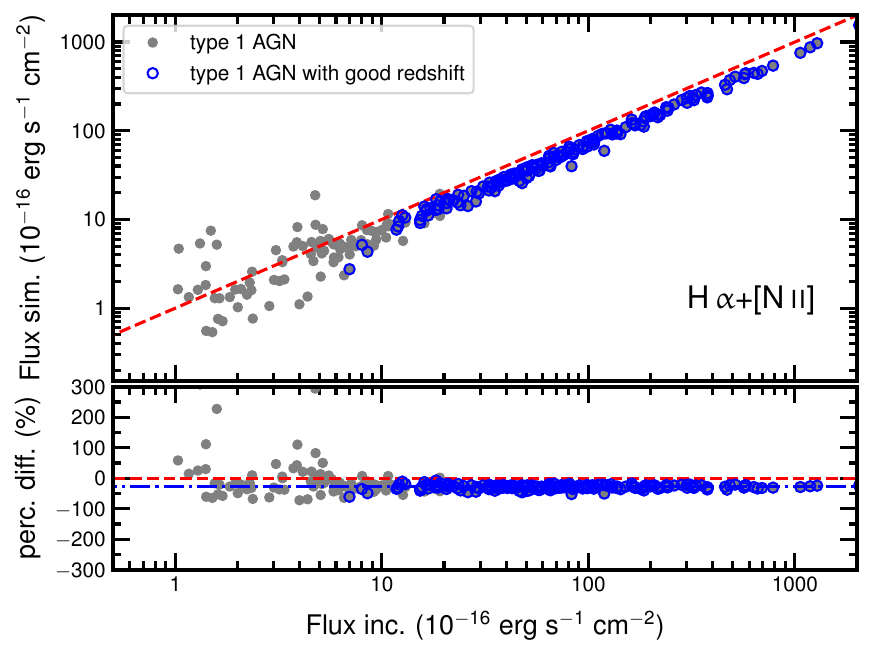}
 \includegraphics[width=0.49\textwidth,clip]{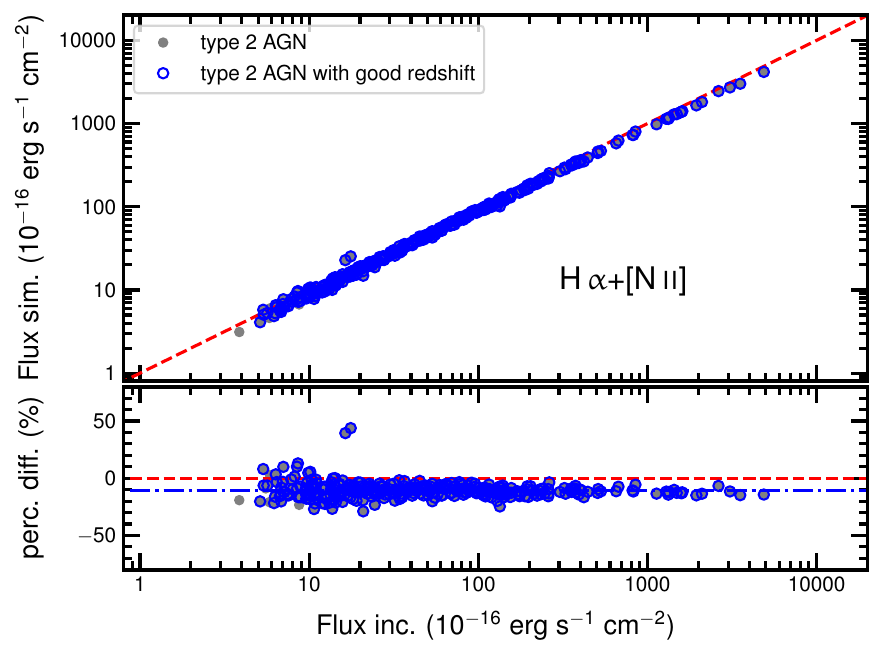}
 \includegraphics[width=0.49\textwidth,clip]{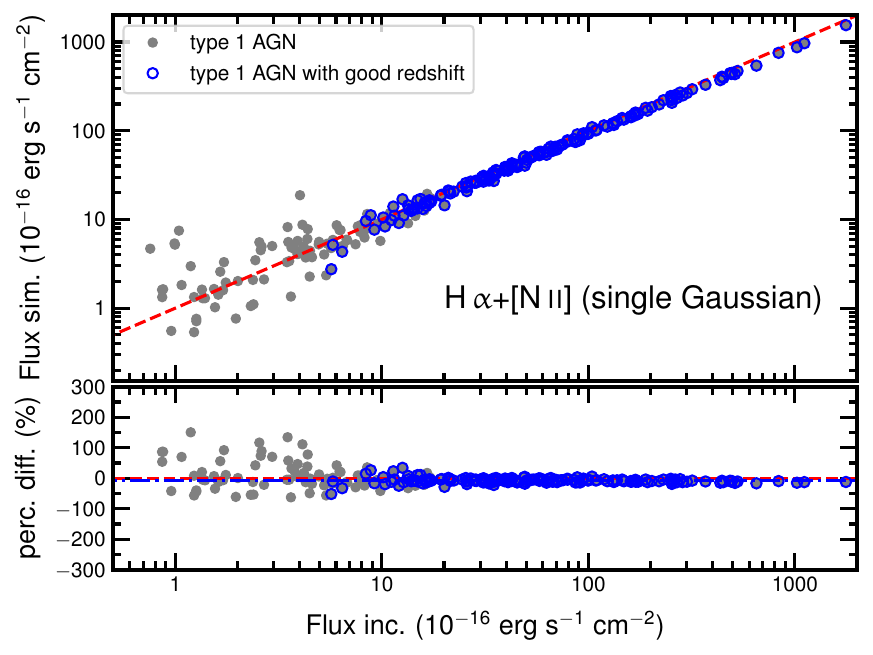}
 \includegraphics[width=0.49\textwidth,clip]{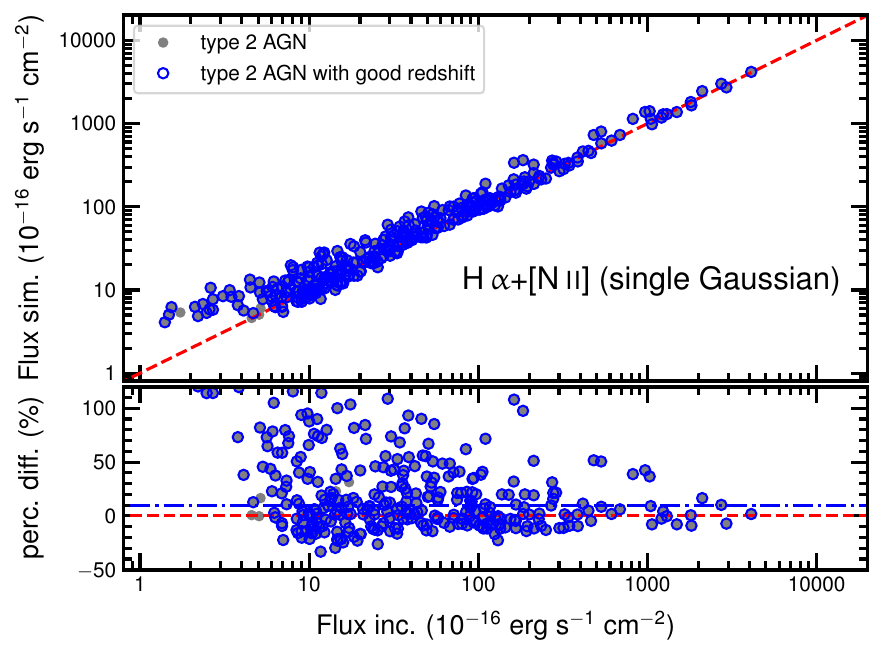}
 \caption{Comparison of the \ion{H}{$\alpha$}$+$[\ion{N}{ii}] flux measurement between the simulated and the incident spectra for the type 1 (\emph{left}) and type 2 (\emph{right}) AGN samples. Sources with a spectroscopic redshift measured from the simulated spectra and flagged as `good' (see Sect.~\ref{Redshift measurements} for details) are marked with blue symbols. The bottom panel in each plot shows the percentage luminosity difference defined as $\Delta F/F=(F_{\rm H\alpha,sim}-F_{\rm H\alpha,inc})/F_{\rm H\alpha,inc}$. The one-to-one flux relation is shown with a red dashed line, which is set to zero in the bottom panel. The dot-dashed blue line in the bottom panel represents the $50^{\rm th}$ percentile of the $\Delta F/F$ distribution. \textit{Top row:} Simulated data: best fit flux obtained from the integration of a single broad component used to model the \ion{H}{$\alpha$}$+$[\ion{N}{ii}] complex. Incident: Best fit flux obtained from the summation of the individual components used to model the \ion{H}{$\alpha$}$+$[\ion{N}{ii}] complex for incident spectra (i.e., broad and narrow component for the \ion{H}{$\alpha$}, [\ion{N}{ii}]$\lambda\lambda$6549, 6583). \textit{Bottom row:} the best fit flux of the \ion{H}{$\alpha$}$+$[\ion{N}{ii}] complex is obtained from the integration of a single broad component for both simulated and incident data.}
 \label{fig:ha_fcomp_12}
\end{figure*}


\section{Results and Discussion}
\label{sec:discussion}
\begin{figure*}[h!]
 \includegraphics[width=0.49\textwidth,clip]{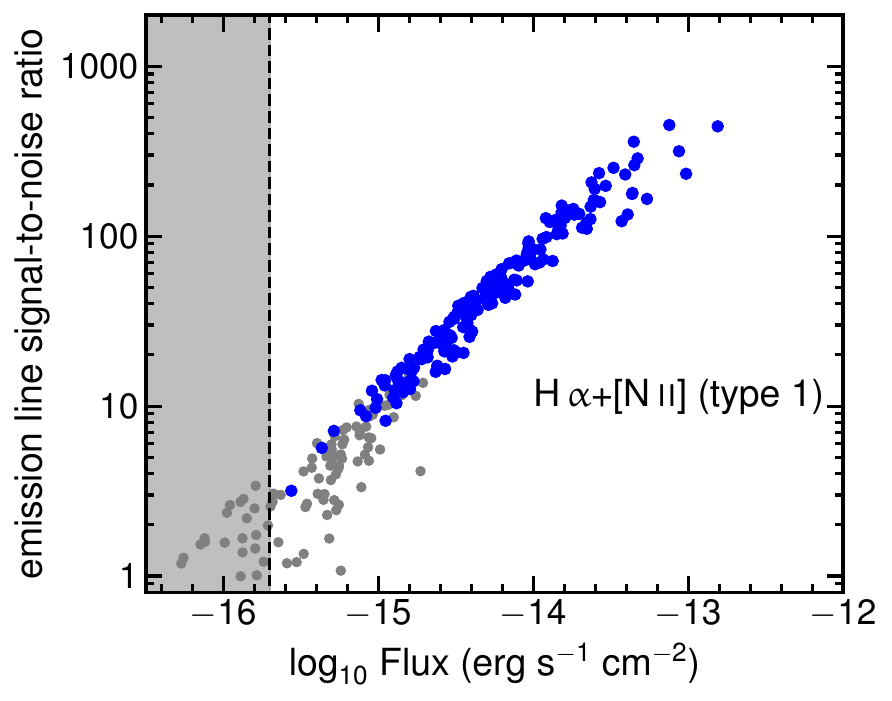}
 \includegraphics[width=0.49\textwidth,clip]{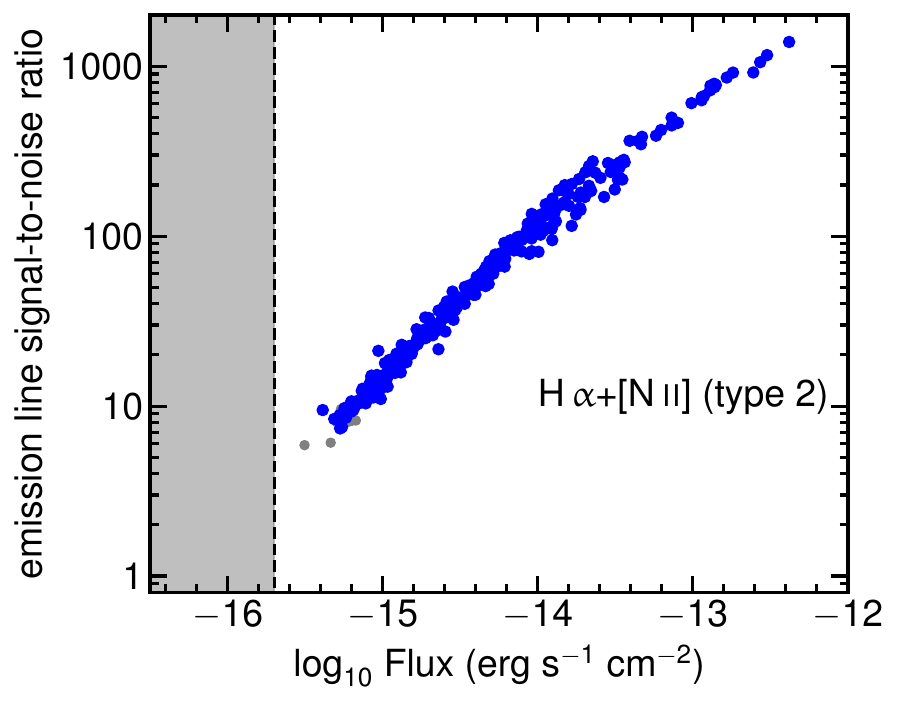}
 \caption{Emission line signal-to-noise ratio as a function of the total integrated flux for \ion{H}{$\alpha$}$+$[\ion{N}{ii}] observed in the simulated spectra in type 1 (\emph{left}) and type 2 (\emph{right}) AGN. The dashed line marks the \ion{H}{$\alpha$} flux limit of the \Euclid survey of $2\times10^{-16}$\,erg s$^{-1}$ cm$^{-2}$. Symbols are as in Fig.~\ref{fig:ha_fcomp_12}.}
 \label{fig:ha_snr_flux_12}
\end{figure*}
\begin{figure*}
 \centering\includegraphics[width=0.8\textwidth,clip]{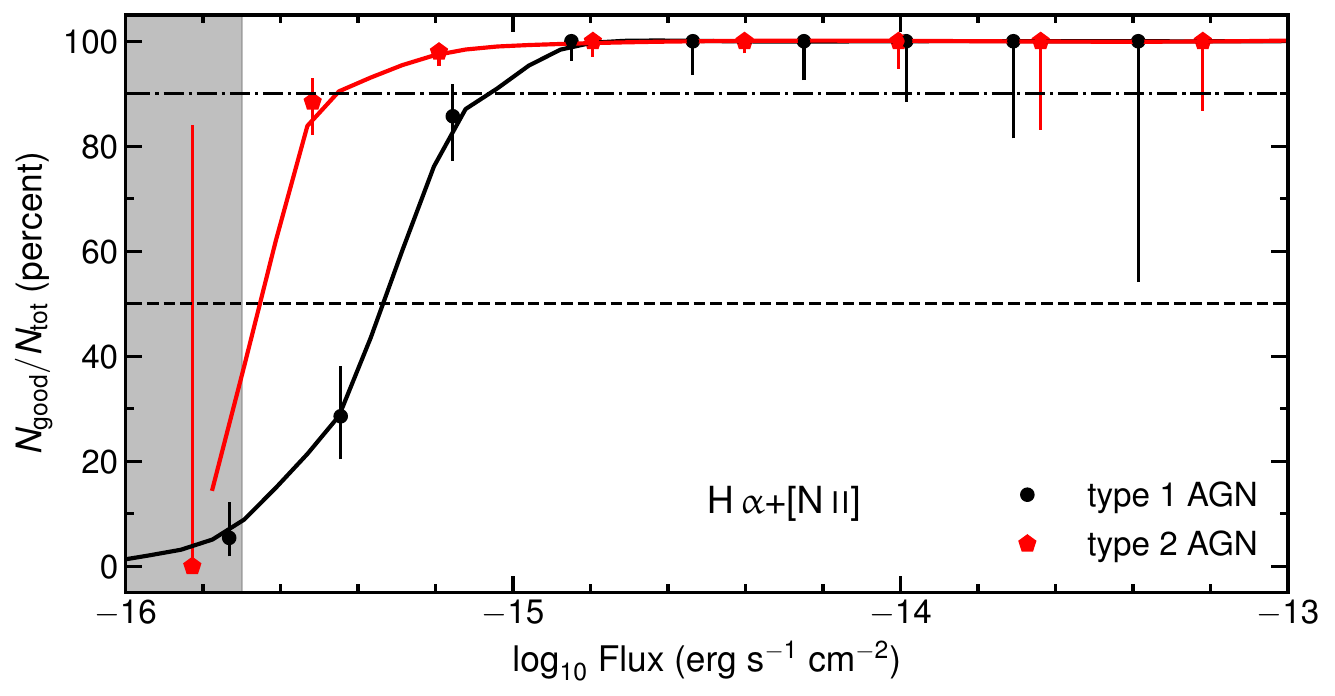}
 \caption{Ratio between the number of sources with a ``good'' redshift measurement (see Sect.~\ref{Redshift measurements} for details) and the total number of objects (in percent) as a function of the integrated flux of the observed \ion{H}{$\alpha$}$+$[\ion{N}{ii}] emission complex in the simulated spectra for type 1 (black filled symbols) and type 2 (red pentagons) AGN. The dashed and dot-dashed lines mark the 50$^{\rm th}$ and 90$^{\rm th}$ percentiles, respectively.}
 \label{fig:ha_percent_12}
\end{figure*}
\subsection{Comparison between incident and simulated data}\label{sec:qsfit_comparisons}

In this Section, we compare emission line fluxes measured with \qsfit{} in both the incident and simulated spectra. 
During the \Euclid mission, all the spectral measurements will be computed by an automatic pipeline that is currently under development by the Euclid collaboration. At the time of writing, the algorithm performance is still under evaluation, and it is beyond the scope of this work to provide an in-depth spectral analysis of the simulated data with this pipeline (details will be provided in Le Brun et al. in preparation). Nonetheless, a preliminary comparison of the emission line flux measurements for the \ion{H}{$\alpha$} complex shows a broad agreement between the values obtained from our user-like analysis and those obtained with the SPE pipeline in its latest available version (see Appendix~\ref{ap-ouspe}).

Since \Euclid is mainly focussed on \ion{H}{$\alpha$} emission as it is expected to be the brighter emission line in the probed redshift range, we concentrated on the comparison of flux measurements for this emission line only. 
This exercise serves to qualify how close the spectral measurements obtained with \Euclid-like spectra are to the truth and to simulate a `user-like' analysis with a spectral fitting pipeline that is not the \Euclid official one. 
Due to the discrete wavelength values used in the construction of mock spectra, the \ion{H}{$\alpha$} complex is only observed in simulated spectra (red grism only) in the redshift range $0.9 <z< 1.8$ for type 1 sources (281 objects) and $0.855 <z< 1.8$ for the type 2 sources (582 objects).  
As the \ion{H}{$\alpha$} complex is blended in simulated spectra (see Sect. \ref{sec:qsfit_sim_analysis}), we compare the flux measured for the single broad component used to model the complex in simulated spectra to the summation of the individual components used to model the \ion{H}{$\alpha$} complex for incident spectra (i.e., broad and narrow component for the \ion{H}{$\alpha$}, [\ion{N}{ii}]$\lambda\lambda$6549, 6583, see Sect. \ref{sec:qsfit_inc_analysis}). To perform the fit and to compute the luminosity values, we considered the input (true) redshift. This is just a (zero-order) working assumption to evaluate the performance of the fitting procedure. 
The \ion{H}{$\alpha$} flux was determined from the simulated spectra of the type 1(2) AGN for 263(353) sources, with 172(344) having a spectroscopic redshift flagged as `good'.   

Figure~\ref{fig:ha_fcomp_12} (top row) presents the comparison of the \ion{H}{$\alpha$}$+$[\ion{N}{ii}] flux measurement between the simulated and the incident spectra for the type 1 (left panel) and type 2 (right panel) AGN samples. We highlighted all objects with a spectroscopic redshift measured from the simulated spectra flagged as `good' (see Sect.~\ref{Redshift measurements}) with blue symbols. We quantified the relative flux difference between the simulated and incident data as $\Delta F/F=(F_{\rm H\alpha,sim}-F_{\rm H\alpha,inc})/F_{\rm H\alpha,inc}$, as shown in the bottom panels of each plot.
The \ion{H}{$\alpha$}+[\ion{N}{ii}] flux values obtained with the simulated and the incident data are in overall good agreement, with the narrower dispersion obtained for the subsample for which we have a good spectroscopic redshift, as expected. 
The median $\Delta F/F$ are about $-28$\% and $-10$\% for the type 1 and type 2 AGN, respectively.
We inspected the best fits of the simulated spectra for both the type 1 and type 2 AGN, finding that the profiles are overall well modelled for both AGN types (see Fig.~\ref{fig:ha_blending}). Nonetheless, a fraction of the flux is lost, especially in the case of type 1 AGN as shown in Fig.~\ref{fig:ha_fcomp_12} (top left panel).

To investigate the reason for this flux loss, we fit all the incident data again with the same prescription employed for the simulated ones: one single broad component to model the whole \ion{H}{$\alpha$}$+$[\ion{N}{ii}] complex. Results are shown in the bottom row of Fig.~\ref{fig:ha_fcomp_12} for type 1 (left) and type 2 (right) AGN. The median $\Delta F/F$ for type 1 is now reduced to about $-8$\%, whilst the flux comparison for the type 2 significantly worsened.
The latter case is explained by the fact that one single Gaussian component is clearly not sufficient to retrieve the \ion{H}{$\alpha$}$+$[\ion{N}{ii}] emission in the incident data. In the case of type 1 AGN, instead, the comparison significantly improved, implying that 1 single component is sufficient to recover most of the \ion{H}{$\alpha$}$+$[\ion{N}{ii}] emission in the simulated data. 

We also checked whether the remaining fraction of the missing flux could be attributed to the slitless spectra extraction procedure. By comparing the median flux values of the incident and simulated spectra, we find that flux loss could be roughly 11\% for type 1 AGN and 2--8\% for type 2 AGN. This implies that flux loss could partly explain the remaining fraction of flux missing in the comparison between simulated and incident flux values (see Appendix~\ref{ap:fluxloss} for details).
We conclude that the low resolution of the NISP instrument will not allow the deblending of the \ion{H}{$\alpha$}$+$[\ion{N}{ii}] complex in AGN of any type, with a single Gaussian profile able to recover $>85$\% of the \ion{H}{$\alpha$}$+$[\ion{N}{ii}] emission in the simulated data.

\subsection{Expected fraction of AGN with spectroscopic redshifts}
As discussed in Sect.~\ref{sec:qsfit_sim_analysis}, the low resolution of the NISP instrument will not deblend the \ion{H}{$\alpha$}$+$[\ion{N}{ii}] complex in AGN of any type. This implies that the number of type 1 AGN for which the redshift will be determined from the observed \Euclid spectra is lower than for type 2 AGN, especially for AGN close to the line flux limit.
The emission line SNR of the \ion{H}{$\alpha$}$+$[\ion{N}{ii}] complex\footnote{The emission line SNR is defined as the ratio between the total \ion{H}{$\alpha$}$+$[\ion{N}{ii}] flux (obtained from the fit of a single Gaussian profile to the line) and the (statistical) uncertainty on this measurement.} observed in the simulated spectra for type 1 and type 2 AGN, for which the spectroscopic redshift is measured, is shown in Fig.~\ref{fig:ha_snr_flux_12} where the emission line SNR is plotted as a function of the integrated observed emission line flux. The extrapolations to the flux limit of the relations fall below the required SNR (3\,$\sigma$), which is more pronounced for type 1 AGN, as expected, given that the noise associated to a broad line is greater than the one associated to the narrow one (i.e. the number of pixels is higher for a broad line and thus the related noise).
We also computed the ratio between the number of sources with a ``good'' redshift measurement and the total number of objects as a function of the integrated \ion{H}{$\alpha$}$+$[\ion{N}{ii}] flux in the simulated spectra for both type 1 and type 2 AGN (Fig.~\ref{fig:ha_percent_12}). 
A spectroscopic redshift is determined for $\sim$90\% of the simulated type 2 AGN down to an emission line flux of roughly $3\times10^{-16}$\,erg s$^{-1}$ cm$^{-2}$, whilst the emission line flux value is more than a factor of 2 higher for type 1 AGN at the same percentage, i.e., $8.5\times10^{-16}$\,erg s$^{-1}$ cm$^{-2}$.

\subsection{Black hole mass estimates using the \ion{H}{$\alpha$}}
\label{BHmass estimates}
The measurement of the black hole mass (\mbh) is challenging, even when reverberation mapping campaigns are available \citep[e.g.,][]{bentz2018}.
Yet, \mbh\ is one of the key parameter for studies focussed on determining the scaling relations between black hole and host galaxy properties. These scaling relations are pivotal to test black hole feedback mechanisms \citep[e.g.,][]{Steinborn2015,Weinberger2017} and cosmological hydrodynamical simulations of structure formation that aim at investigating galaxy-black hole growth \citep[e.g.,][]{DeGraf2015}.
\Euclid will observe hundreds of thousand AGN spectroscopically, thus providing a prime dataset for such studies.

\begin{figure}[h!]
 \includegraphics[width=0.49\textwidth,clip]{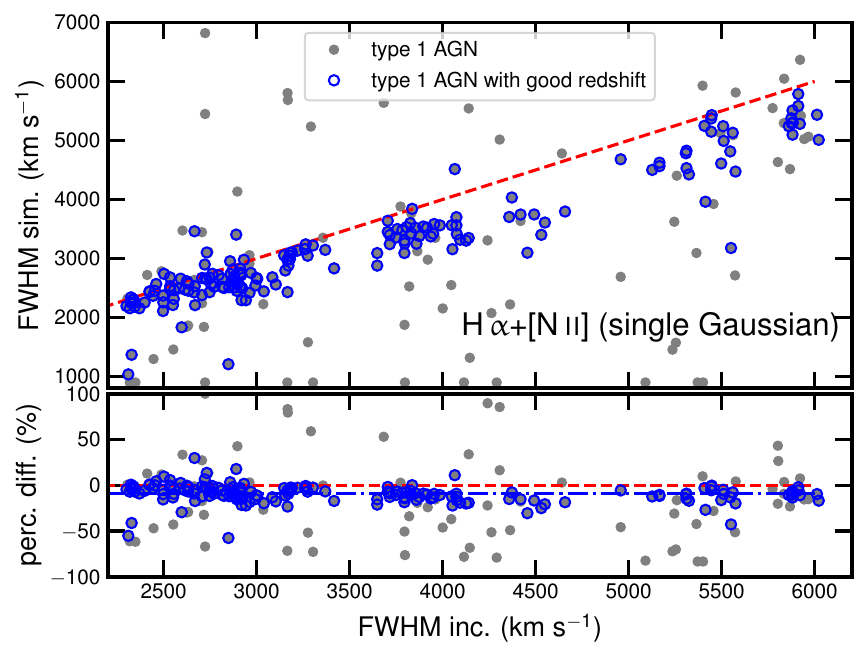}
 \caption{Comparison of the \ion{H}{$\alpha$} FWHM between incident and simulated data. The best fit of incident spectra considers one single Gaussian component of the \ion{H}{$\alpha$} emission line profile (see Sect.~\ref{sec:qsfit_comparisons}). Symbols are as in Fig.~\ref{fig:ha_fcomp_12}.}
 \label{fig:fwhmcomp_singleparam}
\end{figure}
\begin{figure*}[h!]
 \centering\includegraphics[width=0.49\textwidth,clip]{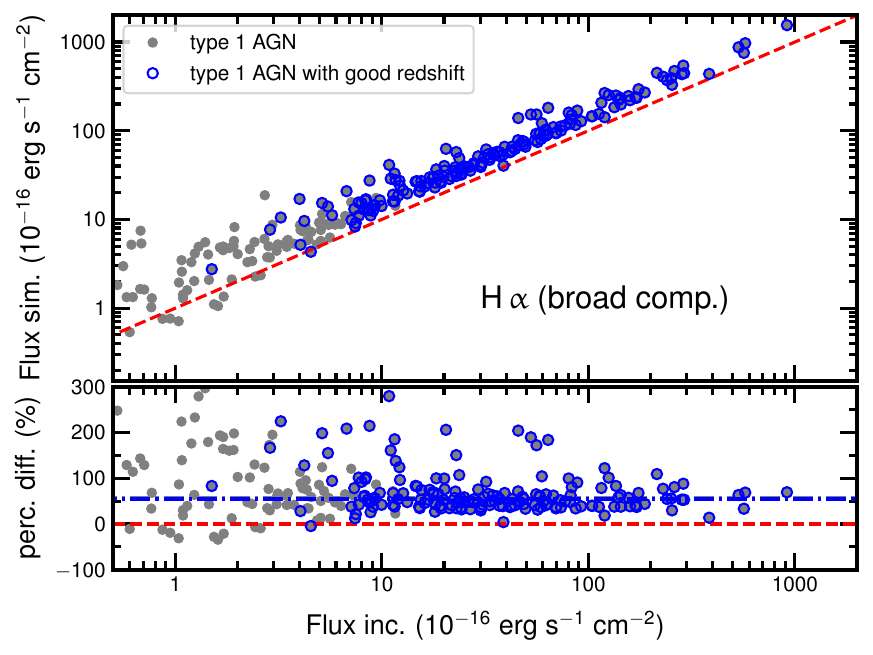}
 \includegraphics[width=0.49\textwidth,clip]{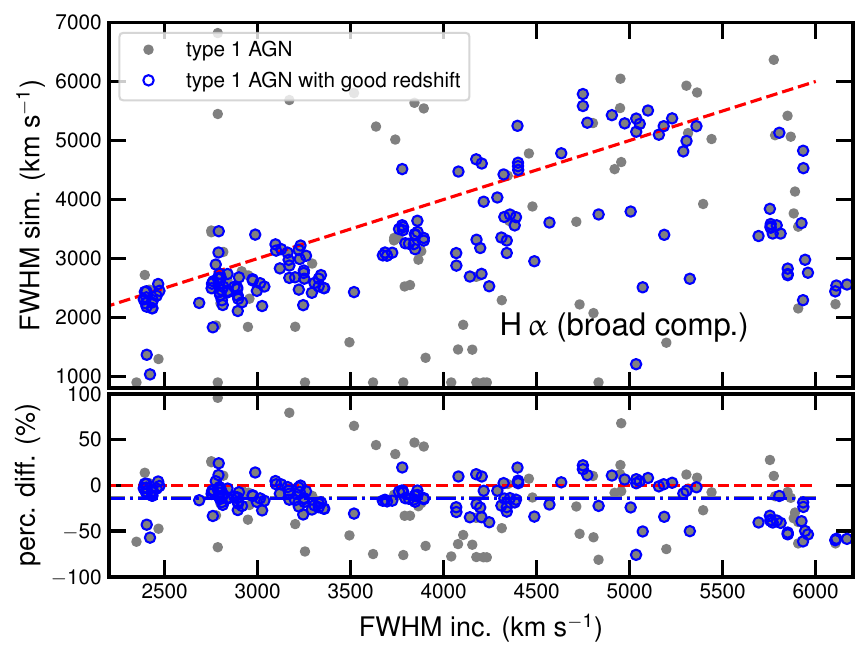}
 \caption{\textit{Left panel}: Comparison of the \ion{H}{$\alpha$} flux between incident and simulated data. \textit{Right panel}: Comparison of the \ion{H}{$\alpha$} FWHM between incident and simulated data. The luminosity and FWHM values derived for incident spectra considers only the broad component of the \ion{H}{$\alpha$} emission line profile (see Sect.~\ref{sec:qsfit_inc_analysis}). Symbols are as in Fig.~\ref{fig:ha_fcomp_12}.}
 \label{fig:bhcomp_brparam}
\end{figure*}
\begin{figure*}
\centering\includegraphics[width=0.49\textwidth,clip]{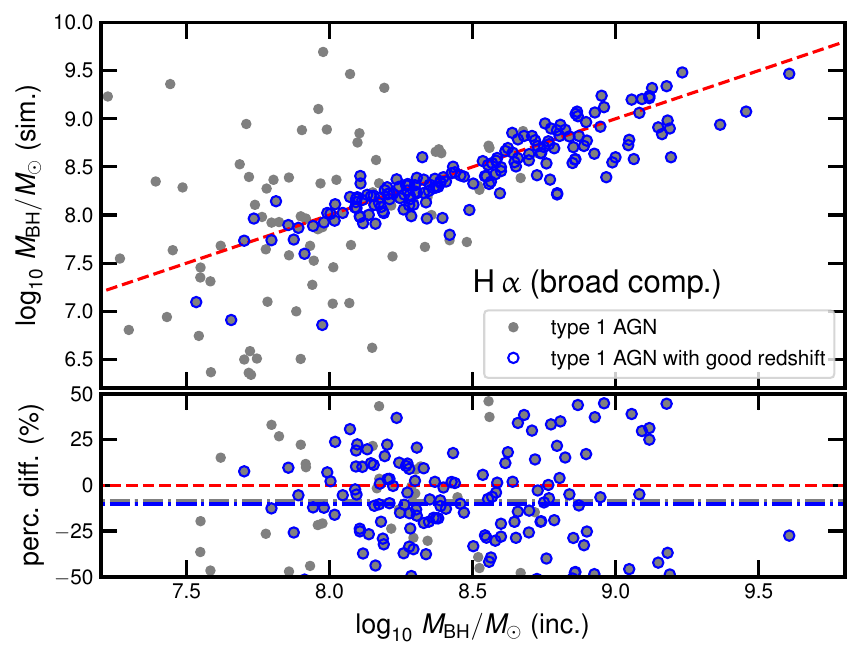}
\includegraphics[width=0.49\textwidth,clip]{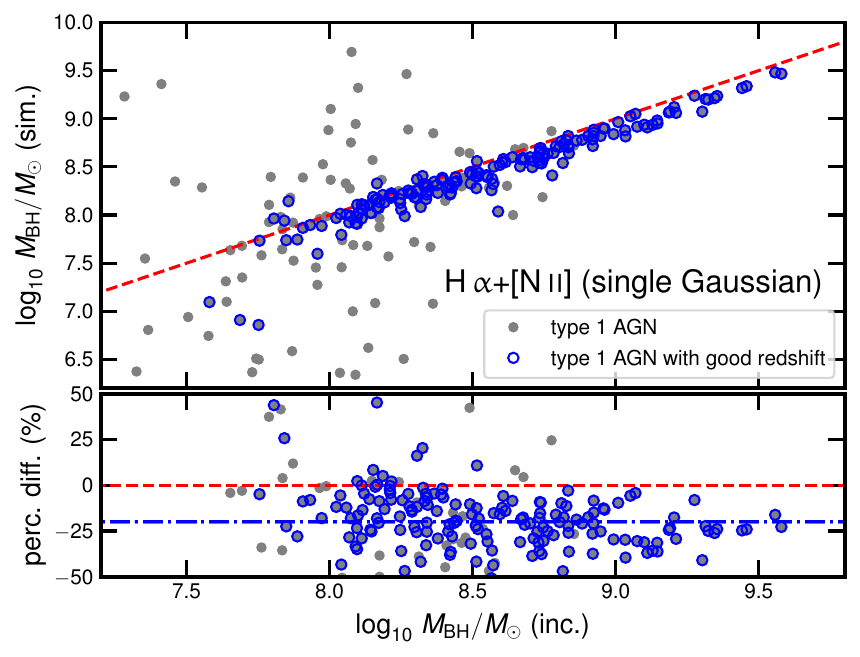}
 \caption{Comparison of \mbh\ estimates between incident and simulated data derived from the \ion{H}{$\alpha$} (see Sect.~\ref{BHmass estimates} for details). \textit{Left panel}: \mbh\ derived from the best fit of the incident data that considers only the broad component of the \ion{H}{$\alpha$} emission line profile (see Sect.~\ref{sec:qsfit_inc_analysis}). \textit{Right panel}: \mbh\ derived from the best fit of the incident data that considers one single Gaussian component of the \ion{H}{$\alpha$} emission line profile (see Sect.~\ref{sec:qsfit_comparisons}). Symbols are as in Fig.~\ref{fig:ha_fcomp_12}.}
 \label{fig:bhcomp}
\end{figure*}

\reve{To estimate \mbh\ for both the incident and simulated datasets, we considered the  recalibrated single-epoch virial relationships recently published by \citet[][see their Table 4]{kynoch2023} that make use of the integrated line flux and FWHM components of the \ion{H}{$\alpha$} emission line. We thus have to investigate first how well the FWHM of the broad \ion{H}{$\alpha$} can be recovered, despite the strong blending expected for the \Euclid NISP spectra.

Fig.~\ref{fig:fwhmcomp_singleparam} compare the \ion{H}{$\alpha$} FWHM between incident and simulated data when the best fit of incident spectra considers one single Gaussian component of the \ion{H}{$\alpha$} emission line profile. Despite the scatter, the values obtained for the type 1 AGN with a robust redshift measurement follow a one-to-one relation, with a median percentage error on the \ion{H}{$\alpha$} FWHM of roughly $-9\%$. We then plotted the comparison of the \ion{H}{$\alpha$} flux and FWHM between incident and simulated data only for the broad component of the \ion{H}{$\alpha$} emission line in Fig.~\ref{fig:bhcomp_brparam}. The median percentage error for \ion{H}{$\alpha$} flux and FWHM are about 55\% and $-14\%$, respectively, with a rather scattered distribution of FWHM values. 

Finally, Fig.~\ref{fig:bhcomp} presents the comparison of \mbh\ estimates between the incident and the simulated data. We quantified the relative difference between \mbh\ for simulated and incident data as $\Delta M_{\rm BH}/M_{\rm BH}=(M_{\rm BH,sim}-M_{\rm BH,inc})/M_{\rm BH,inc}$, which is shown in the bottom panel.
The left panel displays the \mbh\ values obtained from the incident data where only the broad component of the \ion{H}{$\alpha$} emission line profile is considered (see Sect.~\ref{sec:qsfit_inc_analysis}). The agreement is good overall, with a median $\Delta M_{\rm BH}/M_{\rm BH}$ of about $-10\%$. The scatter in the \mbh\ distribution reduces when we consider one single Gaussian component of the \ion{H}{$\alpha$} emission line profile (right panel of Fig.~\ref{fig:bhcomp}) to derive \mbh\ for incident spectra.
Yet, the average shift in the latter case is about $-20\%$.
}

Summarising, although the estimate of both \ion{H}{$\alpha$} flux and FWHM are individually offset with respect to the reference values, the effect of these two measures on the determination of the \mbh\ is opposite. 
The combination of these different shifts in \ion{H}{$\alpha$} flux and FWHM produces the different offsets in \mbh\ discussed above.

\section{Conclusions}
In this work, we present the performance on the spectroscopic characterisation of both unobscured (type 1) and obscured (type 2) AGN with the NISP spectrograph, which is mounted in the ESA \Euclid space telescope. NISP operates in the spectral range 0.9--2\,$\mu$m, opening a new window for spectroscopic AGN studies in the NIR. Several hundred thousand AGN will be observed spectroscopically across a wide redshift range, which will represent the largest AGN sample with NIR spectroscopy thus far. 
We built AGN template libraries that provide realistic spectra for type 1 and type 2 AGN that will be detected by the NISP spectrograph. Starting from these incident spectral models, we then constructed a dedicated set of simulations to create mock AGN spectra, from the rest-frame NIR to the UV, and determined the emission line flux of the \ion{H}{$\alpha$}$+$[\ion{N}{ii}] complex in both the incident and simulated data.
We summarise the main results as follows.
\begin{itemize}
    \item Redshift measurements are robust in the interval where \ion{H}{$\alpha$} is visible within the spectral coverage of the red grism (i.e., $0.89<z<1.83$, probing the peak of AGN activity at $z\simeq 1$--1.5). In this range, our simulation comprises 889 AGN, with 83\% having good redshifts (and an additional 6\% have good redshift but are not reliable, i.e., $R_{\rm xcorr}<5$). \reve{We determined an upper limit on the} purity in this redshift range to be 98\% (with only 11 objects classified as ``false''). Outside this redshift range, the measurement efficiency decreases significantly. At high redshift, where we have fainter objects and a wider redshift interval, the emission lines are weak or absent. At low redshift ($z<0.83$), the high inefficiency in the redshift determination is related to the lack of intense emission lines. 

    \item The low spectral resolution of the NISP instrument does not allow the deblending of the \ion{H}{$\alpha$}$+$[\ion{N}{ii}] complex in AGN of any type, with a single Gaussian profile able to recover $>85$\% of the \ion{H}{$\alpha$}$+$[\ion{N}{ii}] emission in the simulated data (see Sect.~\ref{sec:qsfit_sim_analysis} for details). A small fraction of the observed flux could also be lost at the slitless spectra extraction level, which is expected to be mitigated by the optimal extraction algorithm that is implemented in the upcoming version of the OU-SIR pipeline.

    \item A spectroscopic redshift is determined for about 90\% of the simulated type 2 AGN down to an emission line flux of roughly $3\times10^{-16}$\,erg s$^{-1}$ cm$^{-2}$. The emission line flux value is more than a factor of 2 higher for type 1 AGN at the same percentage, i.e., $8.5\times10^{-16}$\,erg s$^{-1}$ cm$^{-2}$.

    \item We investigated how well luminosity and FWHM of the broad component of \ion{H}{$\alpha$} can be recovered, despite the strong blending expected for the \Euclid NISP spectra. The median percentage error for \ion{H}{$\alpha$} luminosity and FWHM are about 55\% and $-14\%$, respectively, with a rather scattered distribution of FWHM values. Regarding the estimate of the \mbh, we find a good agreement overall, with a median $\Delta M_{\rm BH}/M_{\rm BH}$ of about $-10$\%.
\end{itemize}

\reve{We caution that the simulation used in the present analysis assumed (1) no contamination from neighbouring spectra, (2) a single value for both sky background and out-of-field stray light, and (3) luminosity values are extracted at the true redshift. Regarding the latter, our main results are presented only for the fraction of objects with a good redshift measurement, therefore, even if the the purity will be different in the real dataset \Euclid will observe, this approximation is suitable for sensitivity analysis regarding redshift, \ion{H}{$\alpha$} luminosity measurement, and black hole mass estimation. Extracted NISP-S spectra should have low contamination due to neighbouring objects due to the fact that four orientations of spectra are combined. Thus, the simulations we employed, even though simplified at the level of creation, should, in the end, reflect reasonable NISP-S spectra (see also Sect. 6 in \citealt{Gabarra23}).
The simplification regarding straylight is justified by assuming that straylight from very bright stars outside the current field-of-view would not have a strong gradient along a NISP-S spectrum, so should just represent a raised background. Moreover, straylight from nearby stars should also be somewhat different for the four spectrum orientations and hence be washed-out to some degree when combined. We also caveat that our results for the type 1 AGN are biassed to the bright blue AGN population, and since we have excluded BALs and radio bright AGN from the main AGN sample, \rev{our analysis could thus be extended to include these interesting AGN types, especially the most extreme classes in these two categories (e.g., FeLoBALs, LoBALs)}, perhaps folding-in the AGN luminosity function to quantify the capability of NISP to retrieve the AGN classification from the NISP data.
}

The presented analysis further extends the campaign to assess the performance of the NISP spectrograph in the context of active galaxies. 
Finally, the expected high spectroscopic coverage of AGN at $z<2$ will be of prime importance for studies on AGN demography, scaling relation, and clustering from the epoch of the peak of AGN activity down to the present-day Universe.

\section*{Acknowledgements}
V. A. acknowledges support from INAF-PRIN 1.05.01.85.08. L. B. acknowledges financial support from grant PRIN2017. A. F. acknowledges the support from project "VLT-MOONS" CRAM 1.05.03.07, INAF Large Grant 2022 "The metal circle: a new sharp view of the baryon cycle up to Cosmic Dawn with the latest generation IFU facilities" and INAF Large Grant 2022 "Dual and binary SMBH in the multi-messenger era". F.S. acknowledges partial support
from the European Union’s Horizon 2020 research and innovation programme
under the Marie Skłodowska-Curie grant agreement No. 860744.
The Euclid Consortium acknowledges the European Space Agency and a number of agencies and institutes that have supported the development of \Euclid, in particular the Academy of Finland, the Agenzia Spaziale Italiana, the Belgian Science Policy, the Canadian Euclid Consortium, the French Centre National d’Etudes Spatiales, the Deutsches Zentrum f\"{u}r Luft- und Raumfahrt, the Danish Space Research Institute, the Funda\c{c}\~{a}o para a Ci\^{e}ncia e a Tecnologia, the Ministerio de Ciencia e Innovación, the National Aeronautics and Space Administration, the National Astronomical Observatory of Japan, the Netherlandse Onderzoekschool Voor Astronomie, the Norwegian Space Agency, the Romanian Space Agency, the State Secretariat for Education, Research and Innovation (SERI) at the Swiss Space Office (SSO), and the United Kingdom Space Agency. A complete and detailed list is available on the \Euclid web site (\url{http://www.euclid-ec.org}).



%
   \bibliographystyle{aa} 
   \bibliography{bibl, Euclid} 
%



\appendix

\section{Example of \Euclid-like simulated spectra}
\label{Sim atlas}
\begin{figure*}
 \begin{center}
 \includegraphics[width=0.49\textwidth,clip]{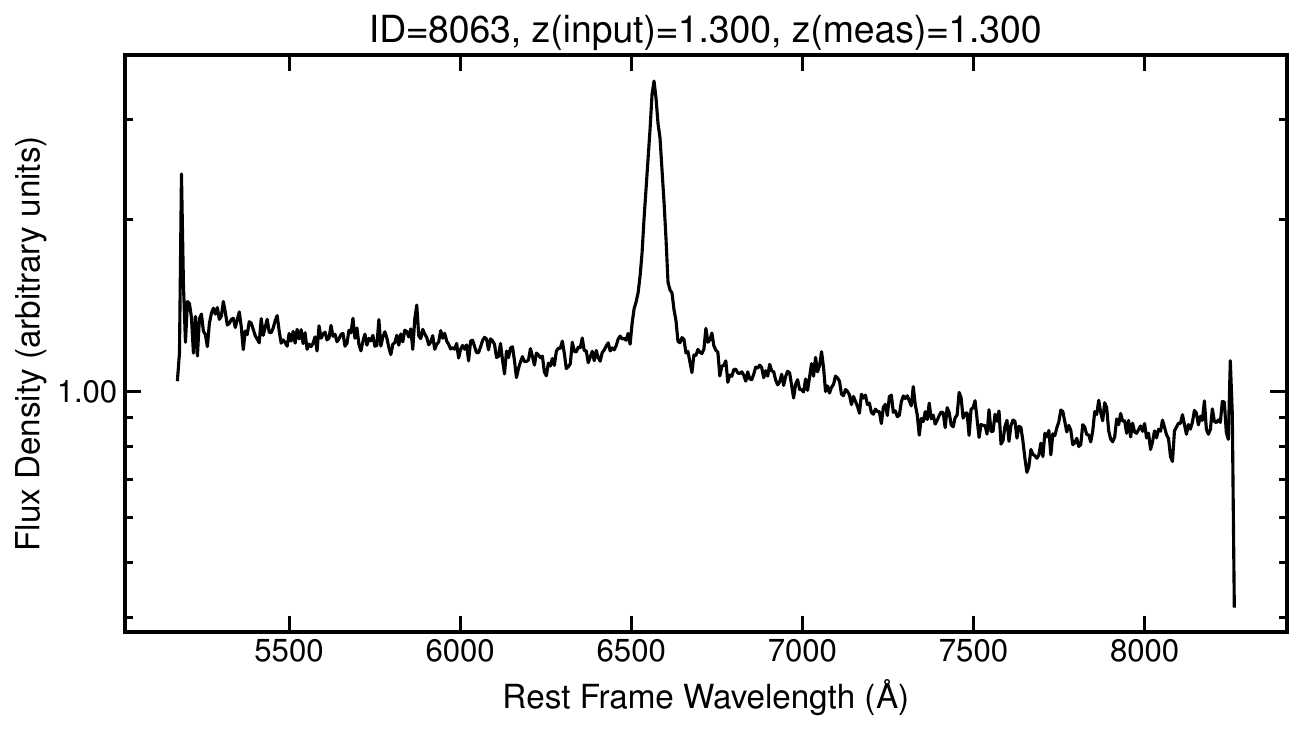}
 \includegraphics[width=0.49\textwidth,clip]{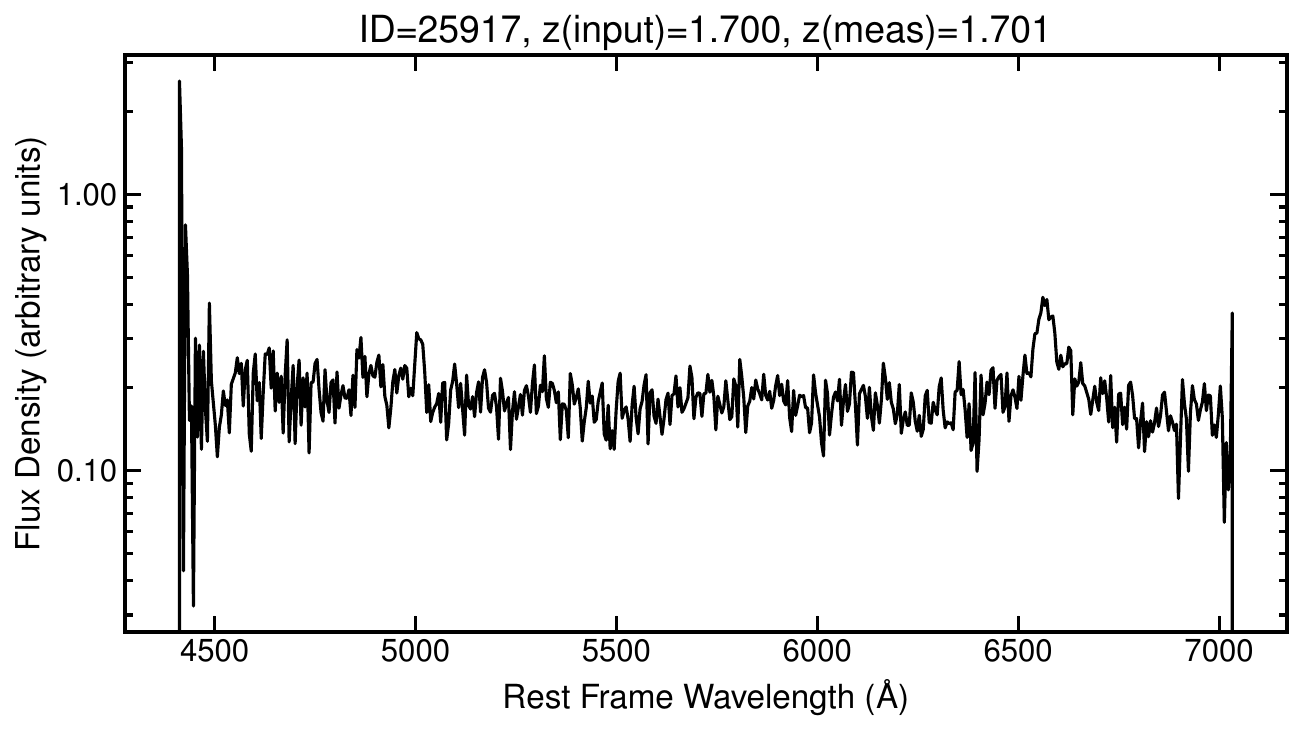}
\includegraphics[width=0.49\textwidth,clip]{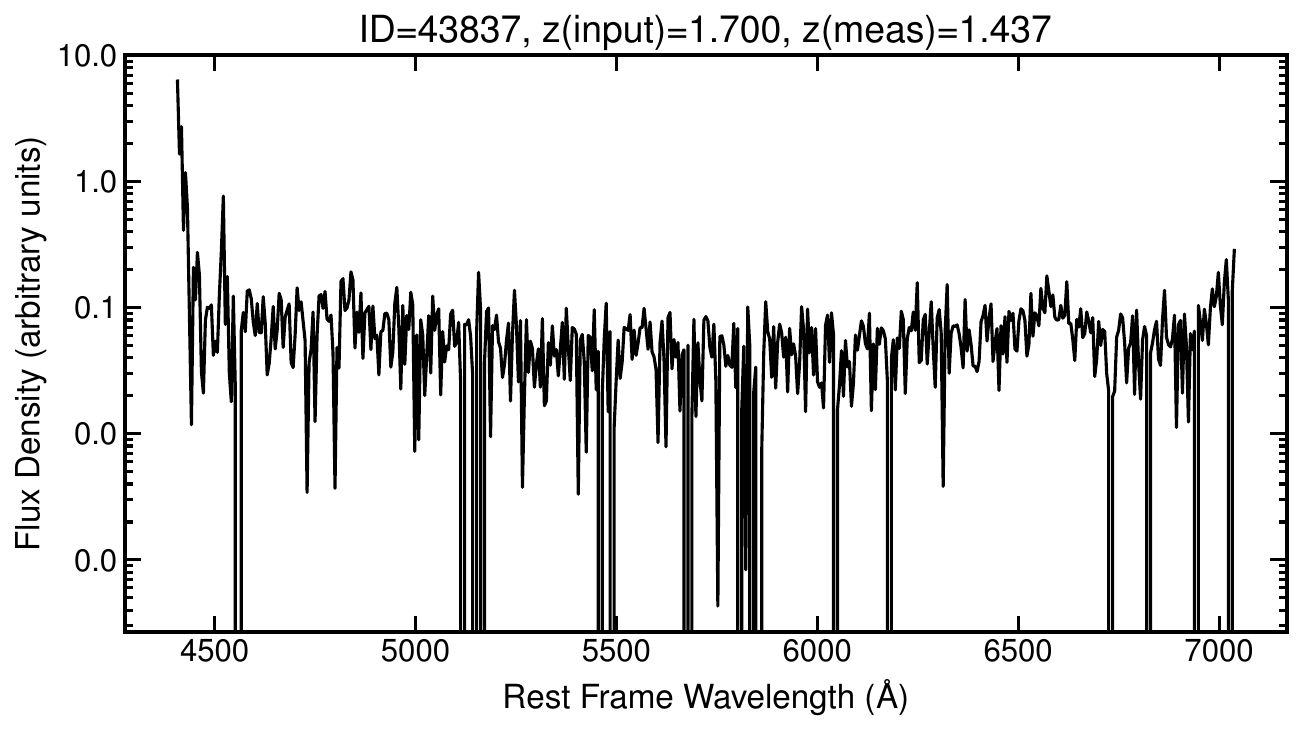}
 \includegraphics[width=0.49\textwidth,clip]{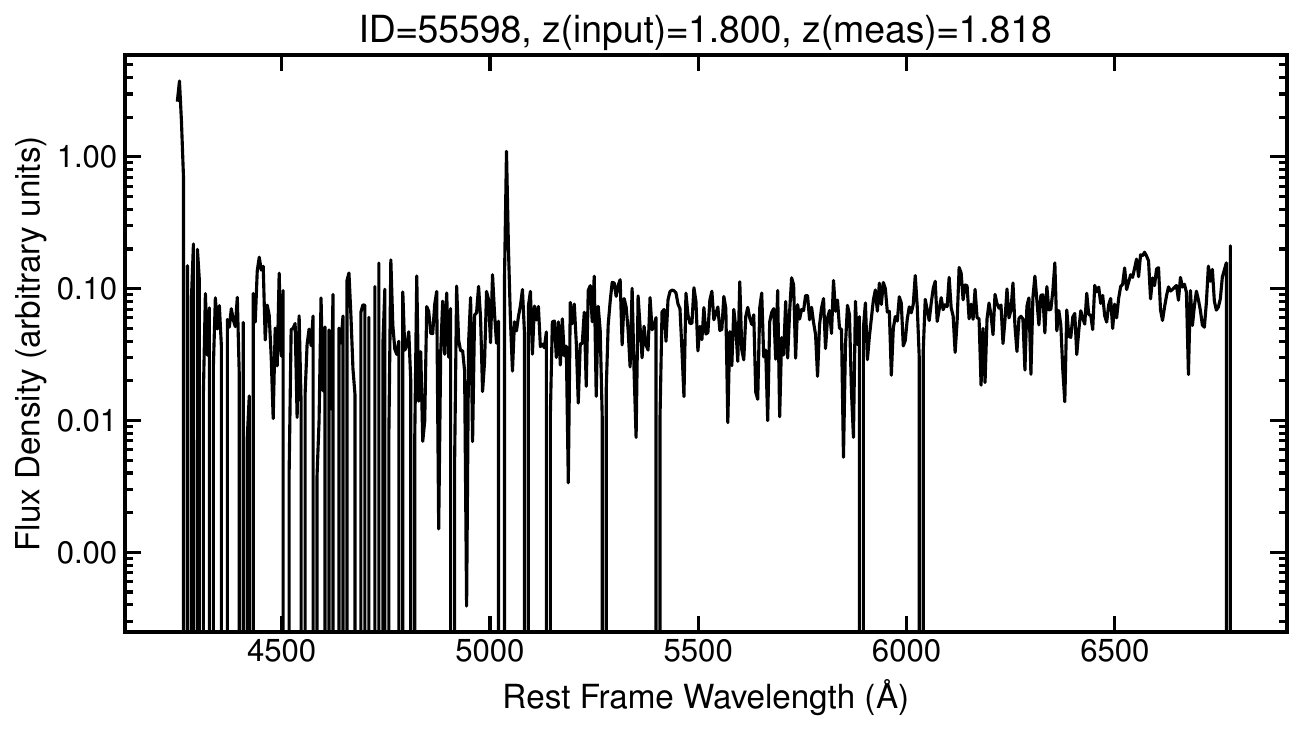} \includegraphics[width=0.49\textwidth,clip]{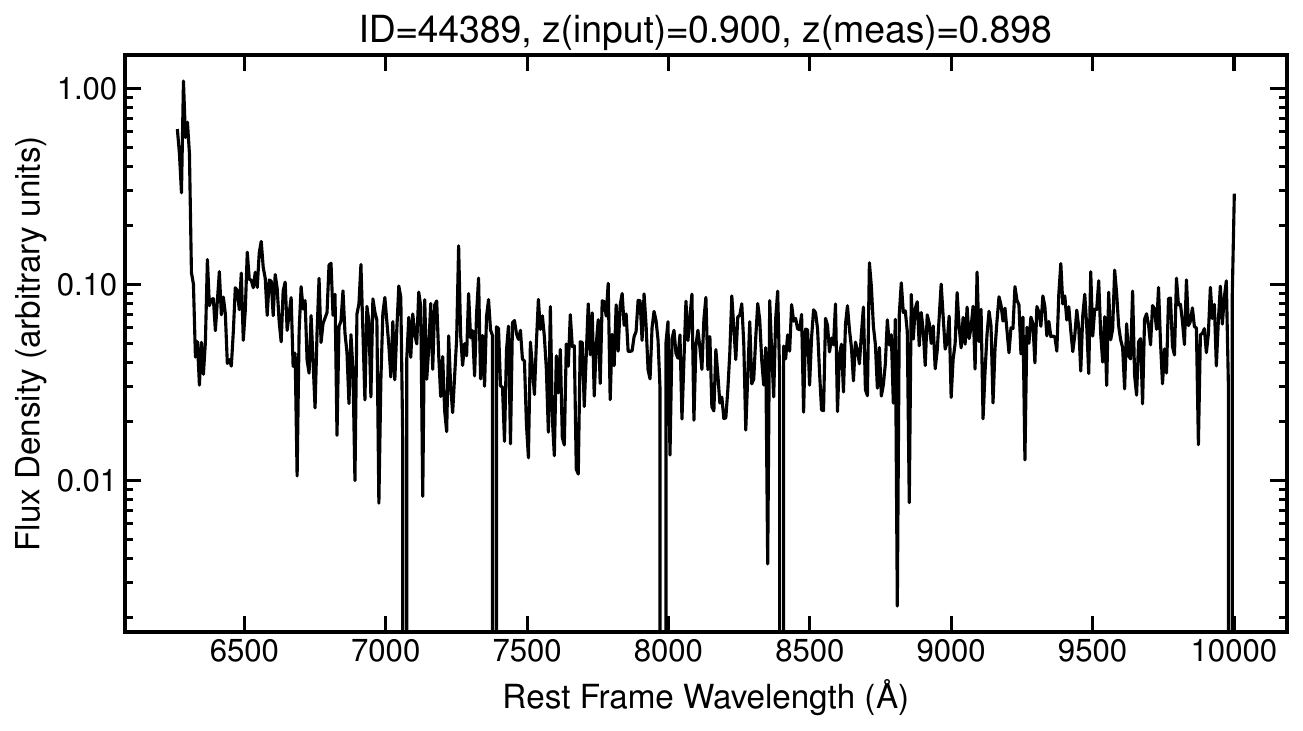}
 \includegraphics[width=0.49\textwidth,clip]{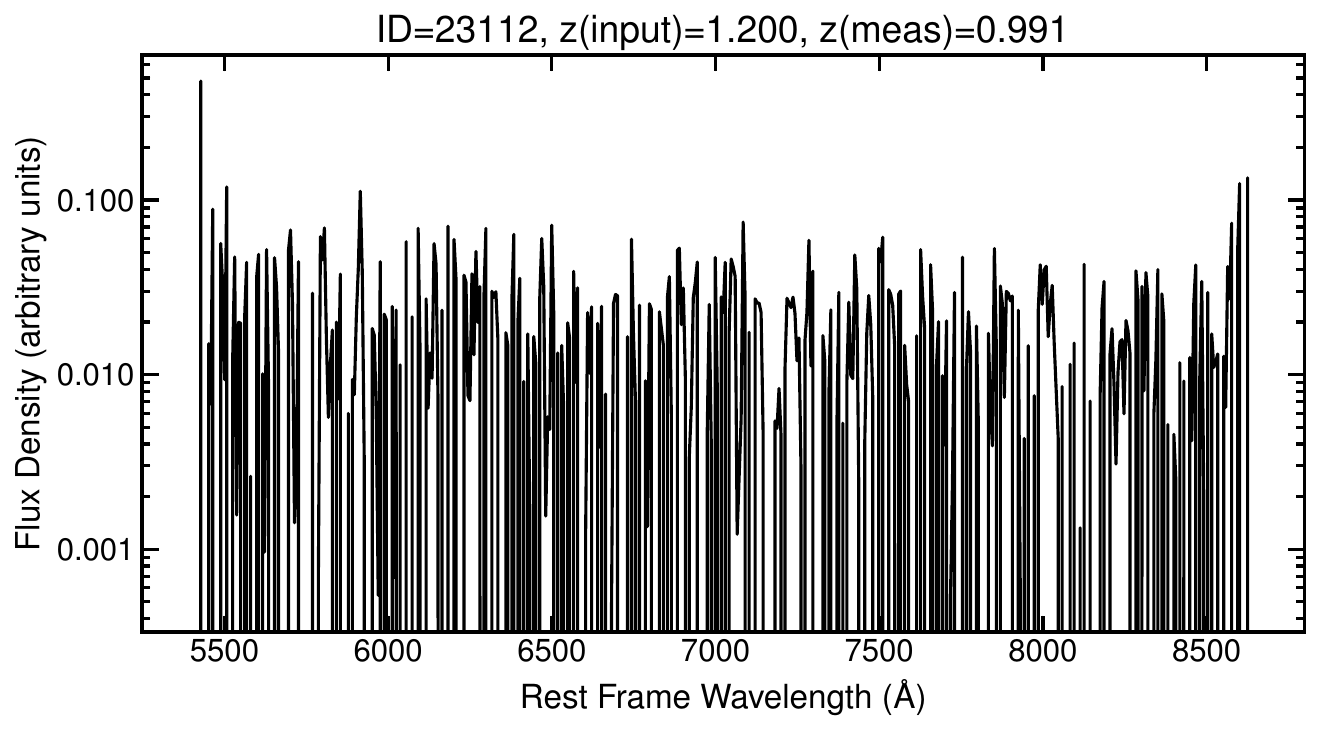}
 \end{center}
 \caption{Example of \Euclid-like simulated spectra for type 1 AGN in the redshift interval $0.89\leq z \leq 1.83$, thus where the \ion{H}{$\alpha$}$+$[\ion{N}{ii}] complex is within the spectral coverage of the red grism. Spectra are shown at different SNR and noise levels (medium/high on the left and low on the right) for the \textit{good} (top row), \textit{false} (mid row), and \textit{lost} (bottom row) redshift classes (see Sect.~\ref{Redshift measurements}).}
 \label{fig:example1}
\end{figure*}

\begin{figure*}
 \begin{center}
 \includegraphics[width=0.49\textwidth,clip]{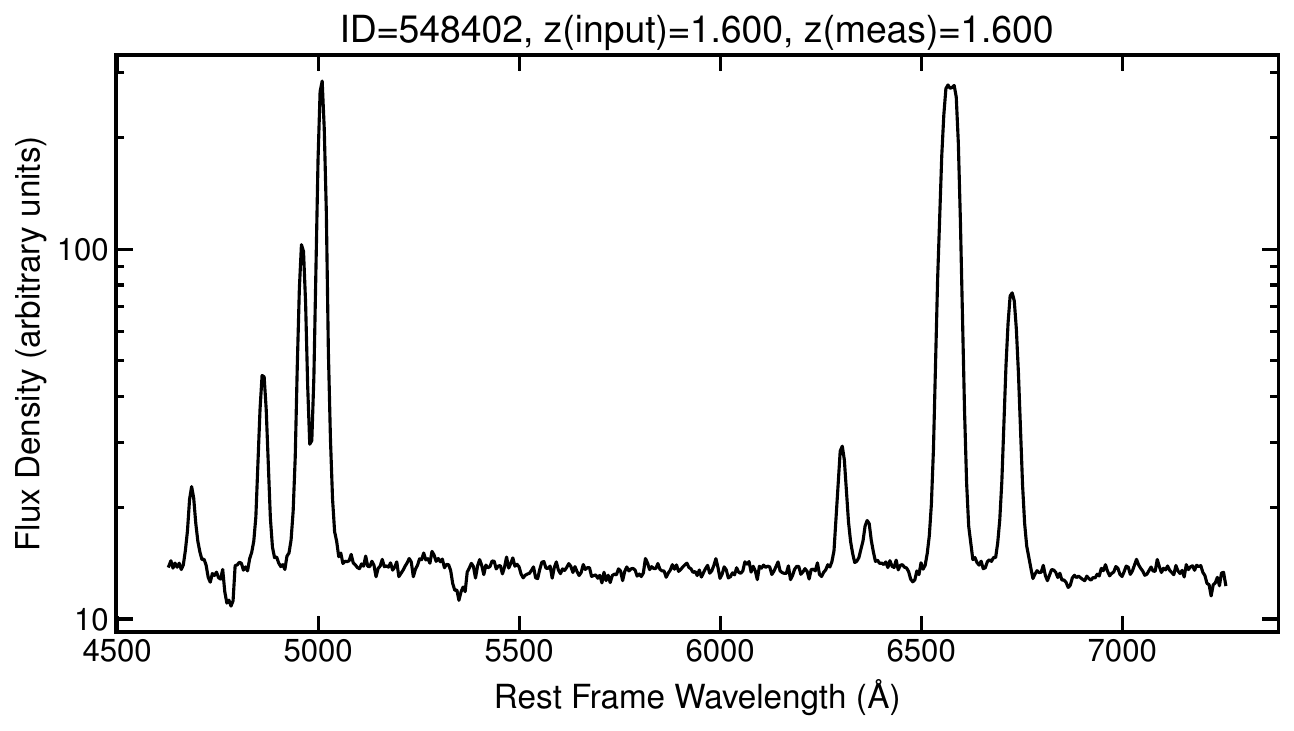}
 \includegraphics[width=0.49\textwidth,clip]{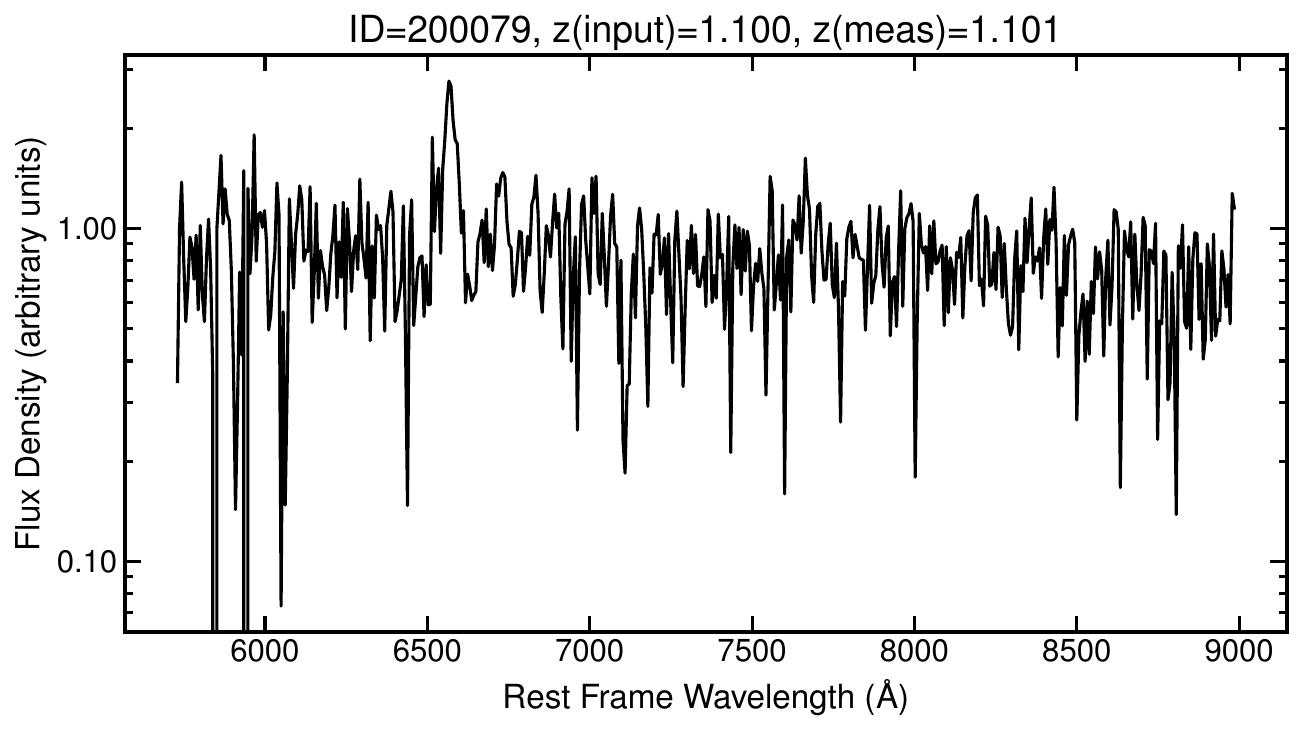}
 \includegraphics[width=0.49\textwidth,clip]{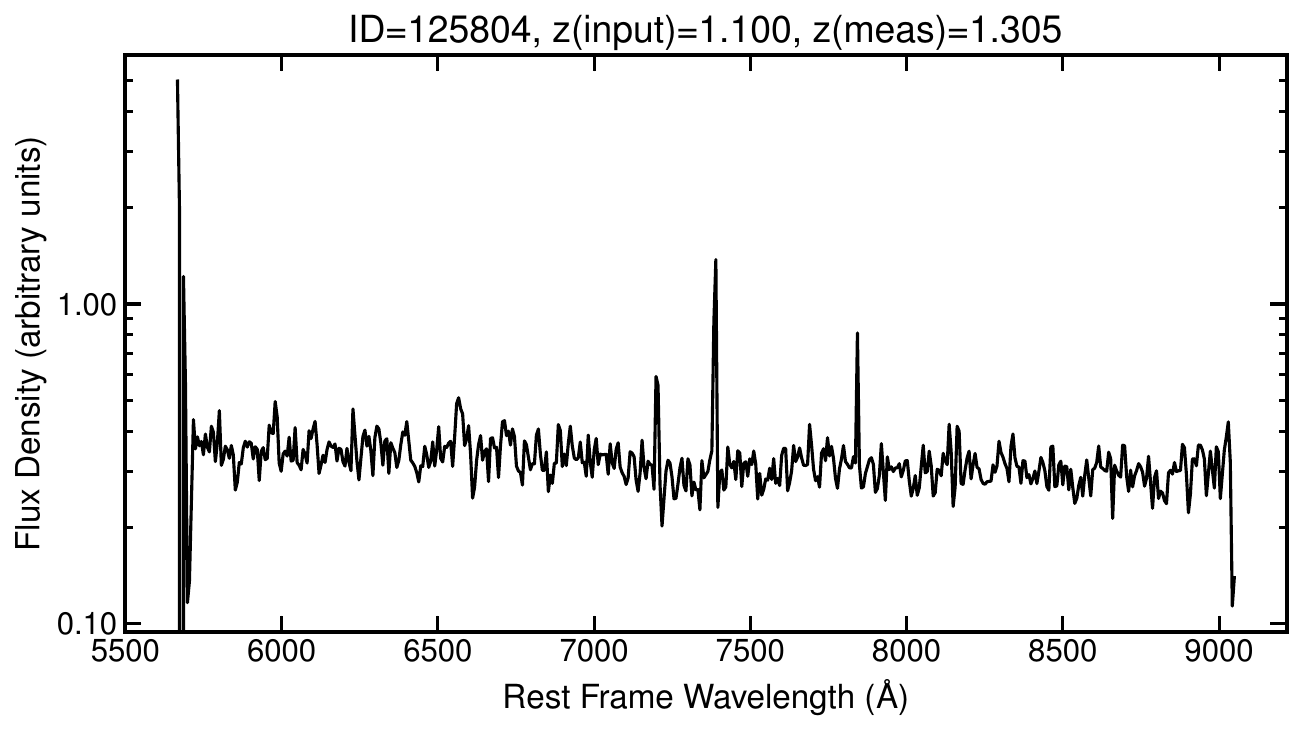}
 \includegraphics[width=0.49\textwidth,clip]{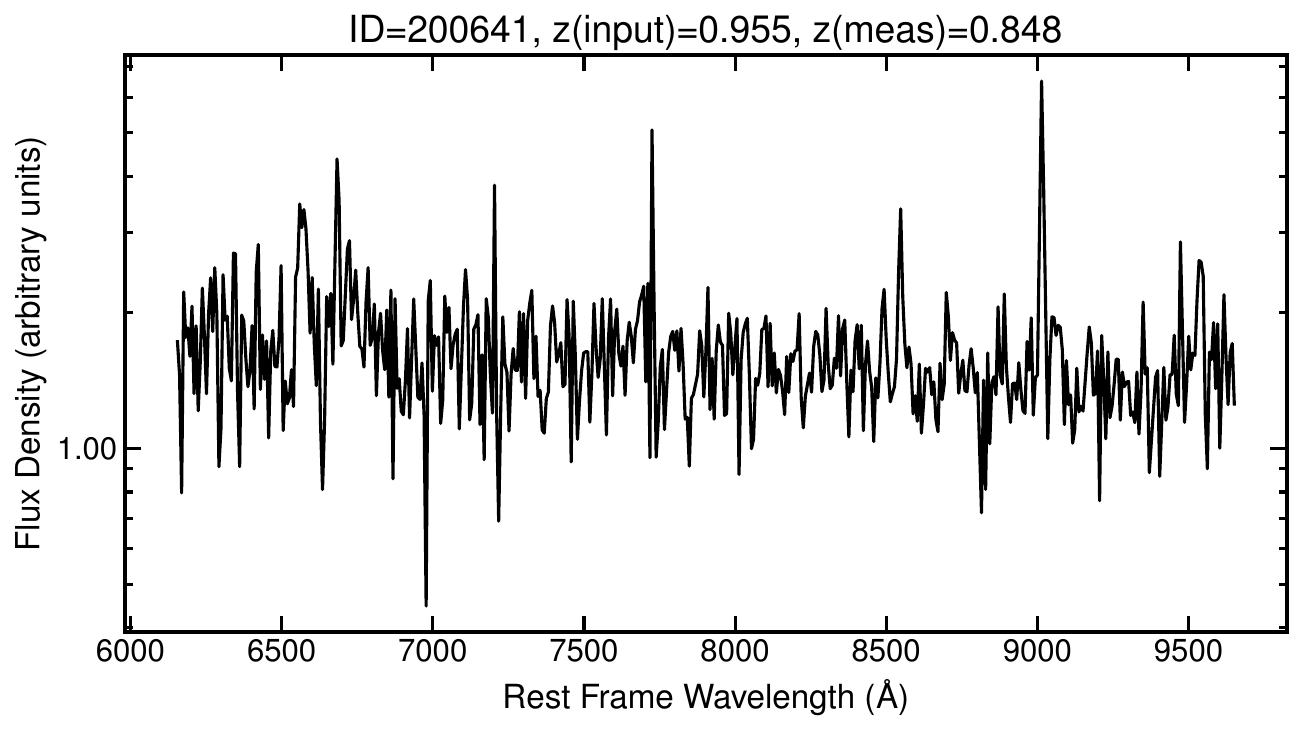}
 \includegraphics[width=0.49\textwidth,clip]{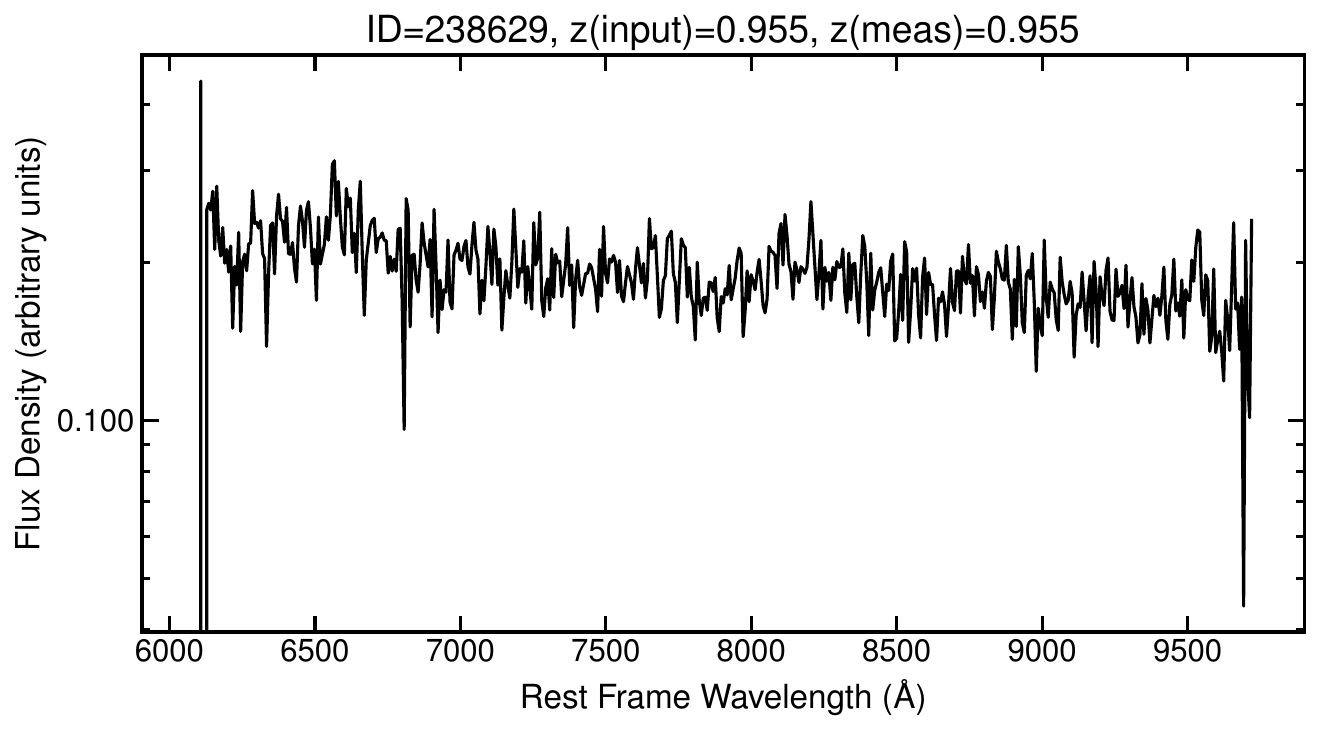}
 \includegraphics[width=0.49\textwidth,clip]{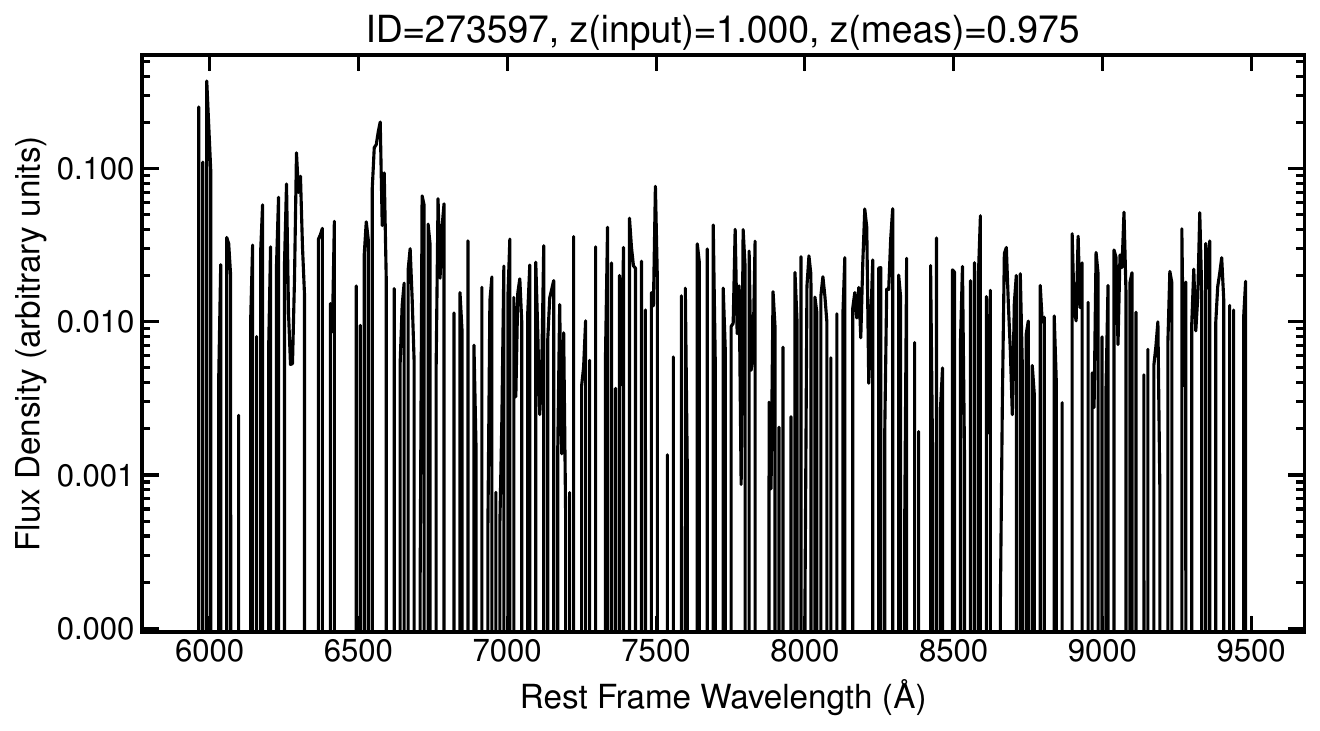}
 \end{center}
 \caption{Example of \Euclid-like simulated spectra for type 2 AGN in the redshift interval $0.89\leq z \leq 1.83$, thus where the \ion{H}{$\alpha$} complex is within the spectral coverage of the red grism. Spectra are shown at different SNR and noise levels (medium/high on the left and low on the right) for the \textit{good} (top row), \textit{false} (mid row), and \textit{lost} (bottom row) redshift classes (see Sect.~\ref{Redshift measurements}).}
 \label{fig:example2}
\end{figure*}

\reve{We present in Figure~\ref{fig:example1} and in Figure~\ref{fig:example2} a few examples of \Euclid-like simulated spectra for type 1 and type 2 AGN, respectively, in the redshift interval $0.89\leq z \leq 1.83$, thus where the \ion{H}{$\alpha$}$+$[\ion{N}{ii}] complex is within the spectral coverage of the red grism. Spectra are shown at different SNR and noise levels for the \textit{good}, \textit{false} and \textit{lost}, redshift classes (see Sect.~\ref{Redshift measurements}).}


\section{Line profile modelling}
\label{Line profile modelling}
\begin{figure*}
 \centering\includegraphics[width=0.49\textwidth,clip]{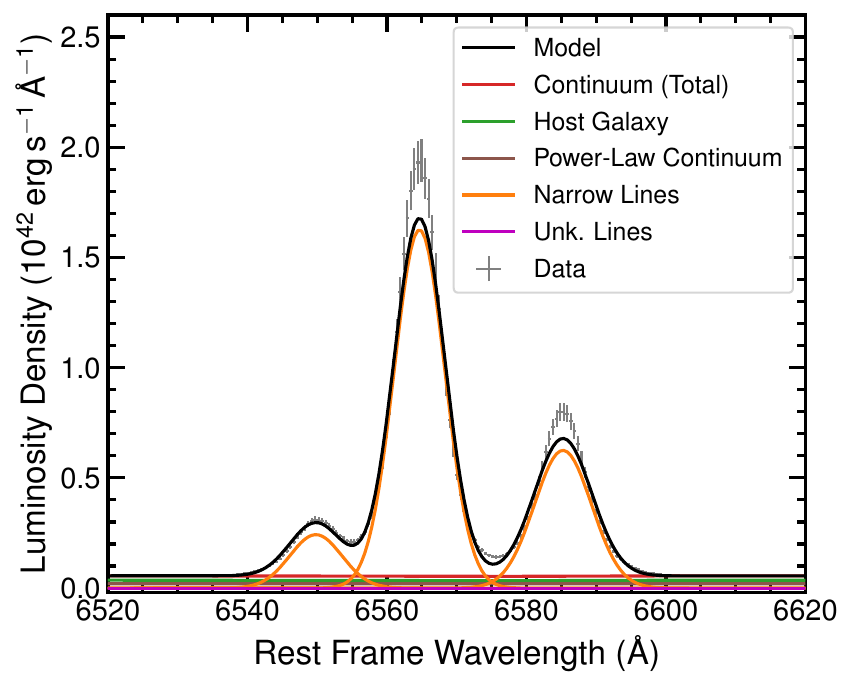}
 \includegraphics[width=0.49\textwidth,clip]{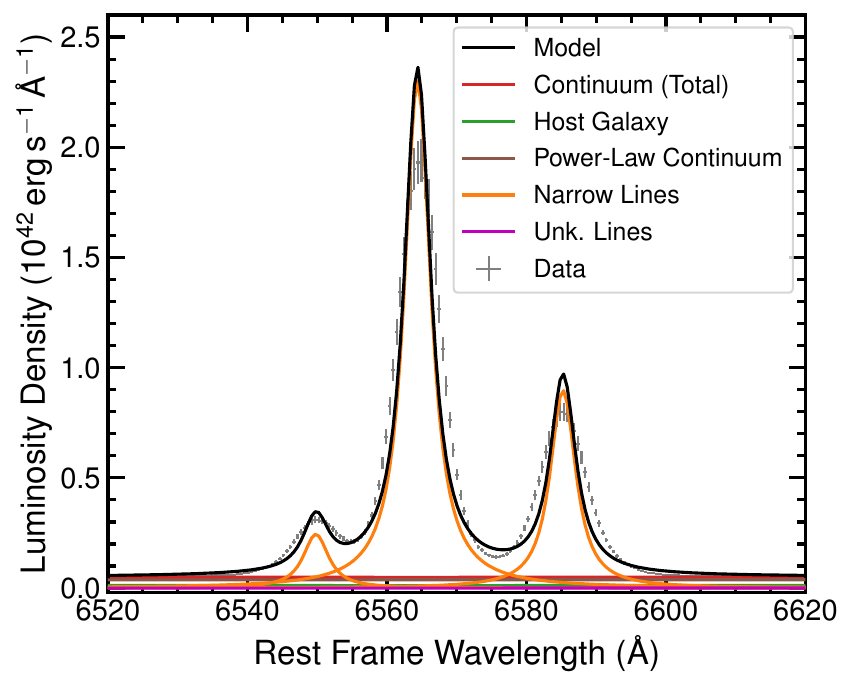}
 \includegraphics[width=0.49\textwidth,clip]{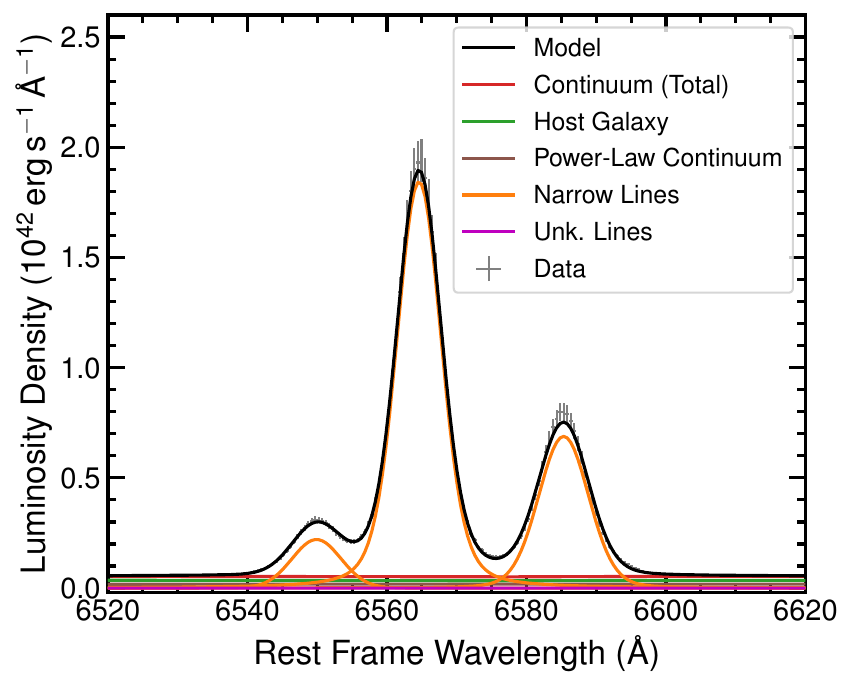}
 \caption{Example of a type 2 AGN incident spectrum \ion{H}{$\alpha$} complex fitted using Gaussian (\emph{left}), Lorentzian (\emph{centre}) and Voigt (\emph{right}) emission line profiles.}
 \label{fig:inc_t2_profiles}
\end{figure*}
\reve{To better understand whether we are capturing the entire line flux in the incident data, we test two spectral profiles (Lorentzian, and Voigt), in addition to the Gaussian one, to model known emission lines. 
Figure \ref{fig:inc_t2_profiles} shows an example for the \ion{H}{$\alpha$} complex fit  in the type 2 incident spectra with each of the profile options. Gaussian profiles underestimate the peak of the emission lines, while effectively fitting the base and core of the lines. Lorentzian profiles overestimate the tips of the emission lines and under-estimate the cores of the lines. Voigt profiles, which are a convolution of Gaussian and Lorentzian profiles, provide a middle ground where the majority of the line flux in the data is captured by the model. Yet, the tips of the emission lines are still not fully modelled with a Voigt profile. We postulate that this is due to the type 2 incident spectra emission lines being constructed with an AGN and starburst component (i.e., two Gaussians) and so cannot be perfectly modelled using a single line profile. We find that Voigt profiles capture the most flux presented in the data and are hence the best trade-off to be used for type 2 incident spectra emission line modelling. To be consistent with the simulated spectra that are insensitive to the profile employed, we restrict the comparison to the results obtained with the Gaussian one only. Yet, we caveat that a (small) fraction of the flux may be already missed at the level of the spectral modelling of the incident data depending upon the profile assumed.}

\section{Emission line flux from the OU-SPE automated spectral fitting}
\label{ap-ouspe}
\begin{figure*}
 \centering\includegraphics[width=0.48\textwidth,clip]{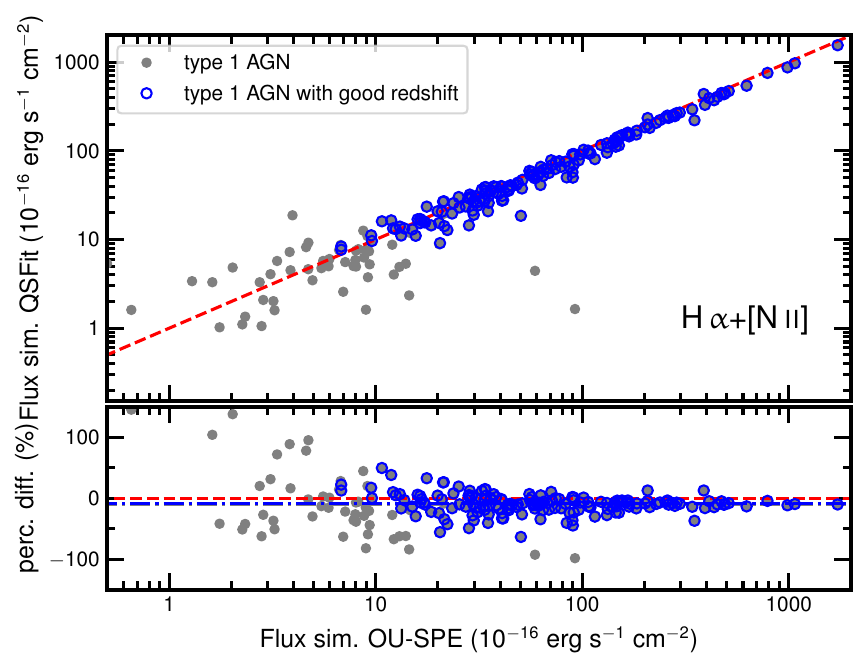}
 \includegraphics[width=0.50\textwidth,clip]{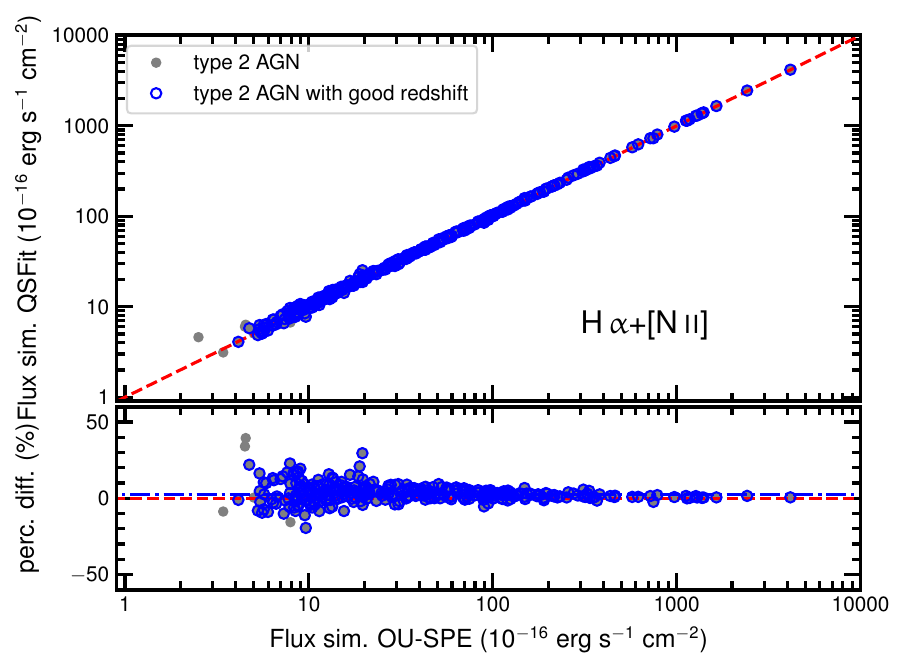}
 \caption{Comparison of the \ion{H}{$\alpha$}$+$[\ion{N}{ii}] flux measurements of the simulated spectra between \qsfit{} and OU-SPE. The results for the type 1 and type 2 AGN are shown in the \emph{left} and \emph{right} panels, respectively. Sources with a spectroscopic redshift measured from the simulated spectra and flagged as \textit{good} (see Sect.~\ref{Redshift measurements} for details) are marked with blue symbols. The bottom panel shows the percentage flux difference defined as $\Delta F/F=(F_{\rm H\alpha,\qsfit{}}-F_{\rm H\alpha,OU-SPE})/F_{\rm H\alpha,OU-SPE}$. The one-to-one flux relation is shown with a red dashed line, which is set to zero in the bottom panel. The dot-dashed blue line in the bottom panel represents the $50^{\rm th}$ percentile of the $\Delta F/F$ distribution.}
 \label{fig:ha_fcomp_12_qsfitouspe}
\end{figure*}
The Euclid collaboration is currently developing an automated pipeline with the main goal to compute spectral line measurements (e.g., emission lines, absorption lines, continuum breaks) in \Euclid spectra.
Briefly, this automatic routine computes line spectral parameters through Gaussian fitting with a non-linear least-squares minimization. 
The \ion{H}{$\alpha$} complex profile is modelled with three Gaussians, one for each feature (i.e., \ion{H}{$\alpha$} and [\ion{N}{ii}]$\lambda\lambda$6548, 6584), on top of a polynomial which accounts for the continuum. For this model, the [\ion{N}{ii}]$\lambda$6584 to [\ion{N}{ii}]$\lambda$6548 ratio is fixed to 3:1 and the FWHMs of the lines are assumed to be the same.
In the case of a broad profile (i.e., FWHM$>2000$\,km s$^{-1}$), the spectral fit is performed with two Gaussians only, one for the broad component of the \ion{H}{$\alpha$} and another for the narrow (i.e., no spectral components are considered for the [\ion{N}{ii}] doublet). At the time of writing, the pipeline provides tested results for the integrated flux of the \ion{H}{$\alpha$} emission line complex (flux uncertainties are currently not implemented). Spectral parameters and relative uncertainties for other emission lines are currently under analysis and will be delivered in the updated version of the pipeline.

Figure~\ref{fig:ha_fcomp_12_qsfitouspe} presents the comparison of the \ion{H}{$\alpha$} complex line flux between the \qsfit{} and the OU-SPE measurements. Fluxes are in very good agreement, with a slightly more scattered dispersion around the one-to-one for the type 1 AGN, as expected.  
The relative flux difference between the \qsfit{} and OU-SPE value is defined as $\Delta F/F=(F_{\rm H\alpha,\qsfit{}}-F_{\rm H\alpha,OU-SPE})/F_{\rm H\alpha,OU-SPE}$. The median $\Delta F/F$ are about $-9$\% and $3$\% for the type 1 and type 2 AGN, respectively.

\section{Comparison of the near-infrared spectral region of the type 1 composite  with the literature}
\label{ap:type1stackcomp}
\begin{figure*}
 \centering\includegraphics[width=0.7\textwidth,clip]{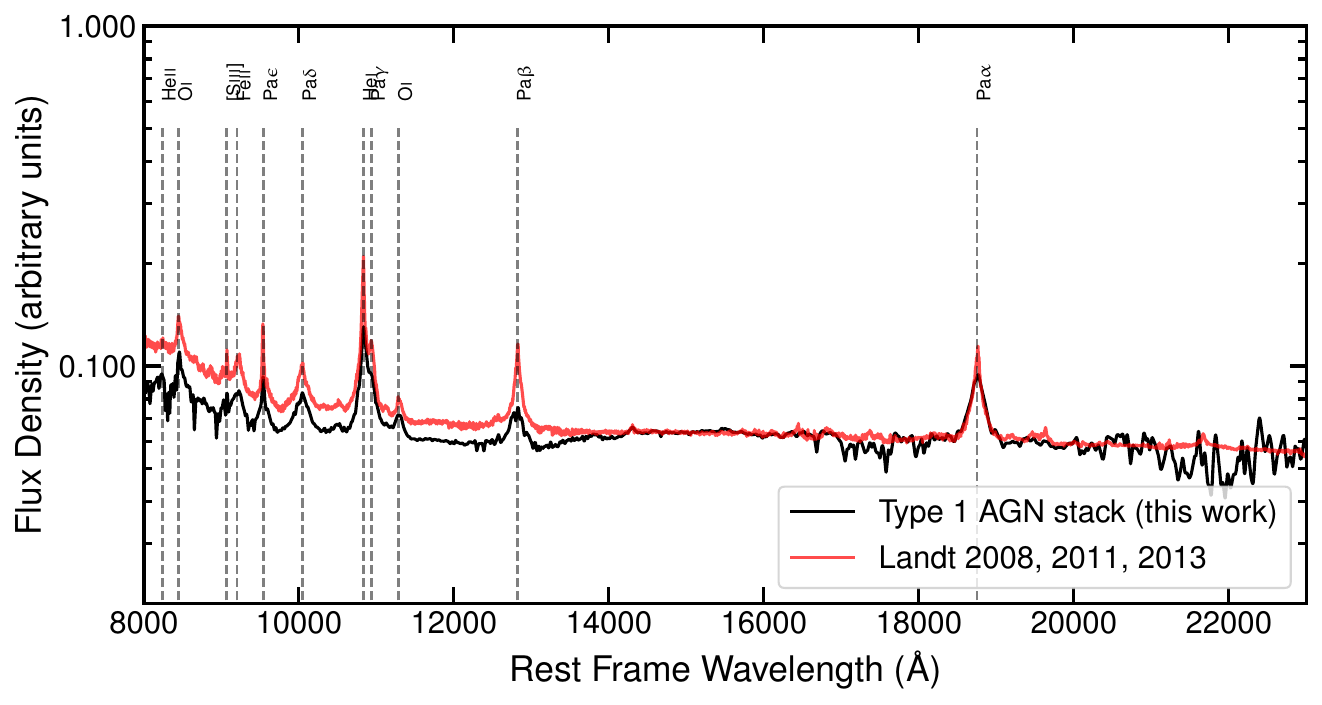}
 \caption{Comparison of the empirical composite \rev{incident} spectrum of type 1 AGN utilised in this work (shown with the black solid line) with the one we constructed from the datasets published by \citet[][in red]{landt2008,landt2011,landt2013}. The Landt et al. composite is normalised at the \num{16000}\,\AA\ flux of our stack. Main emission lines are marked.}
 \label{fig:stack1comp}
\end{figure*}

\begin{table}
\begin{center}
\caption{List of AGN used to construct the NIR composite. The SDSS AGN in the table are drawn from \citet{Glikman2006}, whilst the remaining are BAT AGN from \citet{Ricci22}.}
\label{tab:sources-nir}
\begin{tabular}{ l c }
\hline
Name & Redshift  \\ 
 \hline
   SDSS J000943.1-090839.2  &  0.210    \\ 
   SDSS J001327.3+005231.9  &  0.362   \\ 
   SDSS J005709.9+144610.1  &  0.172   \\ 
   SDSS J005812.8+160201.3  &  0.211   \\ 
   SDSS J010226.3-003904.6  &  0.295   \\ 
   SDSS J011110.0-101631.8  &  0.179   \\ 
   SDSS J015530.0-085704.0  &  0.165   \\ 
   SDSS J015910.0+010514.5  &  0.217   \\ 
   SDSS J015950.2+002340.8  &  0.163   \\ 
   SDSS J021707.8-084743.4  &  0.292   \\ 
   SDSS J024250.8-075914.2  &  0.378   \\ 
   SDSS J031209.2-081013.8  &  0.265   \\ 
   SDSS J032213.8+005513.4  &  0.185   \\ 
   SDSS J034042.93-073125.5 &  0.217   \\ 
   SDSS J073623.12+392617.8 &  0.118   \\ 
   SDSS J125519.6+014412.2  &  0.343   \\ 
   SDSS J150610.5+021649.9  &  0.135   \\ 
   SDSS J170441.3+604430.5  &  0.372   \\ 
   SDSS J172711.8+632241.8  &  0.218   \\ 
   SDSS J211843.2-063618.0  &  0.328   \\ 
   SDSS J212619.6-065408.9  &  0.418   \\ 
   SDSS J232235.6-094438.1  &  0.372   \\ 
   SDSS J234932.7-003645.8  &  0.279   \\ 
   SDSS J235156.1-010913.3  &  0.174   \\
   \hline
   \hline
   419                       &  0.034   \\  
   432                       &  0.060    \\ 
   441                       &  0.039   \\  
   447                       &  0.057   \\  
   488                       &  0.057   \\  
   689                       &  0.084   \\  
   1015                       &  0.056   \\  
   1079                      &  0.144   \\  
   1151                      &  0.060    \\  
\hline
\end{tabular}
\end{center}
\end{table}
\reve{The list of the sources used to build the NIR composite is given in Table~\ref{tab:sources-nir}.} We further investigated whether the NIR composite we included in the type 1 AGN main stack (Sect.~\ref{sec:spec_creation_1}) can be improved by including additional NIR spectra of quasars in the literature. For this purpose we collected all the cross-dispersed NIR spectra published by \citet{landt2008,landt2011,landt2013}. This data set consists on 29
well-known local type 1 AGN observed at the NASA infrared telescope facility (IRTF), a 3\,m telescope, and the 8\,m Gemini North Observatory, both located on Mauna Kea, Hawai’i \rev{(see Table 1 in \citealt{landt2014} for a summary of the sample properties)}. We note that the a few sources in their sample that display a continuum dominated by the host galaxy emission were not considered here. 
The data were corrected for Galactic absorption, shifted to the rest-frame and stacked in the same way as done for the SDSS spectra (see Sect.~\ref{Average spectrum construction}). The resulting composite is shown in Fig.~\ref{fig:stack1comp} with the red line \reve{(available online\footnote{\url{https://drive.google.com/drive/folders/1tjryMUkHhD10NjH_lhfeEBnvYRY4uib7?usp=sharing}})}. Our composite is flatter (redder) and it is more noisy around the Pa$\alpha$ emission line, but there is good agreement overall. We defer a more in-depth analysis of the rest-frame NIR region to future work. 

\section{Comparison of the semi-empirical composite spectrum of type 2 AGN with the literature}
\label{ap:type2stackcomp}
The composite \rev{incident} spectra utilised for the type 1 and the type 2 AGN have been computed in a very different way. 
The former have been derived from observed type 1 spectra of optically selected galaxies from the SDSS, while the latter from simulations (i.e. \spr) with empirical composites of different kind of galaxies (e.g., classical obscured AGN, star-bursts and star forming galaxies with different AGN fractions, see Sect.~\ref{sec:spec_creation_2} for details) considered as an input. The choice of using \spr\ is justified by the fact that it is more difficult to define a clean sample of type 2 AGN. Indeed, even if type 2 AGN display strong narrow high-ionisation emission lines (with ${\rm FWHM}<2000$\,km s$^{-1}$) and weak continua \citep{zakamska2003} in the rest-frame optical spectra, the literature works often refer to type 2 AGN as ``candidates'', as their selection strongly depends on the diagnostic (and thus the available emission lines) considered \citep[e.g.,][]{hao2005,reyes2008,alexandroff2013,mignoli2013}. 

\begin{figure*}
 \centering\includegraphics[width=0.7\textwidth,clip]{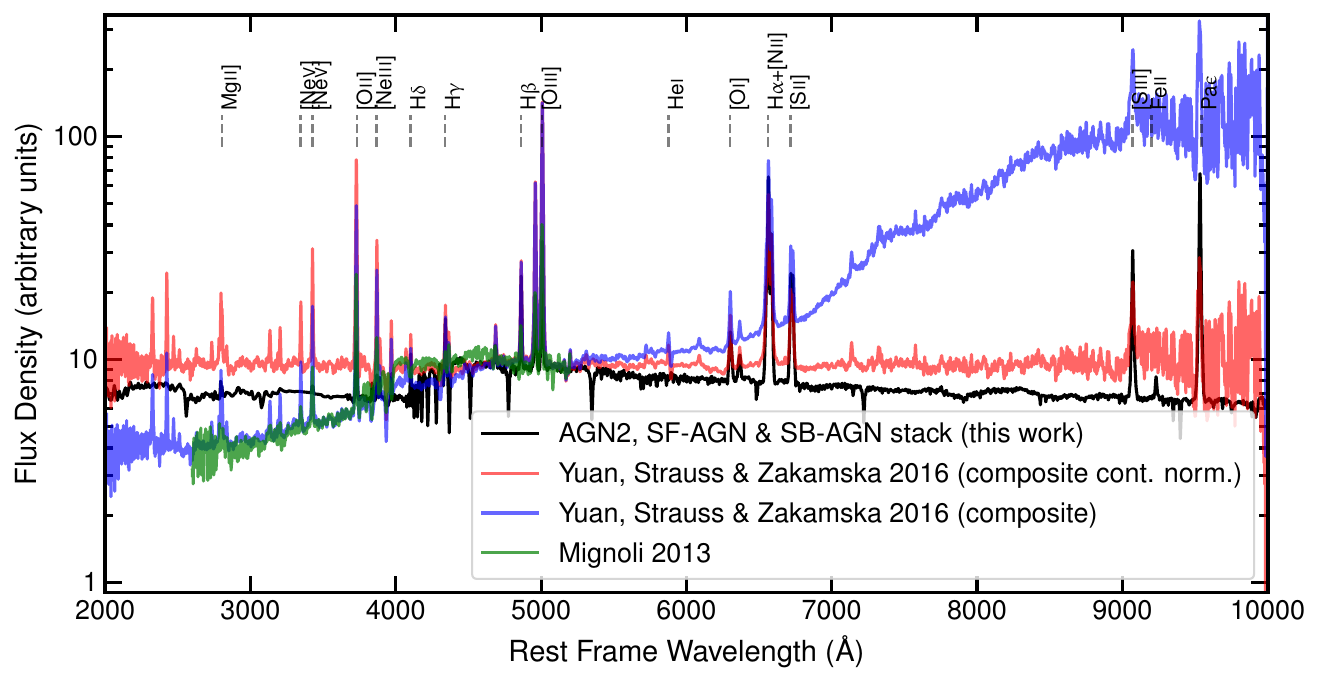}\\
 \centering\includegraphics[width=0.49\textwidth,clip]{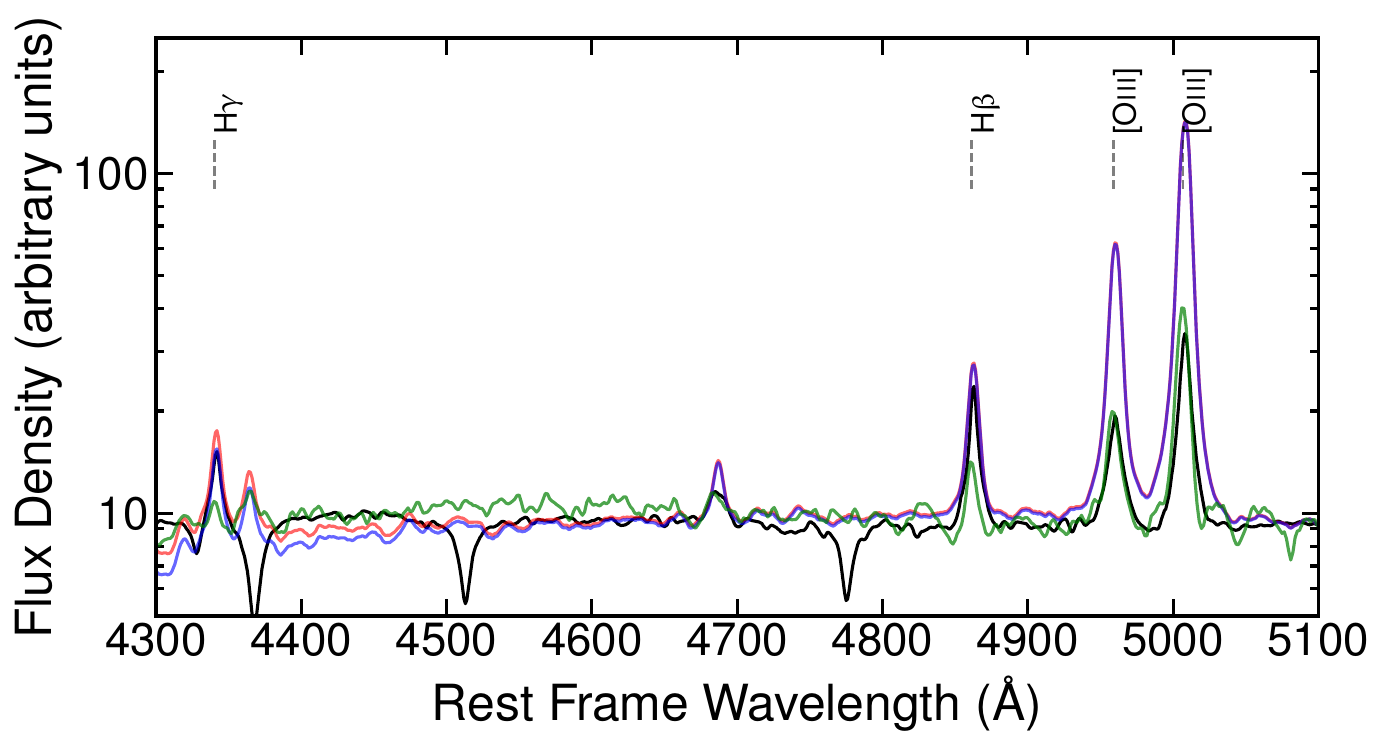}
 \centering\includegraphics[width=0.49\textwidth,clip]{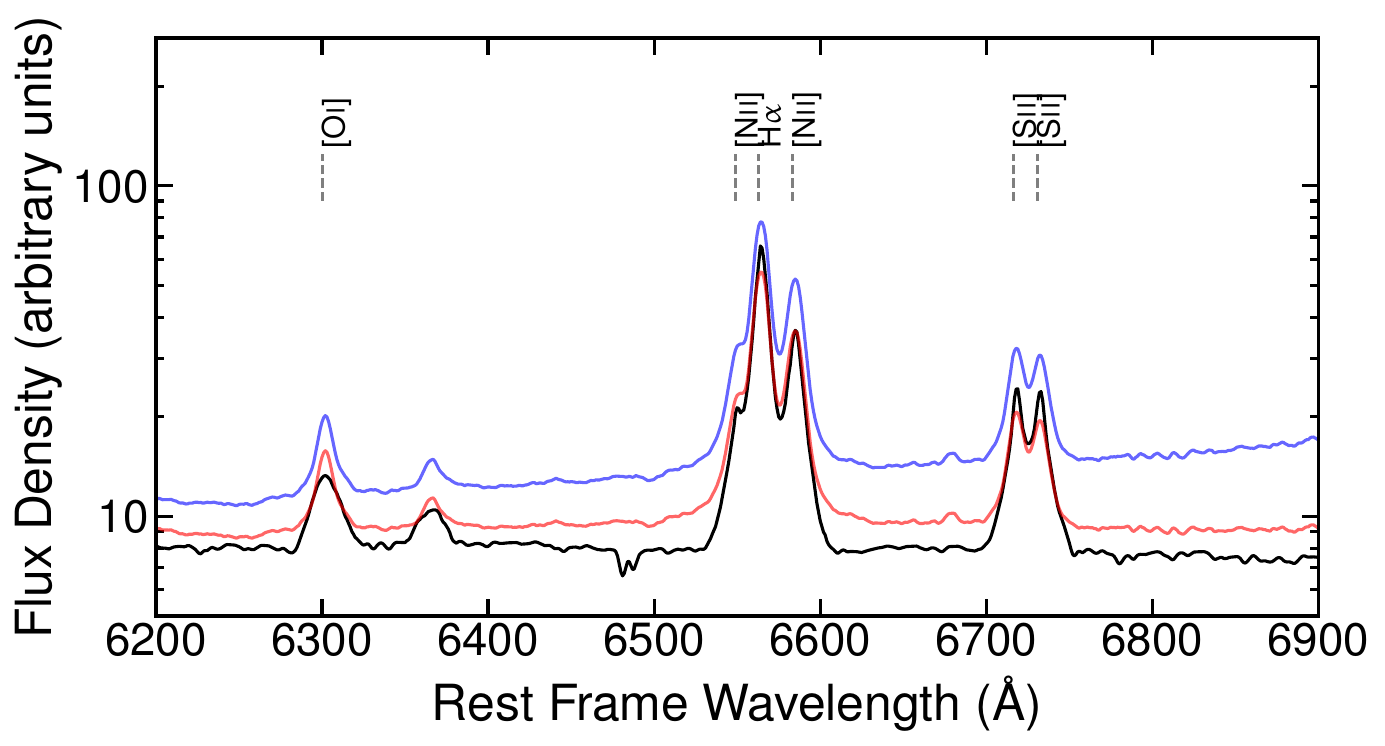}
 \caption{Comparison of the semi-empirical template \reve{incident} spectra of type 2 AGN utilised in this work (shown with the black solid line) with the empirical one published by \citet{ysk2016}. The blue line represents the composite spectrum of all BOSS type 2 quasars in the sample analysed by \citet{ysk2016}, whilst the red line is derived by normalising the composite spectrum to a continuum obtained by spline-interpolating between relatively line-free regions. The green line represents the spectral stack of 94 narrow-line AGN published by \cite{mignoli2013}. For sake of comparison, the literature composites are normalised at the 5100\,\AA\ flux of our stack. Main emission lines are marked. The bottom panels are two zoom-ins around the \ion{H}{$\beta$} (\emph{left}) and the \ion{H}{$\alpha$} (\emph{right}) the spectral regions. \reve{Stellar absorption features are also included and visible in our type 2 AGN spectrum between \ion{H}{$\gamma$} and \ion{H}{$\beta$} (see also Fig. A1 by \citealt{vg2017} and Figs. 9 e 10 by \citealt{bruzalcharlot2003}), which will be blended to the continuum in any moderate (low) resolution spectrum}.}
 \label{fig:stack2comp}
\end{figure*}

We thus want to test whether our type 2 composite derived from a semi-empirical library is consistent with what is currently available in the literature. 
Figure~\ref{fig:stack2comp} shows the comparison of the semi-empirical template spectra of type 2 AGN utilised in this work (shown with the black solid line) with the empirical one published by \citet{ysk2016}. The blue line represents the composite spectrum of all type 2 quasars in the baryon oscillation spectroscopy survey (BOSS) analysed by \citet{ysk2016}. The excess flux at $\lambda>6500$\,\AA\ is due to the host galaxy emission, which also seems to be dominant at bluer wavelengths. Therefore, the difference between the composite by \citet{ysk2016} and ours is mainly due to the different host galaxy. The red line is derived by normalising the composite spectrum by \citet{ysk2016} to a continuum obtained by spline-interpolating between relatively line-free regions. This stack preserves the equivalent widths of the emission lines, although the intrinsic shape of the continuum is lost (see their Fig. 15 and their Sect. 3.7). Overall, there is good consistency between the two composites regarding the \ion{H}{$\alpha$}+[\ion{N}{ii}] complex and the spectral region around the \ion{H}{$\beta$} and [\ion{O}{iii}]. There are instead some differences in the strength of the emission lines in the UV at $\lambda<4000$\,\AA, such as [\ion{O}{ii}], [\ion{Ne}{iii}] and \ion{H}{$\delta$}, with a few spectral features missing in our template (e.g., [\ion{Ne}{v}]). 

To this comparison we also included the composite published by \citet[][green line in Fig.~\ref{fig:stack2comp}]{mignoli2013} obtained from a sample of 94 narrow-line AGN in the redshift range $0.65 < z < 1.20$. Their sample is selected from the 20k-Bright zCOSMOS galaxy sample by the detection of the high-ionisation [\ion{Ne}{v}]$\lambda$3426 line. This composite shows strong, high-ionisation narrow lines, such as [\ion{Ne}{v}], that is not present in our stack, and a very faint \ion{Mg}{ii} emission line consistent with the emission seen in our composite.
We stress that these differences do not affect any result in our analysis, as our comparison is mainly focused on the \ion{H}{$\alpha$} spectral region rather than the UV and the overall shape of the continuum. 
In the future, we will implement a more data-driven composite for the type 2 AGN to improve the incident template library for type 2 AGN in the UV.

\section{Analysis of the flux loss}
\label{ap:fluxloss}
\begin{figure*}
 \includegraphics[width=\textwidth,clip]{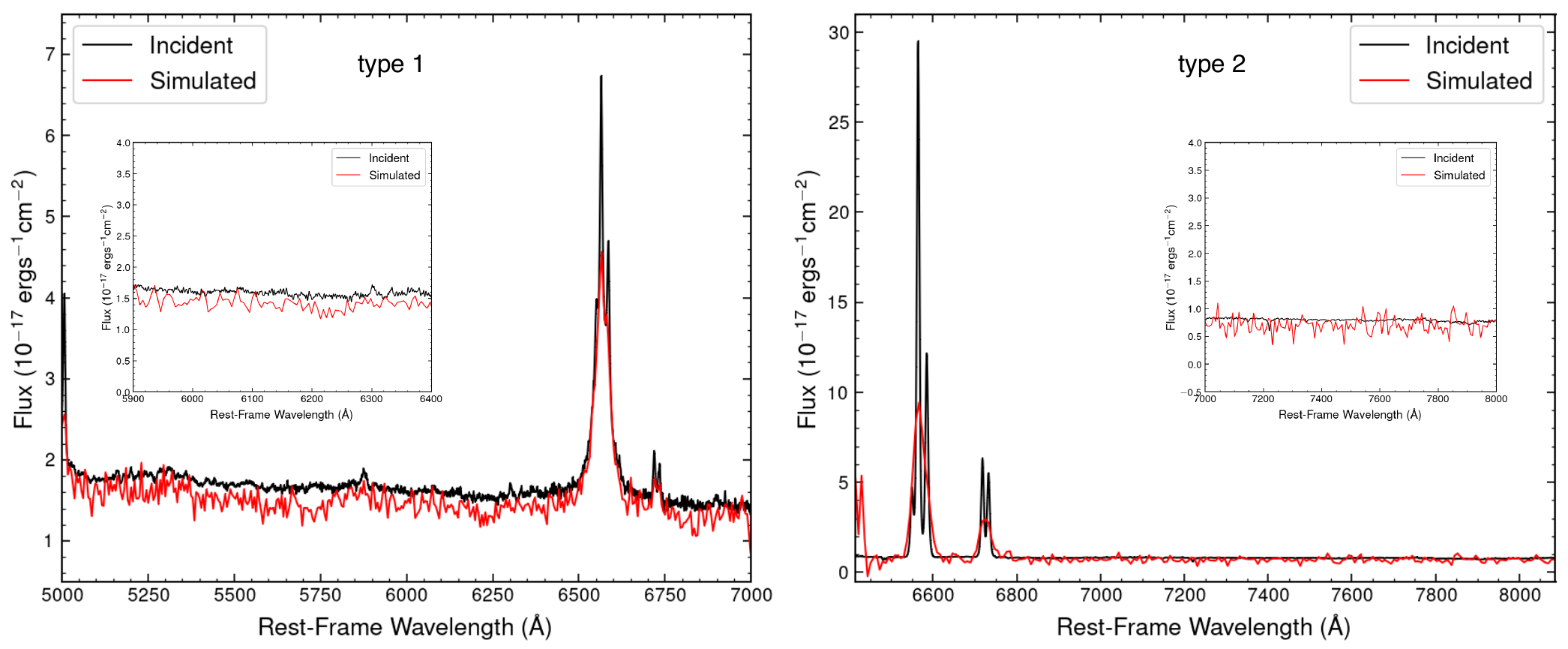}\\
 \centering\includegraphics[width=0.5\textwidth,clip]{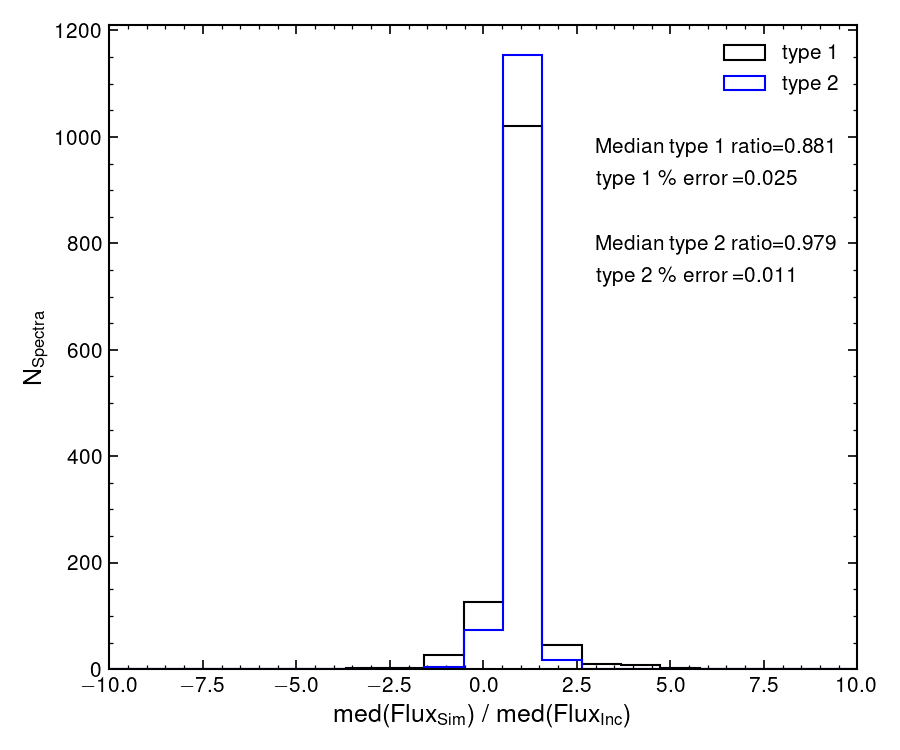}
 \caption{\textit{Top row}: Example of a comparison between the incident (black) and simulated (red) spectra for a type 1 (\emph{left}) and a type 2 (\emph{right}) AGN \reve{around the \ion{H}{$\alpha$} region}. The inset plot is a zoom-in of a continuum (emission line free) region. \textit{Bottom row}: distribution of the ratios of the median flux value (i.e., the median of the observed flux of the total spectrum) of the incident and simulated data for both type 1 (black) and type 2 (blue) AGN. The median values for the two distributions are also reported. The percentage error on the distributions of the spectrum flux median values are computed as $1.253\,\sigma/\sqrt{N}$, where $\sigma$ and $N$ represent the standard dispersion of the distribution and the number of objects, respectively.}
 \label{fig:incsim-spcomparison}
\end{figure*}

Here we present some additional test we performed to investigate whether a fraction of the missing flux could be lost at the slitless spectra extraction level \citep[see Sect. 5.1.3 in][for details]{gabarra2023}. We first compared the distribution of the ratios $r$ of the median observed flux values of the incident spectra $F_{\rm inc}$ and the simulated spectra $F_{\rm sim}$ (i.e., $r=\langle F_{\rm sim}\rangle / \langle F_{\rm inc}\rangle$, where the angle brackets $\langle\rangle$ represent medians);
see Fig.~\ref{fig:incsim-spcomparison}. The median of this distribution is about 12\% for type 1 AGN and 2\% for type 2 AGN, with a percentage error on the distributions of spectrum flux median values (defined as $1.253\,\sigma/\sqrt{N}$, where $\sigma$ and $N$ represent the standard dispersion of the distribution and the number of objects, respectively) at the 2.5\% level for type 1 and 1\% for type 2. As an example, we present a comparison between the incident (black) and simulated (red) spectra for a representative type 1 (left) and a type 2 (right) AGN at the top row of Fig.~\ref{fig:incsim-spcomparison}. The inset plot is a zoom-in of a continuum (emission line free) region. The overall flux of the simulated spectra are systematically lower for both the type 1 and 2 AGN.
For completeness, we also carried out the same exercise of comparing median luminosity density values for the best-fit models of the incident and simulated spectra to their respective data [i.e. ratio = median(data) / median(model)], finding that the \qsfit\ best-fit model is \rev{close to the data, on average, with the biggest discrepancy for type 1 simulated spectra at $\simeq$2\% level}. 
This analysis implies that flux loss could partly explain the remaining fraction of flux missing in the comparison between simulated and incident flux values, although most of the difference on the \ion{H}{$\alpha$}$+$[\ion{N}{ii}] emission between the simulated and incident data is due to the assumptions made in the spectral modelling of the emission lines. Nonetheless, the OU-SIR team is working on the spectral extraction pipeline and flux losses will be alleviated by the optimal extraction routine that is implemented in the upcoming version of the OU-SIR pipeline.

\end{document}